\documentclass[11pt,headings=big,numbers=noenddot,DIV=14,a4paper]{article}

\pdfoutput=1

\usepackage[margin=1 in]{geometry}

\usepackage{graphicx}
\usepackage{dcolumn}
\usepackage{bm,relsize}
\usepackage{amsfonts,amsmath,amssymb,amsthm,mathtools}
\usepackage{slashed}
\usepackage{placeins}
\usepackage{adjustbox}
\usepackage[normalem]{ulem}
\usepackage{lscape}
\usepackage{multirow}
\usepackage{lipsum, color,braket}
\usepackage[usenames,dvipsnames,svgnames,table]{xcolor}
\usepackage[linktoc=page,bookmarks=false,colorlinks=true,linkbordercolor=RoyalBlue,citebordercolor=ForestGreen,urlbordercolor=CornflowerBlue]{hyperref}
\hypersetup{
    colorlinks=true,
    linkcolor=ForestGreen,
    filecolor=magenta,      
    urlcolor=ForestGreen,
    citecolor=ForestGreen,
}
\usepackage[sort&compress,numbers,merge]{natbib}
\usepackage{breakcites}

\usepackage{tikz-feynman}

\definecolor{myBrown}{RGB}{153, 102, 51}
\definecolor{myLightBrown}{RGB}{179, 124, 65}
\definecolor{mygray}{RGB}{133, 124, 155}
\definecolor{mygreen}{RGB}{109, 138, 112}

\def\beq{\begin{equation}}
\def\eeq{\end{equation}}
\def\bea{\begin{eqnarray}}
\def\eea{\end{eqnarray}}

\newcommand{\Tnuc}{T_{\rm nuc}}

\def\LO			{{\mathsmaller{\text{LO}}}}
\def\NLO			{{\mathsmaller{\text{NLO}}}}

\def\LL		{{\mathsmaller{\text{LL}}}}
\def\R		{{\mathsmaller{\text{R}}}}
\def\T		{{\mathsmaller{\text{T}}}}

\begin{document}

\begin{flushright}
\footnotesize
DESY-21-147\\
IFT-UAM/CSIC-21-146
\end{flushright}
\color{black}

\begin{center}

{\LARGE \bf
Friction pressure on relativistic bubble walls
}

\medskip
\bigskip\color{black}\vspace{0.5cm}

{
{\large Yann Gouttenoire},$^{a,b}$
{\large Ryusuke Jinno},$^{a,c}$
{\large Filippo Sala}$^{d}$
}
\\[7mm]

{\it \small $^a$ Deutsches Elektronen-Synchrotron DESY, Notkestr. 85, 22607 Hamburg, Germany}\\
{\it \small $^b$ School of Physics and Astronomy, Tel-Aviv University, Tel-Aviv 69978, Israel}\\
{\it \small $^c$ Instituto de F\'{\i}sica Te\'orica UAM/CSIC, C/ Nicol\'as Cabrera 13-15, Campus de Cantoblanco,}\\
{\it \small 28049, Madrid, Spain}\\
{\it \small $^d$ LPTHE,  CNRS \& Sorbonne Universit\'e, 4 Place Jussieu, F-75252, Paris, France}\\
\end{center}

\bigskip

\centerline{\bf Abstract}
\begin{quote}
\color{black}

During a cosmological first-order phase transition, particles of the plasma crossing the bubble walls can radiate a gauge boson.
The resulting pressure cannot be computed perturbatively for large coupling constant and/or large supercooling. We resum the real and virtual emissions at all leading-log orders, both analytically and numerically using a Monte-Carlo simulation.
We find that radiated bosons are dominantly soft and that the resulting retarding pressure on relativistic bubble walls is linear both in the Lorentz boost and in the order parameter, up to a log.
We further quantitatively discuss IR cut-offs, wall thickness effects, the impact of various approximations entering the calculation, and comment on the fate of radiated bosons that are reflected.

\end{quote}

\clearpage
\noindent\makebox[\linewidth]{\rule{\textwidth}{1pt}} 
\tableofcontents
\noindent\makebox[\linewidth]{\rule{\textwidth}{1pt}} 	

\section{Introduction}

Cosmological 1st-order Phase Transitions (1stOPTs) have been the subject of a lot of interest because they can generate Gravitational Waves (GWs) \cite{Witten:1984rs,Hogan:1986qda,Kamionkowski:1993fg}, set the abundance of dark matter~\cite{Konstandin:2011dr,Falkowski:2012fb,Hambye:2013dgv,Hambye:2018qjv,Bai:2018dxf,Baker:2019ndr,Azatov:2021ifm,Baldes:2020kam,Hong:2020est,Asadi:2021pwo,Baldes:2021aph}, explain the baryon asymmetry \cite{Kuzmin:1985mm,Cohen:1993nk,Konstandin:2011ds,Servant:2014bla,Katz:2016adq,Cline:2020jre,Dorsch:2021ubz,Azatov:2021irb,Baldes:2021vyz,Baker:2021zsf}, generate primordial black holes~\cite{Kodama:1982sf,Hawking:1982ga,Gross:2021qgx,Baker:2021nyl,Baker:2021sno,Kawana:2021tde}, primordial magnetic fields~\cite{Vachaspati:1991nm}, or topological defects~\cite{Kibble:1976sj,Kibble:1995aa,Borrill:1995gu} and offer a new access to supersymmetry breaking~\cite{Craig:2020jfv}. 
A key quantity for the physics of these transitions, and in turn for most of their applications, is the wall velocity $v_w $, which is set by the friction pressure on the bubble walls. 
For example, the primordial GW spectrum resulting from a 1stOPT~\cite{Caprini:2015zlo,Cai:2017cbj,Caprini:2019egz,Caprini:2018mtu,Mazumdar:2018dfl,Hindmarsh:2020hop} strongly depends on whether the walls run away or reach a constant terminal velocity before colliding:
in the first case GWs are sourced by scalar field gradient, while in the latter they are sourced by fluid motion~\cite{Caprini:2015zlo,Caprini:2019egz}. 

In the non-relativistic regime, the bubble wall velocity is usually calculated by assuming local thermal equilibrium in a thermal field theoretic or hydrodynamic approach \cite{Ignatius:1993qn, Moore:1995si,Moore:2000wx,John:2000zq,Megevand:2009ut,Megevand:2009gh,Espinosa:2010hh,Huber:2013kj, Dorsch:2018pat,Friedlander:2020tnq, Balaji:2020yrx, BarrosoMancha:2020fay,Ai:2021kak}. In the opposite limit where bubble walls are ultra-relativistic, $\gamma = 1/\sqrt{1-v_w^2} \gg 1$, interactions between particles crossing the wall can be safely neglected \cite{Bodeker:2009qy,Bodeker:2017cim,Vanvlasselaer:2020niz,Hoche:2020ysm} when computing the friction pressure.
This limit is most relevant for 1stOPTs with large supercooling, which are natural predictions of nearly-conformal potentials, and which have attracted a lot of interest in recent literature (e.g. \cite{Kolb:1979bt, Creminelli:2001th,Randall:2006py,Nardini:2007me, Espinosa:2008kw,Konstandin:2010cd, Konstandin:2011dr, Jinno:2016knw,Marzola:2017jzl,Iso:2017uuu,Chiang:2017zbz,vonHarling:2017yew,Fujikura:2019oyi,Megias:2018sxv, Arunasalam:2017ajm,Bruggisser:2018mus,Bruggisser:2018mrt,Baratella:2018pxi,Hambye:2018qjv,Baldes:2018emh,Prokopec:2018tnq,Brdar:2018num,Marzo:2018nov,Ellis:2019oqb,Agashe:2019lhy,Agashe:2020lfz,DelleRose:2019pgi,vonHarling:2019gme,Bloch:2019bvc,Baldes:2020kam,Azatov:2021ifm,Azatov:2021irb,Baldes:2021aph} and \cite{Gouttenoire:2022gwi} for a review).

In this work we make progress on the computation of the friction pressure on relativistic bubble walls. 
In Sec.~\ref{sec:LO_friction_pressure}, as a warm-up we compute the retarding pressure coming from particles acquiring a mass in the broken phase, as initially derived in BM '09 \cite{Bodeker:2009qy}, and recover their results.
In Sec.~\ref{sec:NLO_pressure}, we review the possibility for the incoming particle to radiate one gauge boson acquiring a mass $m$ in the broken phase, and again recover the result from BM '17 \cite{Bodeker:2017cim}, $\mathcal{P}_{\NLO} \propto  g^2\,m\,\gamma \, T_{\rm nuc}^3$, where $\Tnuc$ is the bubble nucleation temperature, $g$ is the gauge coupling constant and $\mu$ is the IR cut-off on the transverse momentum of the emitted boson.
In Sec.~\ref{sec:IR_cut_off}, we discuss various possibilities to cure the IR logarithmic divergence.
In Tab.~\ref{tab:tablePE}, we show that for large supercooling $m /T_{\rm nuc} \gg 1$ and/or large gauge coupling constant $\alpha$, perturbativity breaks down and we must account for the possibility to radiate multiple vector bosons.

In Sec.~\ref{sec:all_order}, we perform an analytical Sudakov resummation of virtual and real emissions at all leading-log orders (LL). The perturbative splitting probability can then be used to express the mean exchanged momentum, and thus the pressure, as a resummed quantity that includes leading-log real and virtual corrections to all orders, see Eq.~\eqref{eq:Deltap_resummation}.
In the limit where the initial energy is large, $E_a/m \gg 1$, such that the kinematics of the multiple emissions can be considered as independent, we find that the pressure is linear in both the wall Lorentz factor $\gamma$ and in the order parameter of the transition $m$, up to a log.
In Sec.~\ref{sec:MC}, we simulate a particle shower using a Monte-Carlo algorithm, and we confirm the analytical resummation in the limit $E_a/m \gg 1$, in which energy depletion due to multiple emissions can be neglected. 
In Sec.~\ref{sec:bubble_velocity}, we deduce the bubble wall Lorentz factor at collision time and discuss the consequences on the nature of the GW source.
We conclude in Sec.~\ref{sec:conclusion}.

To keep our paper easier to read, we defer a lot of technical details and complementary - yet interesting - calculations to a series of appendices.
In App.~\ref{app:vertex} we compute the vertex function for boson emission, in App.~\ref{app:validity_approximations}  we compute the particle mode functions precisely, and discuss the validity of the thin-wall and relativistic-soft-collinear limits, in App.~\ref{sec:3to2_scattering} we compute the $3\to 2$ boson scattering rate, in App.~\ref{app:fate_reflected_particles} we sketch the effect of particles reflected multiple times, in App.~\ref{app:massless_scenario} we treat the case of a massless vector boson, in App~\ref{app:comment_hoeche} we discuss the discrepancy of our result with the one of \cite{Hoche:2020ysm} that found a scaling $\mathcal{P}_{\LL} \propto \gamma^2 T_{\rm nuc}^2$, in App.~\ref{app:backreaction} we compute the backreaction of successive boson emissons on the kinematics of the parent particle, and in App.~\ref{app:run_away_Lorentz_factor} we compute the Lorentz factor of a constantly accelerating bubble wall.
 
\section{LO friction pressure}
\label{sec:LO_friction_pressure}

In this work, we assume that the wall is moving at relativistic velocities $\gamma \gg 1$, such that we can work in the so-called ballistic regime in which we can neglect the interaction between neighboring particles during the time when they cross the wall, see e.g. \cite{BarrosoMancha:2020fay}.
The leading-order pressure comes from the particle getting a mass in the broken phase \cite{Bodeker:2009qy}. 
\begin{equation}
\label{eq:PLO_def}
\mathcal{P}_{\LO} =\sum_a g_a \int \frac{d^3 p}{(2\pi)^3}\,f_a(p) \times \Delta p_{\LO},
\end{equation}
where $g_a$ is the number of internal degrees of freedom. The momentum change is given by
\begin{equation}
\Delta p_{\LO} = E  -\sqrt{E-\Delta m^2} \simeq  \frac{\Delta m^2}{2E},
\label{eq:pLO_momentum}
\end{equation}
where we assumed energy conservation and the relativistic limit $E \gg \Delta m$.
We compute 
\begin{equation}
\int _0^\infty \frac{4\pi p^2 dp}{(2\pi)^3} \frac{1}{e^{p/T_{\rm nuc}} \pm 1} \frac{\Delta m^2}{2p} = \frac{\Delta m^2\,T_{\rm nuc}^2}{24} \times \left\{
                \begin{array}{ll}
                 1 \quad \text{boson} \\
                  \frac{1}{2}\quad \text{fermion}
                \end{array}
              \right.
\end{equation}
where $T_{\rm nuc}$ is the nucleation temperature.
Therefore 
\begin{equation}
\mathcal{P}_{\LO} =\sum_a g_a c_a \, \frac{\Delta m^2\,T_{\rm nuc}^2}{24},\label{eq:PLO_1}
\end{equation}
where $c_a = 1~(1/2)$ for bosons (fermions).  If the negative pressure due to the vacuum energy difference of the phase transition $\Delta V$ is larger than the LO retarding pressure in Eq.~\eqref{eq:pLO_momentum},
\begin{equation}
\Delta V > \mathcal{P}_{\LO},
\end{equation}
then the bubble wall is supposed to be accelerated to larger and larger $\gamma$ factors until either it collides with other walls or until soft particle radiation further contributes to the pressure and eventually stops the wall from accelerating.
The rest of the paper is dedicated to the computation of the latter contribution.

\section{The splitting probability at first order}
\label{sec:NLO_pressure}

\subsection{Transition splitting}

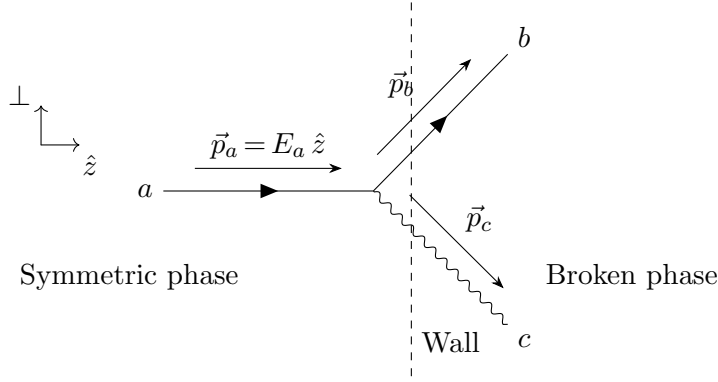
\begin{figure}[t]
\centering
\begin{adjustbox}{max width=1\linewidth,scale=1,center}
\begin{tikzpicture}

\begin{feynman}
\vertex (a) {\(a\)};
\vertex [right=3 cm of a] (b);
\vertex [above right=2.5cm of b] (c) {\(b\)};
\vertex [below right=2.5cm of b] (d)  {\(c\)};
\vertex [right=0.5cm of b] (w);

\diagram* {
 (a) -- [fermion, momentum=\( \vec{p}_{a}\, {=} \,E_{a}\,\hat{z}\)] (b),
 (b) -- [fermion, momentum=\(\vec{p}_b\)] (c),
 (b) -- [boson, momentum=\(\vec{p}_c\)] (d),
};
\end{feynman}

\node [left=1cm of a] (a0) {};
\node [above=1cm of a0] (e) {};
\node [below=0.5cm of e, minimum width=0 pt,inner sep=0pt] (f) {};
\node [right=0.5 of f] (g) {};
\draw [->,shorten <= -0.8pt] (f) -- (e) node[left] {$\perp$};
\draw [->,shorten <= -0.3pt] (f) -- (g) node[below] {$\hat{z}$};

\node [above=2.5cm of w] (b1) {};
\node [below=2.5cm of w] (b2) {};
\draw [dashed] (b1) -- (b2) node[pos=0.9,right] {Wall};

\node [below=1cm of b] (b3) {};
\node [left=1.5cm of b3] (ps) {Symmetric phase};
\node [right=2cm of b3] (ph) {Broken phase};

\end{tikzpicture}
\end{adjustbox}
\caption{\it \small \label{fig:transition_splitting}
NLO contribution to the retarding pressure: while approaching the bubble wall, an incoming particle $a$ radiates a vector boson $c$ which gets a mass in the confined phase.
}
\end{figure}

In this section we discuss another contribution to the pressure, which arises when a particle entering the wall radiates a gauge boson which gets mass in the broken phase \cite{Bodeker:2017cim}.
The resulting NLO pressure reads
\begin{equation}
\mathcal{P}_\NLO =\sum_a g_a \int \frac{d^3 p_a}{(2\pi)^3}\,f_a(p_a)\,\frac{p_{a}^z}{p_{a}^0} \times \sum_{b,c} \int dP_{a \to bc} \times \Delta p \times [1\pm f_c(p_c)][1\pm f_{b}(p_b)],
\label{eq:PNLO_formula}
\end{equation}
with
\begin{equation}
\Delta p = p_a^z - p_b^z - p_c^z,
\end{equation}
where $dP_{a \to bc}$ is the differential splitting probability, $p_a$ and $p_b$ are the momenta of the incoming particle before and after the splitting while $p_c$ is the momentum of the radiated boson,\footnote{
Note that our notation for `$a$', `$b$' and `$c$' is different from \cite{Bodeker:2017cim} where the roles of `$b$' and `$c$' are interchanged.
}
see Fig.~\ref{fig:transition_splitting}.
The momentum $p_a$ is the momentum in the far past and thus in the symmetric phase, while $p_b$ and $p_c$ are the momenta in the far future and thus in the broken phase if they are transmitted, or in the symmetric phase if they are reflected by the wall boundary.
We summed over all the species $a$ likely to participate in the process, $g_a$ being their number of degrees of freedom.
The Pauli blocking or Bose enhancing factors $1 \pm f_{b}$ are $\simeq 1$, while $1\pm f_c$ sum to $1$ when considering both absorption and emission processes.\footnote{\label{footnote:Caputo}
The interaction Hamiltonian can be written as (see e.g.~\cite{Caputo:2018vmy})
\begin{equation}
H_{\rm int} = \mathcal{M}_0 a_c^\dagger a_b^\dagger a_a + \rm h.c.,
\end{equation}
where $a_x$ are the usual creation operators in Fock space. Then the transition amplitudes for emission and absorption read, respectively
\begin{align}
&\mathcal{M}_{a \to bc}=\left<f_a-1, f_b +1, f_c+1| H_{\rm int} | f_a, f_b, f_c \right> = \mathcal{M}_0 \sqrt{f_a}\sqrt{1\pm f_b} \sqrt{1 + f_c}, \\
&\mathcal{M}_{bc\to a}=\left<f_a+1, f_b -1, f_c-1| H_{\rm int} | f_a, f_b, f_c \right> = \mathcal{M}_0 \sqrt{1\pm f_a}\sqrt{ f_b} \sqrt{f_c},
\end{align}
where $+/-$ refers to boson/fermion statistic.
We deduce the interaction rate accounting for both  emission and absorption
\begin{equation}
|\mathcal{M}_{a \to bc}|^2 - |\mathcal{M}_{bc\to a}|^2 = |\mathcal{M}_0|^2 \left[ f_a(1 \pm f_b) + f_c(f_a -f_b) \right].
\end{equation}
Hence we see that as long as $(f_a - f_b) f_c \ll f_a$, we have
\begin{equation}
|\mathcal{M}_{a \to bc}|^2 - |\mathcal{M}_{bc\to a}|^2 \simeq |\mathcal{M}_0|^2 f_a.
\end{equation}
As we will see, for non-abelian theories we will cut $f_c \sim 1/g^2$, so our calculation in those cases holds good as long as $|f_a - f_b| \lesssim g^2$. We leave a more quantitative study for future works.
}

\subsection{The momentum exchange}

\paragraph{Kinematics.}

Upon introducing 
\begin{equation}
x \equiv \frac{E_{c}}{E_a},
\end{equation}
we can write
\begin{align}
&p_a = \left( E_{a}, \, 0 \,\hat{x},\, \sqrt{E_{a}^2-m_a(z)^2}\,\hat{z} \right),\label{eq:pa}\\
&p_b = \left((1-x) E_{a}, \,-k_{\perp}\,\hat{x},\, \sqrt{(1-x)^2E_{a}^2 -m_b(z)^2- k_{\perp}^2}\,\hat{z} \right),\label{eq:pb}\\
&p_c = \left( x\,E_{a}, \,k_{\perp}\,\hat{x},\, \sqrt{x^2E_{a}^2 -m_c(z)^2- k_{\rm \perp}^2}\,\hat{z} \right),\label{eq:pc}
\end{align}
where $m_a(z)$, $m_b(z)$ and $m_c(z)$ are the masses of the three particles involved in the vertex as a function of $z$, and $k_\perp \equiv |\vec{k}_\perp|$. 
As the wall breaks $z$-translation invariance, the momentum along $\hat{z}$ is not conserved. We approximate the wall by a Heaviside function at $z=0$ 
\begin{equation}
m_a(z) =\left\{
                \begin{array}{ll}
                 m_{a,s} \quad ~\text{if }z<0, \\[0.1cm]
                 m_{a,h} \quad ~\text{if }z\geq 0,
                \end{array}
              \right.\quad
              m_b(z) =\left\{
                \begin{array}{ll}
               m_{b,s} \quad ~\text{if }z<0, \\[0.1cm]
               m_{b,h} \quad ~\text{if }z\geq 0,
                \end{array},
              \right.\quad
              m_c(z) =\left\{
                \begin{array}{ll}
                 m_{c,s} \quad ~\text{if }z<0, \\[0.1cm]
                m_{c,h} \quad ~\text{if }z\geq 0,
                \end{array}
              \right.
\end{equation}
where the masses with `$s$' and `$h$' denote the ones at infinity in the symmetric and Higgs phase, respectively. 
As long as the masses in the symmetric phase are small compared to the ones in the broken phase, we can safely assume that 
\begin{equation}
m_{a,s}=m_{b,s} =0.
\end{equation}
We can not make the same simplification for $m_c$ because $m_{c,s} =0$ implies the existence of a double (soft and collinear) divergence, see Eq.~\eqref{eq:dP_abc_mcs}.
Hence, $m_{c,s}$ plays the role of an IR cut-off which regulates the double singularity, whose possible values are discussed in Sec.~\ref{sec:IR_cut_off}.

\begin{figure}
\begin{center}
\includegraphics[width=0.7\columnwidth]{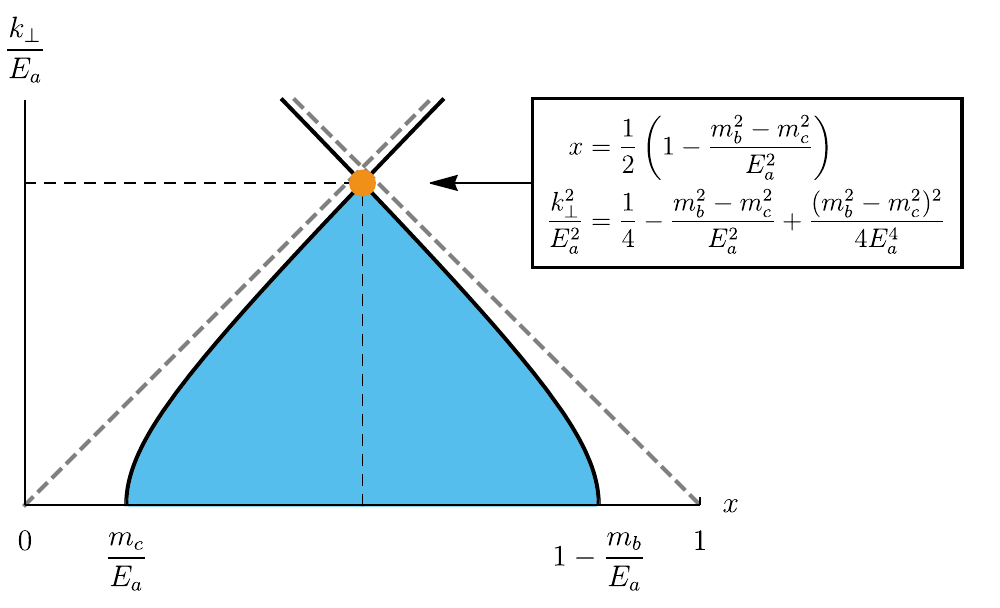}
\caption{\small
Kinematically allowed region (blue) for $x$ and $k_{\perp} / E_a$ for one emitted boson.
}
\label{fig:allowed}
\end{center}
\end{figure}

Note that the parameterization of the kinematics in Eq.~\eqref{eq:pa}, \eqref{eq:pb} and \eqref{eq:pc}, imposes
\begin{align}
\frac{\sqrt{k_{\perp}^2 + m_c(z)^2}}{E_{a}}~ \leq ~ &x ~ \leq ~1- \frac{\sqrt{k_{\perp}^2 + m_b(z)^2}}{E_{a}} , \label{eq:range_x_0}\\
0~\leq ~& k_{\perp}^2 ~ \leq~ \frac{E_a^2}{4} - \frac{m_b(z)^2+m_c(z)^2}{2} + \frac{(m_b(z)^2 - m_c(z)^2)^2}{4E_a^2},
\label{eq:range_K_perp_0}
\end{align} 
and the corresponding allowed region for $x$ and $k_{\perp}$ is shown in Fig.~\ref{fig:allowed}.
Since all the results derived in this work are UV insensitive, we can simplify the upper boundaries in Eq.~\eqref{eq:range_x_0} and \eqref{eq:range_K_perp_0} as
\begin{align}
\label{eq:range_x}
\frac{\sqrt{k_{\perp}^2 + m_c(z)^2}}{E_{a}}~ \leq ~ &x ~ \leq ~1 ,\\
0 ~\leq ~& k_{\perp} ~ \leq~ E_{a}.
\label{eq:range_x_K_perp}
\end{align} 
The associated correction terms are anyway beyond the soft-collinear approximation, that we will assume when deriving the phase of the mode function in Eq.~\eqref{eq:A_def} and the vertex function in Eq.~\eqref{eq:vertex_BM}.

\subsection{The splitting probability}
\paragraph{Relation between splitting probability and matrix element.}

The differential splitting probability is given by \cite{Bodeker:2017cim}
\begin{equation}
\int dP_{a \to bc} \equiv \int  \frac{d^3 p_b}{(2\pi)^3 2E_b}  \frac{d^3 p_c}{(2\pi)^3 2E_c} \left< \phi | \mathcal{T} |p_b,p_c\right> \left< p_b,p_c | \mathcal{T} |\phi \right>,
\end{equation}
with the properly normalized state $\left|\phi\right>$ for the incoming particle $a$ being defined as
\begin{equation}
\left|\phi\right> \equiv \int \frac{d^3\vec{p}}{(2\pi)^3} \frac{\phi(\vec{p})}{2E}\left|\vec{p}\right>, \qquad \int \frac{d^3\vec{p}}{(2\pi)^3} \frac{|\phi(\vec{p})|^2}{2E} = 1,
\end{equation}
with $\left|\vec{p}\right> = \sqrt{2E_{\vec{p}}}\,a_{\vec{p}}^\dagger \left|0\right>$.
We assume $\phi(\vec{p})$ to be sharply localized around $\vec{p} = \vec{p}_a$.
The transition element ${\cal T}$ can be written in terms of the matrix element ${\cal M}$ as
\begin{equation}
\left< \vec{p}_b, \vec{p}_c | {\cal T} | \vec{p}_a \right>
=\int d^4x  \left< p_b,p_c | \mathcal{H}_{\rm int} |p_a \right>=
\delta^{(2)} \left( \sum \vec{p}_\perp \right)
\delta \left( \sum E \right)
{\cal M}.
\end{equation}
Substituting this in the above expression, we get
\begin{align}
\int dP_{a \to bc}
&=
\int_{\vec{p}_b} \frac{1}{2E_b}
\int_{\vec{p}_c} \frac{1}{2E_c}
\int_{\vec{p}_a'} \frac{1}{2E_a'}
\int_{\vec{p}_a''} \frac{1}{2E_a''}
~
\phi (\vec{p}_a') \phi^*(\vec{p}_a'')
\,\notag\\
&\hspace{0.5cm}
\times
\delta^{(2)} \left( \sum \vec{p}_\perp' \right)
\delta \left( \sum E' \right)
~
\delta^{(2)} \left( \sum \vec{p}_\perp'' \right)
\delta \left( \sum E'' \right)
~
{\cal M} {\cal M}^*.
\end{align}
Here $\delta^{(2)} \left( \sum \vec{p}_\perp'' \right)$ can be eliminated by the perpendicular direction in $\vec{p}_a''$ integration.
Also, $\delta \left( \sum E'' \right)$ can be eliminated by the remaining direction in $\vec{p}_a''$ integration, yielding an extra factor of $(E_a'' / p_a''^z)$
\begin{align}
\int dP_{a \to bc}
&=
\int_{\vec{p}_b} \frac{1}{2E_b}
\int_{\vec{p}_c} \frac{1}{2E_c}
\int_{\vec{p}_a'} \frac{|\phi (\vec{p}_a')|^2}{2E_a'}
~
\frac{1}{2p_a'^z}
~
\delta^{(2)} \left( \sum \vec{p}_\perp' \right)
\delta \left( \sum E' \right)
~
|{\cal M}|^2.
\end{align}
Next we use the fact that $\phi$ is sharply localized around $\vec{p}_a$
\begin{align}
\int dP_{a \to bc}
&=
\int_{\vec{p}_b} \frac{1}{2E_b}
\int_{\vec{p}_c} \frac{1}{2E_c}
~
\frac{1}{2p_a^z}
~
\delta^{(2)} \left( \sum \vec{p}_\perp \right)
\delta \left( \sum E \right)
~
|{\cal M}|^2. \label{eq:M2_final_LIPS}
\end{align}
We finally perform $\vec{p}_b$ integration.
Eliminating the perpendicular $\delta$ function is trivial.
The remaining $\delta$ function can be performed by using $p_b^z dp_b^z = E_b dE_b$.
We also decompose $\vec{c}$ integration into $\vec{k}_\perp$ and $E_c$ integrations. 
As a result, we get
\begin{align}
\int dP_{a \to bc}
&=
\int \frac{d^2 k_{\perp}}{(2 \pi)^2} \int \frac{dE_c}{2 \pi}
~
\frac{1}{2p_a^z}
\frac{1}{2p_b^z}
\frac{1}{2p_c^z}
~
|{\cal M}|^2.
\label{eq:dP_abc_0}
\end{align}
The sums over momenta assume asymptotic initial and final states, which thus are far from the wall and do not see the Lorentz violation it induces. This Lorentz violation instead enters the computation of the amplitude in the next paragraph.

\paragraph{Matrix element and vertex function.}

We next evaluate the matrix element 
\begin{equation}
\label{eq:M_matrix}
\mathcal{M} \equiv \int dz ~ \chi_{a}(z) V(z) \chi_{b}^*(z) \chi_{c}^*(z),
\end{equation}
where $\chi_{a,\,b,\,c}$ are the mode functions of particles $a$, $b$, and $c$, respectively, and $V(z)$ is the vertex function. 
It has been pointed out \cite{Bodeker:2017cim} that the most important process contributing to the pressure at large $E_a$ is $X(p_a) \to V_T(p_c)~ X(p_b)$ where $V_T$ is a transverse vector boson and $X$ can be a fermion, a scalar and a boson.\footnote{The computation of the vertex function for the emission of a longitudinal vector boson possibly involves subtleties related to the breaking of Lorentz invariance at the wall boundary and the non-applicability of the Ward Identity. We then leave it for future work, and focus on transverse components in this paper. \label{footnote:longi}}
The corresponding vertex function, which we re-derive in App.~\ref{app:vertex}, is phase-independent, $V_h=V_s$, and equal to
\begin{equation}
|V|^2 = 4\,g^2\,C_{abc}\,\frac{k_{\perp}^2}{x^2},\label{eq:vertex_BM}
\end{equation}
where $g$ is the gauge coupling constant and $C_{abc}$ is the corresponding charge factor \cite{Hoche:2020ysm}. 

\paragraph{Mode functions.}
\label{par:mode_functions}

In App.~\ref{app:mode_function}, we show that in the high-energy limit $p_c^z \gg m_{c,h}$ and thin-wall limit, we can approximate the mode function of $c$ by 
\begin{equation}
\chi_{c}(z) \simeq  {\rm exp} \left(i \int_0^z p_c^z(z') dz' \right) \simeq e^{i E_{c} z} \exp \left(  - \frac{i}{2E_{c}} \int_0^z(m_{c}^2(z')+k^2_\perp)~ dz' \right),
\label{eq:chi_z}
\end{equation}
and idem for $a$ and $b$, which allows to re-write the triple wave function overlap as a function of a phase-dependent quantity $A$,
\begin{equation}
\chi_a(z)\chi_b^*(z)\chi_{c}^*(z) = {\rm  exp} \left( \frac{i}{2E_a} \int_0^z ~A(z')~dz' \right)\,,
\label{eq:M_matrix_1p5}
\end{equation}
with 
\begin{equation}
-A \simeq m_a(z)^2 -\frac{m_b(z)^2 + k_{\perp}^2}{1-x}- \frac{m_c(z)^2 + k_{\perp}^2}{x}  \simeq -\frac{m_c(z)^2 + k_{\perp}^2}{x}.
\label{eq:A_def}
\end{equation}
We have assumed relativistic and collinear final momenta $p_b^z \gg \sqrt{m_b^2+k_\perp^2}$, $p_c^z \gg \sqrt{m_c^2+k_\perp^2}$ in the first equality and soft emission energy $x \equiv E_{c} /E_a \ll 1$ in the last equality. 

We can now separate the integral over $z$ across the wall in Eq.~\eqref{eq:M_matrix} into a contribution from the broken phase and a contribution from the symmetric phase. 
Therefore, we assume that the vertices $V$ and phases $A$ on each side of the wall are Heaviside functions and we denote them by ($V_h$, $A_h$) and ($V_s$, $A_s$), such that we obtain
\begin{equation}
\mathcal{M} \simeq V_s \int_{-\infty}^0 dz\,{\rm exp} \left( iz\frac{A_s}{2E_a} +\epsilon z \right) + V_h \int_{0}^{\infty} dz\,{\rm exp} \left( iz\frac{A_h}{2E_a}- \epsilon z \right) = 2iE_a \left(  \frac{V_h}{A_h} - \frac{V_s}{A_s} \right).
\label{eq:M_matrix_2}
\end{equation}
We have regulated the behavior at infinity by introducing an imaginary momentum $\pm i\epsilon$, which we can safely set to $0$ in the right-hand side of Eq.~\eqref{eq:M_matrix_2} since $A_h, A_{s}\neq 0$.

In App.~\ref{app:finite_wall_thickness}, we derive the mode function $\chi_c(z)$ in the presence of a finite wall thickness $L_{\rm w}$ and we show that it can be neglected up to logarithmic corrections in the limit, cf. Eq.~\eqref{eq:wall_thickness_matters}
\begin{equation}
L_{\rm w} \ll   m_{c,h}/m_{c,s}^2.
\label{eq:wall_thickness_matters_main}
\end{equation}
Since we expect $L_{\rm w}  \sim m_{c,h}^{-1}$, cf. Eq.~\eqref{eq:wall_thickness}, we conclude that the wall thickness can be neglected as soon as $m_{c,s} \ll m_{c,h}$.
Note that the validity of using the Heaviside function is not given by the comparison between the (inverse) wall thickness and the momentum of individual $a$, $b$, or $c$ particles.

In App.~\ref{sec:step_potential_1D}, we compute the full mode function valid for any $p_c^z$.
Particularly, $\chi_c(z)$ contains a second wave propagating in the opposite direction which implies that in the limit $p_c^z \lesssim m_{c,h}$, the particle $c$ is reflected.
The full matrix   $\mathcal{M}$ accounting for reflection and transmission coefficients is given in Eq.~\eqref{eq:M_matrix_mode_function}.

\paragraph{Perturbative splitting probability.}
\label{sec:Perturbative_splitting_probability}

Therefore, the matrix element in Eq.~\eqref{eq:M_matrix_2} becomes
\begin{align}
|\mathcal{M}|^2 &\simeq 4E_a^2\times 4\,g^2\,C_{abc}\,\frac{k_{\perp}^2}{x^2} \times \frac{x^2\,(m_{c,h}^2-m_{c,s}^2)^2}{(k_{\perp}^2+m_{c,s}^2)^2(k_{\perp}^2+m_{c,h}^2)^2}\notag \\
&\simeq 16\,g^2 \,C_{abc} \,E_{a}^2 \,\frac{k_{\perp}^2\,(m_{c,h}^2-m_{c,s}^2)^2}{(k_{\perp}^2+m_{c,s}^2)^2(k_{\perp}^2+m_{c,h}^2)^2}. \label{eq:MSquare_main}
\end{align}
The splitting probability in Eq.~\eqref{eq:dP_abc_0} reduces to
\begin{align}
dP_{a \to bc} &=\frac{d^2 k_{\perp}}{(2\pi)^2 }  \frac{d x}{(2\pi) 2x} \frac{1}{(2E_a)^2}16\,g^2 \,C_{abc} \,E_{a}^2 \,\frac{k_{\perp}^2\,(m_{c,h}^2-m_{c,s}^2)^2}{(k_{\perp}^2+m_{c,s}^2)^2(k_{\perp}^2+m_{c,h}^2)^2}\notag \\
&=\zeta_a\frac{d k_{\perp}^2}{k_{\perp}^2}  \frac{d x}{x}~\Pi(k_\perp),
\label{eq:dP_abc_mcs}
\end{align}
where $\Pi(k_{\perp})$ contains the IR and UV suppression factors
\begin{equation}
\Pi(k_{\perp}) \equiv \left(\frac{k_{\perp}^2}{k_{\perp}^2+m_{c,s}^2}\right)^2 \left(\frac{m_{c,h}^2-m_{c,s}^2}{k_{\perp}^2+m_{c,h}^2}\right)^2, \label{eq:IR_UV_suppression_factor}
\end{equation}
and
\begin{equation}
\zeta_a \equiv \frac{\alpha}{\pi} \sum_{b,c} C_{abc}, \qquad  \qquad \alpha  \equiv \frac{g^2}{4\pi}. \label{eq:zeta_a_def}
\end{equation}
Here note that we integrated the radial direction of $\vec{k}_\perp$ as $d^2 k_\perp = \pi d k_\perp^2$. The charge factors $C_{abc}$ in the SM can be found in \cite{Hoche:2020ysm}. In the rest of the paper we assume $m_{c,h}\gg m_{c,s}$. 

In Eq.~\eqref{eq:dP_abc_mcs}, we have replaced $p_i^z$ with $i=a,b,c$ in Eq.~\eqref{eq:dP_abc_0} by $E_i$.
In App.~\ref{app:beyond-relativistic-limit}, we investigate the validity of the relativistic-soft-collinear approximation for the phase space factor $1/p_i^z$, the vertex function $V$, the phase $A$ of the mode function and the momentum exchange $\Delta p$ and we show that this underestimates the final result by a few percents only for $T_{\rm nuc} = 10^{-2} T_{\rm start}$.

\paragraph{Symmetry restoration.}

The UV suppression factor $\left(\dfrac{m_{c,h}^2-m_{c,s}^2}{k_{\perp}^2+m_{c,h}^2}\right)^2$ in Eq.~\eqref{eq:IR_UV_suppression_factor}, compared to usual splitting functions in collider context \cite{Altarelli:1977zs}, vanishes in the limit where the symmetry is restored $m_{c,h} \to m_{c,s}$ or $k_{\perp}^2 \gg m_{c,h}^2$.
In contrast, as first claimed by \cite{Vanvlasselaer:2020niz} the splitting function used by \cite{Hoche:2020ysm} does not go to zero in the limit where the distinction between the two phases disappears.
We give more details on the origin of this discrepancy in App.~\ref{app:comment_hoeche}.

\subsection{The IR cut-off}
\label{sec:IR_cut_off}

The splitting probability Eq.~(\ref{eq:dP_abc_mcs}) is divergent in the IR, for $k_\perp^2 \to 0$.
That divergence is regulated by values of $m_{c,s} > 0$, via the factor $\left(k_{\perp}^2/(k_{\perp}^2+m_{c,s}^2)\right)^2$, or by some other physical process.
To encompass both possibilities, we find it convenient to define a general IR cutoff $\mu$ and rewrite Eq.~(\ref{eq:dP_abc_mcs}) as
\begin{align}
dP_{a \to bc} &= \zeta_a\frac{d k_{\perp}^2}{k_{\perp}^2}  \frac{d x}{x} \left(\frac{k_{\perp}^2}{k_{\perp}^2+\mu^2}\right)^2\left(\frac{m_{c,h}^2-m_{c,s}^2}{k_{\perp}^2+m_{c,h}^2}\right)^2,
\label{eq:dP_abc}
\end{align}
where by definition
\beq
\mu \geq m_{c,s}.
\eeq
This definition gives us also the possibility to treat $\mu$ as a free parameter, so that one could later account for IR cut-offs which are not relevant for the physical situation of interest for this paper, or which we simply miss.

Let us start by discussing the case $\mu = m_{c,s}$, for which Eq.~(\ref{eq:dP_abc}) goes back to Eq.~(\ref{eq:dP_abc_mcs}).
Values of $m_{c,s} > 0$ are guaranteed by screening effects in the plasma \cite{Arnold:2002zm,Arnold:2003zc},
\begin{equation}
m_{c,s}^2 \simeq \sum_i 2 g_i \frac{g^2 C_i}{d_A} \int \frac{d^3 p_i}{2|\vec{p}_i|(2\pi)^3} f_i(\vec{p}_i),\label{eq:screeening_mass_def}
\end{equation}
with the sum running over all species in the plasma $i$ that couple to $c$, $f_i$ being the occupation number of the $i$ particle, $g_i$ the number of relativistic degrees of freedom of species $i$, $C_i$ the quadratic Casimir ($g^2 C_i$ = charge squared for abelian theories) and $d_A$ the dimension of the adjoint representation.
We discuss two possible screening effects in the next two paragraph, and later discuss other possible cut-offs for $k_\perp$.

\paragraph{Thermal mass.}

A contribution to $f$ in Eq.~(\ref{eq:screeening_mass_def}) is always given by the particles in thermal equilibrium. Taking a $U(1)$ and an $SU(N)$ gauge theories as examples, we obtain
\begin{align}
&U(1):\qquad \quad m_{c,s}^2 =\frac{N_f}{6} g^2 T_{\rm nuc}^2, \label{eq:mcs_U1}\\
&SU(N):\qquad m_{c,s}^2 =\frac{1}{6}\left(N + \frac{N_f}{2}  \right) g^2 T_{\rm nuc}^2, \label{eq:mcs_SUN}
\end{align}
where $N_f$ is the number of Dirac fermion flavors in the fundamental representation.
The thermal mass in Eq.~\eqref{eq:mcs_U1} and \eqref{eq:mcs_SUN}, which for the sake of simplicity we write as
\begin{equation}
m_{c,s} \simeq m_{\rm th} \equiv \alpha^{1/2} T_{\rm nuc}, \label{eq:mcs_thermal}
\end{equation}
constitutes the minimal IR cut-off for abelian and non-abelian gauge theories.

\paragraph{Phase-space saturation.}
\label{sec:phase_space_sat}

As the occupation number of emitted vector bosons grows in the IR, it must exist a scale $m_{\rm sat}$ below which perturbation theory breaks down and vector bosons start to act collectively \cite{Bodeker:2017cim}.\footnote{
We thank Dietrich Bodeker and Guy Moore for very useful discussions which helped us writing this section.
}
Due to the soft-collinear divergence of the splitting function, the occupation number $f_c(p_c)$ grows like
\begin{align}
f_c(p_c) &=\sum_a g_a \int \frac{d^3 p_a}{(2\pi)^3}f_a(\vec{p}_a)\,\frac{dP_{a \to bc}}{d^2k_{\perp} dp_c^0} \notag\\
& \simeq \sum_a \nu_a g_a\zeta_a \frac{\zeta(3)}{\pi^2} \gamma T_{\rm nuc}^3 \frac{1}{\pi k_{\perp}^2}  \frac{1}{xE_a}~\Pi(k_{\perp}),
\label{eq:occupation_number}
\end{align}
where we have again used $dp_c^0 \simeq dp_c^z$  in the first equality and Eq.~\eqref{eq:dP_abc_mcs} in the second one, $g_a$ is the number of relativistic degrees of freedom of particle $a$ and $\nu_a = 1~(3/4)$ for bosons (fermions) assuming $f_a$ to be a thermal distribution.
From Eq.~\eqref{eq:screeening_mass_def} and \eqref{eq:occupation_number}, we can see that in the case of non-abelian gauge theories (without loss of generality, we focus on $SU(N)$ here and for numerical applications we fix $N=3$), in the large $\gamma$ limit, the emitted vector bosons back-react self-consistently on the screening mass
\begin{align}
m_{c,s}^2 &\simeq 2N g^2 \int \frac{d^3 p_c}{2|\vec{p}_c|(2\pi)^3} f_c(\vec{p}_c)  \label{eq:mcs_first_eq}\\
&\simeq 2N g^2\sum_a \nu_a g_a\zeta_a \frac{\zeta(3)}{\pi^2} \gamma T_{\rm nuc}^3 \frac{1}{(2\pi)^32E_a}\int_0^{E_a^2} \frac{d k_{\perp}^2}{ k_{\perp}^2}  \int_{\frac{\sqrt{k_{\perp}^2 +m_{\rm sat}^2}}{E_a}}^{1}\frac{d x}{x^2}~\Pi(k_{\perp}),
\end{align}
which implies
\begin{equation}
m_{c,s}^3 \simeq m_{\rm sat}^3
\equiv \frac{2\zeta(3)}{3\pi^5}N\alpha^2 \gamma T_{\rm nuc}^3 \sum_{a,b,c} \nu_a g_a \,C_{abc} .
\end{equation}
Note that $m_{\rm sat}$ increases as $\gamma^{1/3}$.
Its value at the time of bubble collision depends on whether the wall runs away or reaches a terminal velocity.
Upon plugging typical values of the parameters, we obtain\footnote{
Note that if the screeening mass due to gluon collective behavior, $m_{\rm sat}$ in Eq.~\eqref{eq:kperp_satur}, becomes larger than the mass of free gluons in the broken phase $m_{c,h}$, then the logarithmic divergence in the splitting probability in Eq.~\eqref{eq:dP_abc_mcs} is replaced by a $1/k_\perp^4$ divergence and the friction pressure becomes \cite{Bodeker:2017cim}
\begin{equation}
m_{\rm sat} \gtrsim m_{c,h} \quad \implies \quad  \left< \Delta p \right> \simeq 0.3\,\zeta_a\, m_{\rm sat}\left(\frac{m_{c,h}}{m_{\rm sat}}\right)^4 \propto \gamma^{-3/7} \quad \implies \quad \mathcal{P}_{\rm NLO} \propto \gamma^{4/7}.
\end{equation}
This change of scaling $\gamma \to \gamma^{4/7}$, which is enough to prevent bubble walls to run-away, should only occurs for large latent heat $\Delta V \gtrsim  \left< \phi \right>^4$, small gauge coupling $\alpha \lesssim 0.01$, and at the end of the bubble expansion stage in the friction dominated regime $\gamma \simeq \gamma_{\LL}$.
}
\begin{equation}
m_{\rm sat} \simeq 
\left\{
                \begin{array}{ll}
                  \displaystyle 0.023~\,m_{c,h}\,\left(
                  \frac{\gamma}{\gamma_{\rm run}}\frac{1}{c_w}
                  \frac{\rm TeV}{\left< \phi \right>}
                  \frac{10}{\beta/H_*}
                  \frac{N \sum_{a,b,c} g_a \,C_{abc}}{10 g_{*}}
                  \right)^{\!\!1/3} \left(
                  \frac{\alpha}{1/30}
                  \frac{\Delta V}{0.1\left< \phi\right>^4}
                                    \right)^{\!\!1/6}\left(
                  \dfrac{T_{\rm nuc}}{10^{-4}T_{\rm start}}
                   \right)^{\!\!4/3}\\[0.5cm] \quad \text{(run-away~walls)},
                   \\[0.2cm]
                 \displaystyle 0.12~\,m_{c,h}\,\left(
                 \frac{\gamma}{\gamma_{\LL}}
                 \frac{\Delta V}{0.1\left< \phi\right>^4}
                 \frac{1/30}{\alpha}
                 \frac{4}{\kappa}\,
                 \frac{N \, \ln{10}}{3 \, \ln (m_{c,h}/m_{\rm sat})}
                 \right)^{\!1/3} \\[0.5cm] \quad \text{(terminal-velocity~walls)},
                \end{array}
              \right.
\label{eq:kperp_satur}
\end{equation}
where we have introduced parameters that will be discussed later in the paper: $T_{\rm start}$ is the temperature where vacuum domination starts (Eq.~(\ref{eq:Tstart})), $H_*$ is the Hubble parameter at the time of the phase transition, $\beta$ is the time variation of the nucleation rate (Eq.~\eqref{eq:beta}), $\gamma_{\rm run}$ is the Lorentz factor of bubble walls that run away evaluated at the time of collision (Eq.~(\ref{eq:gamma_run})), $\gamma_{\LL}$ is that of bubble walls where the external pressure compensates the internal one $\Delta V$ (Eq.~(\ref{eq:gamma_infty})), $c_w$ is the bubble radius at nucleation in unit of $T_{\rm nuc}^{-1}$  (Eq.~\eqref{eq:Rnuc}), and $\kappa$ is the exchanged momentum in unit of $\zeta_a m_{c,h}\ln(m_{c,h}/\mu)$ (Eq.~\eqref{eq:Delta_p_analytical}).

The modification of the dispersion relation of vector bosons, that we just derived, relies on a perturbative particle description.
However, as we now sketch, perturbativity breaks down for  $p_c$ so low to give roughly the same occupation numbers $f_c$ that lead to our cut-off $m_{\rm sat}$ above.
Eq.~\eqref{eq:mcs_first_eq} can be rewritten as
\begin{equation}
m_{c,s}^2 \sim g^2 f_c(p_*)p_*^2,
\end{equation}
where we define $p_*$ as some IR cut-off of the momentum of the $c$ particle, because the integrand in Eq.~\eqref{eq:mcs_first_eq} is peaked in the IR.
where $p_*$ is the typical momentum of the $c$ particle, which is identified as the IR cut-off because the integrand in Eq.~\eqref{eq:mcs_first_eq} is peaked in the IR. Modes with $p_* \lesssim m_{c,s}$ are screened, hence the condition for the vector bosons to back-react on the dispersion relation can be recast as \cite{Bodeker:2017cim}
\begin{equation}
g^2 f_c(p_*) > 1. \label{eq:g2f}
\end{equation}
The occupation number $ f_c(p_*) $ of vector bosons  is related to  the vector boson wave function $A_\mu$ by
\begin{equation}
(\partial A)^2 \sim
p^2 A^2
\sim \int_{p_*}\! d^3p\,p\, f_c(p)
\sim
p_*^4  \, f_c(p_*).
\end{equation}
 
We deduce that the hierarchy between the 3 terms in the Lagrangian
\begin{equation}
\mathcal{L} \supset \partial A \partial A + g A A \partial A + g^2 A A A A\,,
\label{eq:lagragian_vector_boson}
\end{equation}
which is essential for perturbation theory to apply, breaks down as soon as we enter the regime of Eq.~\eqref{eq:g2f}, or equivalently, as soon as $p_* < m_{\rm sat}$ in Eq.~\eqref{eq:kperp_satur}.
Therefore, a treatment beyond perturbation theory would be needed to determine the friction pressure for momenta below $m_{\rm sat}$. 
We leave such an interesting study for future work, and here for simplicity we conservatively choose to still interpret $m_{\rm sat}$ as an IR cut-off.

Let us finally comment on abelian gauge theories.
In those cases both previous arguments do not apply, because they rely on the existence of gauge self-interactions, so that one cannot use the cutoff $m_{\rm sat}$ in Eq.~(\ref{eq:kperp_satur}).
This motivates the use of the naive thermal cutoff in Eq.~\eqref{eq:mcs_thermal}, with one potential limitation that we now comment upon.
In the abelian case we expect an IR cut-off to arise from the presence of fermions and scalars originating from the splitting of the soft gauge bosons, themselves radiated from the incoming energetic quanta.
This splitting of soft gauge bosons into softer fermions or scalar pairs is not enhanced, however one has many soft gauge bosons to start with, so that the IR cutoff induced by the produced fermions and scalars may perhaps be larger than the thermal mass.
Determining this IR cut-off goes beyond the purpose of this paper, where we just content ourselves with pointing out this potential limitation of using the naive thermal mass of Eq.~\eqref{eq:mcs_thermal} in the abelian case.

\paragraph{Other cutoffs.}

Below we discuss processes that may regulate the IR divergence of Eq.~(\ref{eq:dP_abc_mcs}) at values $\mu > m_{c,s}$.
We anticipate that,  in the physics case of our interest, we find that the thermal mass and the phase-space saturation discussed above provide IR cutoffs that are stronger than the effects we discuss next.

\begin{itemize}
\item[$\diamond$] \textbf{3-to-2 boson scattering.}
In the IR limit $k_\perp^2 \to 0$, the occupation number of emitted vector bosons diverges.
At some point, in the case of non-abelian gauge theories, we could expect the occupation number of emitted vector bosons to be large enough to trigger 3-to-2 scatterings.
In that case, the population of emitted vector bosons is depleted and the pressure stops growing.
In App.~\ref{sec:3to2_scattering}, we compute the resulting value of the IR cut-off $\mu_{3\to 2}$ at tree level, and find that $\mu_{3\to 2} < m_{\rm sat}$ for typical values of the parameters.
We then consider $m_{\rm sat}$ as the effective IR cut-off of our emissions.\footnote{
Moreover, for $f \gtrsim 1$, one would expect a Bose enhancement of $2\to3$ transitions, further jeopardising the validity of the cut-off $\mu_{3\to 2}$.
Given that $m_{\rm sat} > \mu_{3\to 2}$, we do not need to quantitatively investigate this issue.
}

\item[$\diamond$] \textbf{Energy-momentum conservation.}
Successive boson emissions lower the energy and momentum of the parent particle $(E_b,\,p_b)$.
The impossibility to radiate more energy and momentum than what is initially available, therefore, tames the IR divergence even when the cut-offs discussed so far go to zero.
To describe this effect, we rely on a numerical Monte-Carlo simulation which we describe in Sec.~\ref{sec:MC} and App.~\ref{app:backreaction}.
We find that the number of emissions and the resulting $\left<\Delta p \right>$ saturate in the IR, see Fig.~\ref{fig:MC_simu_backreaction} of App.~\ref{app:backreaction}.
This leads to an effective IR cut-off which depends on the initial energy $E_a$. 
However we find that the corresponding IR cut-off is smaller than the thermal mass and backreaction can be neglected. 

\end{itemize}

\paragraph{Prescription for calculations.}

In the remaining part of the text, we consider the IR cut-off $\mu$ in Eq.~\eqref{eq:dP_abc} as a free parameter.
For numerical applications, we will consider two benchmark scenarios according to whether the IR cut-off is set, in the case of abelian gauge theory, by the thermal mass $\mu = \alpha^{1/2} T_{\rm nuc}$ in Eq.~\eqref{eq:mcs_thermal} or, in the case of non-abelian gauge theory, by the screening length $\mu = m_{\rm sat}$ in Eq.~\eqref{eq:kperp_satur} resulting from phase space saturation
\begin{equation}
\mu = m_{c,s} \simeq \left\{
                \begin{array}{ll}
                  m_{\rm th}=\alpha^{1/2} \Tnuc \qquad \qquad   \quad~ \text{(abelian gauge theory)}, \\[0.2cm]
                  \textrm{Max}\left[ m_{\rm th}, \,m_{\rm sat} \right] \qquad\quad \qquad \text{(non-abelian gauge theory)}.
                \end{array}
              \right. \label{eq:mu_msc}
\end{equation}

\subsection{Perturbativity breakdown}
\label{sec:pertub_breakdown}

\paragraph{Integrated perturbative splitting probability.}

We integrate $x$ in Eq.~\eqref{eq:dP_abc_mcs} over the range defined in Eq.~\eqref{eq:range_x} with $m_{c}(z) \to  m_{c,s}$. We obtain the probability to emit a vector boson with transverse momentum $k_{\perp}$ in the collinear limit
\begin{equation}
dP_E(k_{\perp})
\simeq \frac{\zeta_a}{2} \,\frac{dk_{\perp}^2}{k_{\perp}^2}\,  \left(\frac{k_{\perp}^2}{k_{\perp}^2+\mu^2}\right)^2 \,\left(\frac{m_{c,h}^2-m_{c,s}^2}{k_{\perp}^2+m_{c,h}^2}\right)^2\ln{\frac{E_{a}^2}{k_{\perp}^2+m_{c,s}^2}}.
\label{eq:dPE}
\end{equation}
The last integration over $k_{\perp}^2$, Eq.~(\ref{eq:range_x_K_perp}), is logarithmically dependent on the IR cut-off $\mu$.
Upon identifying $\mu = m_{c,s} $ and in the limit $\mu \ll m_{c,h} \ll E_{a}$, the integrated vector boson emission probability becomes 
\begin{equation}
\label{eq:P_E_2}
P_E(\mu) \simeq 2\zeta_a\,\textrm{ln} \frac{m_{c,h}}{\mu} \,\textrm{ln} \frac{E_{a}}{{m_{c,h}}}+ \zeta_a\,\textrm{ln}^2 \frac{m_{c,h}}{\mu} .
\end{equation}
We recover the standard double logarithm for integrated probability of vector boson radiation (see e.g.~\cite{Ellis:1991qj,Peskin:1995ev,Schwartz:2013pla,Larkoski:2017fip}).
The soft-collinear divergence is here regularized by the IR cut-off $\mu$, which is not the energy threshold of some particle detector like in a collider context, because here all emitted vector bosons contribute to the observable $\left< \Delta p \right>$ and so are physical. 
We take the IR cut-off in our context from processes due to the existence of (high) particle densities, see Sec.~\ref{sec:IR_cut_off} for more details.

\paragraph{Numerical estimates in physical scenarios.}
\label{par:scenario}

We now evaluate the emission probability $P_E$, for different benchmark values in Table \ref{tab:tablePE}.
The incoming energy is given by the thermal energy $3T_{\rm nuc}$ \cite{Kolb:1990vq} boosted in the wall frame 
\begin{equation}
\label{eq:thermal_initial_energy}
E_{a} \simeq 3 \gamma T_{\rm nuc}. 
\end{equation}
We assume that the bubble wall Lorentz factor $\gamma$ is set by its value evaluated at the time of bubble collision, see Eq.~\eqref{eq:gamma_final}
\begin{equation}
\gamma=\gamma_{\rm coll}.
\end{equation}
Without much loss of generality, we choose
\begin{equation}
C_{abc} =1.
\end{equation}
We assume that the vector boson mass in the broken phase is set to
\begin{equation}
m_{c,h} \simeq \sqrt{2\pi\alpha} \left<\phi\right>.
\end{equation}
We choose to characterize the amount of supercooling by the ratio $T_{\rm nuc}/T_{\rm start}$ where $T_{\rm start}$ is the temperature when the universe becomes vacuum-dominated\footnote{
In the low-temperature expansion of the thermal potential, i.e. when the abundance of the particles $i$ thermally correcting the potential of $\phi$ is Boltzmann-suppressed, $m_i(\phi) \gg T_{\rm start}$, one has
\begin{equation}
T_c \simeq 3^{1/4} \left(\frac{g_{*,\rm i}}{g_{*,\rm f}}\right)^{1/4} T_{\rm start} ~\gtrsim ~T_c,
\end{equation}
where $T_c$ is the critical temperature when the two minima of the potential are degenerate, cf. App.~A in \cite{Baldes:2020kam}, and where $g_{*,\rm i}$ and $g_{*,\rm f}$ are the number of relativistic degrees of freedom before and after the phase transition.
}
\begin{equation}
\frac{\pi^2}{30}g_* T_{\rm start}^4 \simeq \Delta V \qquad \implies  \qquad T_{\rm start} \simeq \left(\frac{30 \,c_{\rm vac}}{\pi^2\,g_*}  \right)^{\!1/4} \left<\phi\right>, \qquad \text{with} \quad c_{\rm vac}  \equiv \Delta V/ \left<\phi\right>^4.
\label{eq:Tstart}
\end{equation}
Here $N_e \equiv \ln\left(T_{\rm start}/T_{\rm nuc}\right)$ gives the number of e-folds of inflation generated during the supercooled phase transition.
As shown in Table \ref{tab:tablePE}, at large supercooling or large coupling constant, e.g. $\alpha \gtrsim 0.3$ and $T_{\rm nuc} \lesssim 10^{-3}\,T_{\rm start}$, the perturbative calculation in Eq.~\eqref{eq:P_E_2} cannot be trusted and we must account for the possibility to radiate multiple vector bosons.

\begin{table}[]
{
\centering
\renewcommand{\arraystretch}{1.2}
\setlength\extrarowheight{2pt}
\begin{tabular}{|c|c|ccc|}
\hline
\multicolumn{2}{|c|}{\begin{tabular}[c]{@{}c@{}}Emission probability\\  $P_{\rm E}$ at LO in $\alpha$ \end{tabular}}& \begin{tabular}[c]{@{}c@{}} $\dfrac{T_{\rm nuc}}{T_{\rm start}} = 0.1 $\end{tabular} & \begin{tabular}[c]{@{}c@{}}  $\dfrac{T_{\rm nuc}}{T_{\rm start}} = 10^{-3} $\end{tabular} & \begin{tabular}[c]{@{}c@{}}  $\dfrac{T_{\rm nuc}}{T_{\rm start}} = 10^{-6} $ \end{tabular}\\[15pt]\hline
 \begin{tabular}[c]{@{}c@{}} 
$\mu \simeq \alpha^{1/2} T_{\rm nuc}$ \\(thermal mass)  \end{tabular}                 & \begin{tabular}[c]{@{}c@{}} $\alpha =0.03$ \\$\alpha =0.3$           \end{tabular}                  & \begin{tabular}[c]{@{}c@{}} $0.6$ \\$3.2 \gtrsim 1$       \end{tabular}                                       & \begin{tabular}[c]{@{}c@{}} $2.8 \gtrsim 1$ \\$24.5 \gg 1$       \end{tabular}  & \begin{tabular}[c]{@{}c@{}} $4.1 \gg 1$ \\$38.3 \gg 1$       \end{tabular}\\[15pt]\hline
 \begin{tabular}[c]{@{}c@{}} 
$\mu = m_{\rm sat}$ \\ (phase space saturation)\end{tabular}             
&      
\begin{tabular}[c]{@{}c@{}} $\alpha =0.03$ \\$\alpha =0.3$       \end{tabular}                  & \begin{tabular}[c]{@{}c@{}} $0.2$ \\  $1.7$   \end{tabular}                                       & \begin{tabular}[c]{@{}c@{}} $0.5$ \\$5.5 \gg 1$       \end{tabular}  & \begin{tabular}[c]{@{}c@{}} $2.0 \gtrsim 1$ \\$17.3 \gg 1$       \end{tabular}                               \\[15pt] \hline
\end{tabular}
\caption{\it \small
\label{tab:tablePE}
Values of  the probability for radiating a single soft vector boson $P_{\rm E}$ in Eq.~\eqref{eq:P_E_2}, for different amount of supercooling, coupling constant $\alpha$ and IR cut-off $\mu$, see Sec.~\ref{sec:IR_cut_off}.
We can see that for large supercooling $T_{\rm start}/T_{\rm nuc}$ or large coupling constant $\alpha$, perturbativity, which we define by $P_{\rm E} \lesssim 1$, breaks down and the Sudakov logarithms of the fixed-order calculation in Eq.~\eqref{eq:P_E_2} must be resummed.
We have fixed $c_{\rm vac}= 0.1$ and $\gamma$ equal to the wall Lorentz factor $\gamma_{\rm coll}$ when bubbles collide, see Eq.~\eqref{eq:gamma_final}.
}
}
\end{table}

\section{The splitting probability at all leading-log orders}
\label{sec:all_order}

\subsection{Multiple boson emission}

\paragraph{Exchanged momentum.}

For multiple $c$ emission, the exchange momentum along $z$ is
\begin{equation}
\Delta p  = E_{a} - \sqrt{(1-X)^2E_{a}^2 - m_{b,h}^2 -K_{\perp}^{2}} - \sum_{i=1}^n p_{c_i}^z \label{eq:Delta_p}
\end{equation}
where we have denoted by $X$ and $K_{\perp}$ the sum of energies and transverse momenta of the $n$ emitted vector bosons
\begin{equation}
X =  \sum_{i=1}^n x_i, \qquad \text{and} \qquad K_{\perp} = \sum_{i=1}^n k_{\perp,i}. \label{eq:def_X_Kperp}
\end{equation}
The values of the $z$ momenta of the emitted vector bosons $p_{c_i}^z$ depend on whether they are transmitted or reflected by the wall
\begin{equation}
p_{c_i}^z= \sqrt{x_i^2 E_{a}^2 - m_{c,h}^2 - k_{\perp, \, i}^{2}}\,\Theta(p_{c_i,h}^2)  - \sqrt{x_i^2 E_{a}^2 - m_{c,s}^2 - k_{\perp, \, i}^{2}} \,\Theta(-p_{c_i,h}^2). \label{eq:pci_exact}
\end{equation}
with 
\begin{equation}
p_{c_i,h}^2 = x_i^2 E_{a}^2 - m_{c,h}^2- k_{\perp, \, i}^{2}.\label{eq:pci_h}
\end{equation}
Note that Eq.~\eqref{eq:Delta_p} assumes that successive emissions take place in the same $(xz)$ plane and Eq.~\eqref{eq:pci_exact} neglects the transverse recoils of the successive emissions on the momentum of the parent particle.
In App.~\ref{app:transverse_recoil}, we show that those two approximations are very good.

\paragraph{Soft-collinear limit.}

To first order in the soft $x_i \ll 1$ and collinear $k_{\perp,i}\ll 1$ limit, we obtain
\begin{align}
&\Delta p= \sum_{i=1}^n \Delta p_i, \label{eq_Delta_p_sum_Delta_pi}
\qquad{\rm with}\qquad
\Delta p_i=\frac{m_{c,h}^2+k_{\perp,i}^2}{2x_i E_a}\Theta(p_{c_i,h}^2) +  2 x_i E_{a}  \Theta(-p_{c_i,h}^2). 
\end{align}

\subsection{Sudakov resummation}
\label{eq:sudakov_resummation}

\paragraph{Poisson distribution.}

At leading-log order, the many-boson emission distribution follows a Poisson distribution, see e.g. Eq.~(2.30) of \cite{Banfi:2004yd} or Eq.~(6.86) of \cite{Peskin:1995ev}, and the mean value of an observable  $\mathcal{O}$ can be computed from
\begin{equation}
\left< \mathcal{O} \right> = \sum_{n=0}^{\infty}\frac{1}{n!}\left[ \prod_{j=1}^{n} \int dP_{E,\,j}\right] \mathcal{O}  \exp\left[-\int dP_{E} \right], \label{eq:mean_O}
\end{equation}
where
\begin{equation}
\int dP_{E,\,j} =  \int_{0}^{E_{a}^2} dK_{\perp,\, j}^2\int^1_{\frac{\sqrt{\mu^2+K_\perp^2}}{E_{a}}} dx_j \; \frac{dP_{\rm E}}{d{K_{\perp,\,j}^2}dx_j}. \label{eq:def_dPE}
\end{equation}
The product factors in Eq.~\eqref{eq:mean_O} account for the $n$ indistinguishable leading-log real emissions while the exponential resums the leading-log virtual corrections.
The matching between the matrix element of the virtual correction in the argument of the exponential and the perturbative splitting probability stems from unitarity
\begin{equation}
\sum_{n=0}^{\infty}\frac{1}{n!}\left[\int dP_{E}\right]^{n}  \exp\left[- \int dP_{E} \right]= 1. \label{eq:unitarity_poisson}
\end{equation}
The real emissions are assumed to be independent and we neglect correlations which are higher order effects  \cite{Banfi:2004yd}.

\paragraph{Mean exchange momentum.}

We can write
\begin{align}
\left< \Delta p \right> &=  \sum_{n=0}^{\infty}\frac{1}{n!}\left[ \prod_{j=1}^{n} \int dP_{E,\,j}\right] \sum_{i=1}^n \Delta p_i ~\exp\left[-\int dP_{E} \right] \notag\\
&= \left[\int dP_{E}\, \Delta p_i \right]  \sum_{n=1}^{\infty}\frac{1}{(n-1)!}\left[\int dP_{E}\right]^{n-1} \exp\left[- \int dP_{E} \right] \notag\\
&= \int dP_{E}\, \Delta p_i \label{eq:Deltap_resummation}
\end{align}
where we have made use of Eq.~\eqref{eq:unitarity_poisson}.
This resembles the average for a Poisson distribution $\sum n \times \lambda^n e^{-\lambda} / n! = \lambda \sum \lambda^{n - 1} e^{-\lambda} / (n - 1)! = \lambda$.
We conclude that in the soft-collinear limit in which $\Delta p$ is an additive observable, see Eq.~\eqref{eq_Delta_p_sum_Delta_pi}, the average exchanged momentum in which both real and virtual leading-log emissions are resummed is identical to the naive expectation using the perturbative splitting probability as in the original article \cite{Bodeker:2017cim}. 

Note however that in the steps leading to Eq.~\eqref{eq:Deltap_resummation}, we have neglected the dependence of the kinematics boundaries on earlier emissions, i.e. we have kept fixed integration boundaries in Eq.~(\ref{eq:def_dPE}).
Instead, the energy and momentum of the parent particle should get depleted with the number of emissions.
The proper average exchanged momentum $\left< \Delta p \right>$ taking into account these backreaction effects is given in Eq.~\eqref{eq:Deltap_resummation_backreaction} of App.~\ref{app:backreaction}.
Since this analytical formula is not easily tractable, we propose to include the effect of backreaction with a Monte-Carlo simulation in the next section, Sec.~\ref{sec:MC}.

\subsection{Analytical estimate}
\label{sec:analytical_estimate}

\begin{figure}[t]
\centering
\begin{adjustbox}{max width=1.\linewidth,center}
\raisebox{0cm}{\makebox{\includegraphics[ width=0.55\textwidth, scale=1]{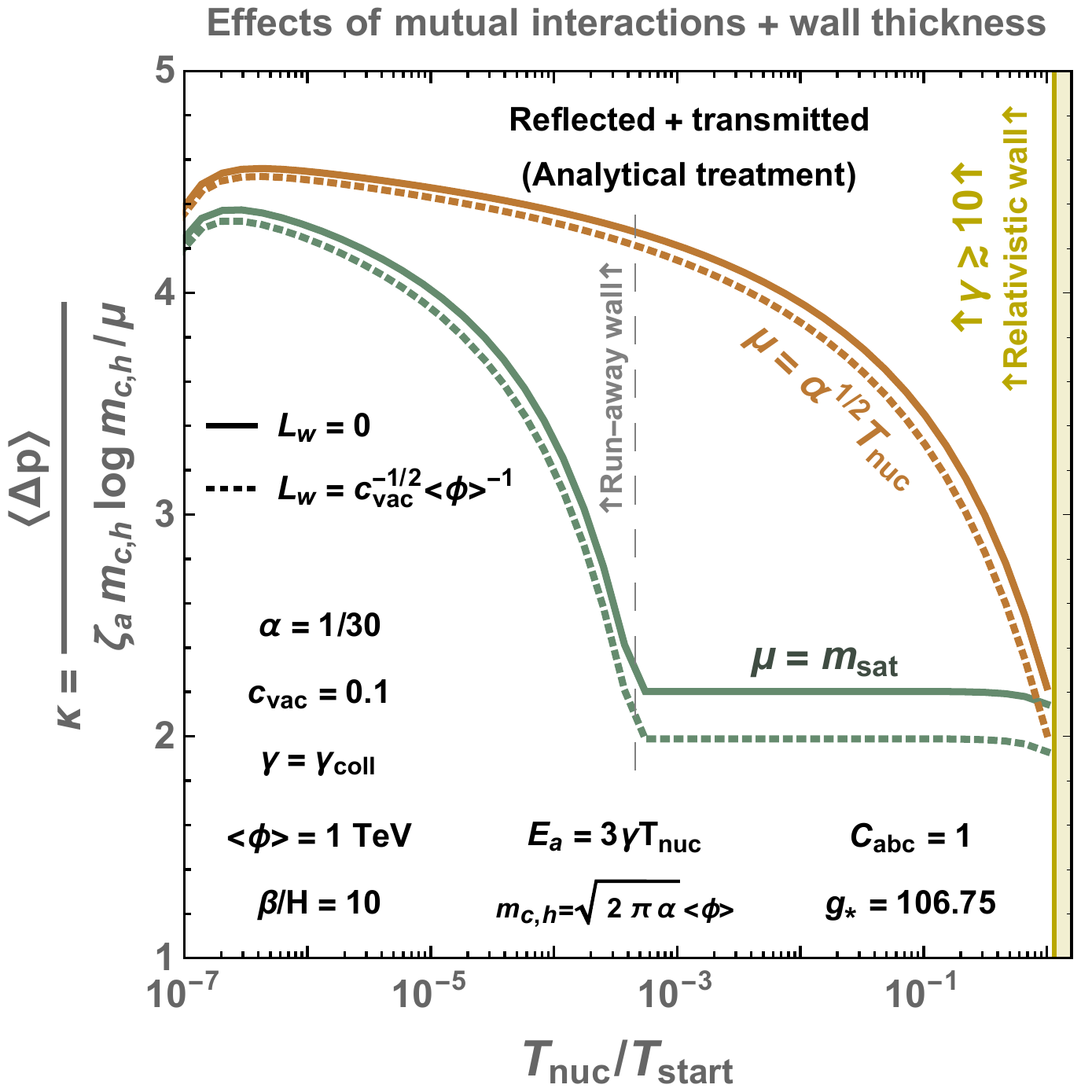}}}
\end{adjustbox}
\caption{\it \small  \label{fig:DeltaP_mu}
\textbf{Analytical average exchanged momentum}
$\left< \Delta p \right> = \left< \Delta p_\R \right> + \left<\Delta p_\T \right> $, where the resummed pressure $\mathcal{P}_\LL$ is proportional to $ \gamma\, \Tnuc^3 \left<\Delta p\right>$, see Eq.~\eqref{eq:PNLO}.
For the solid lines we used the full analytical expressions in Eqs.~(\ref{eq:Delta_p_R}) and Eqs.~(\ref{eq:Delta_p_T}), which assumed negligible wall thickness, and for the dotted ones we use the full numerical result Eq.~\eqref{eq:Deltap_finite_wall_thickness} for a finite wall thickness $L_{\rm w} = c_{\rm vac}^{-1/2}\left<\phi\right>^{-1}$.
The IR cut-off $\mu$ is set to (brown lines) the thermal mass $\alpha^{1/2} T_{\rm nuc}$ and to (green lines) the mass $m_{\rm sat}$ from the collective behaviour of the phase-space-saturated emitted vector bosons (valid for non-abelian gauge theories only), see Eq.~\eqref{eq:mu_msc}.
The presence of the plateau can be understood by the fact that $m_{\rm sat}$ does not depend on $T_{\rm nuc}/T_{\rm start}$ for terminal-velocity walls, cf. Eq.~\eqref{eq:kperp_satur}.
Inside the yellow region, we compute $\gamma \lesssim 10$, cf. Eq.~\eqref{eq:gamma_infty}, such that interactions between neighboring incoming particles can not be safely neglected during the time of wall crossing and our analysis may break down.
}
\end{figure}

\begin{figure}[t]
\centering
\begin{adjustbox}{max width=1\linewidth,center}
\raisebox{0cm}{\makebox{\includegraphics[ width=0.55\textwidth, scale=1]{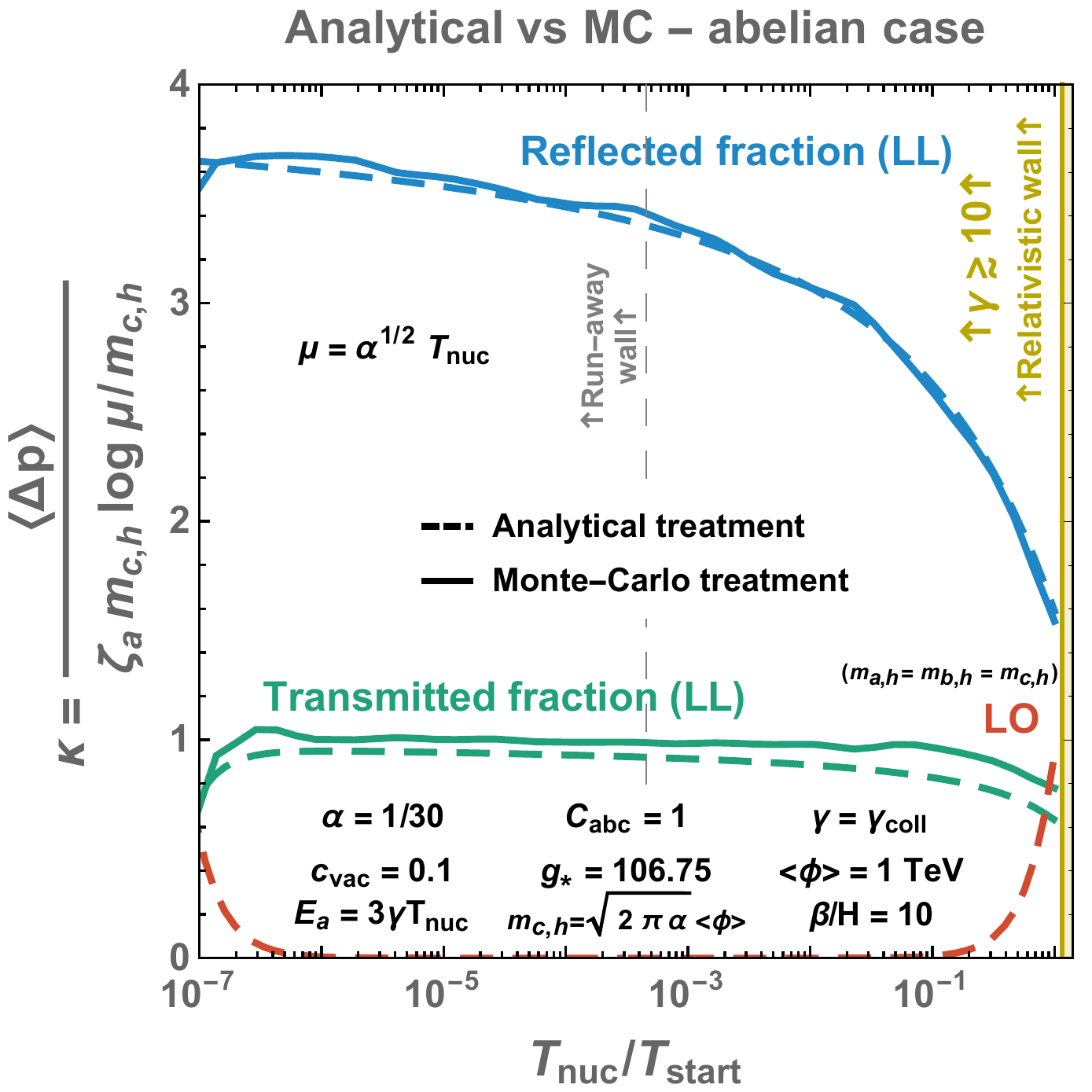}}}
\raisebox{0cm}{\makebox{\includegraphics[ width=0.55\textwidth, scale=1]{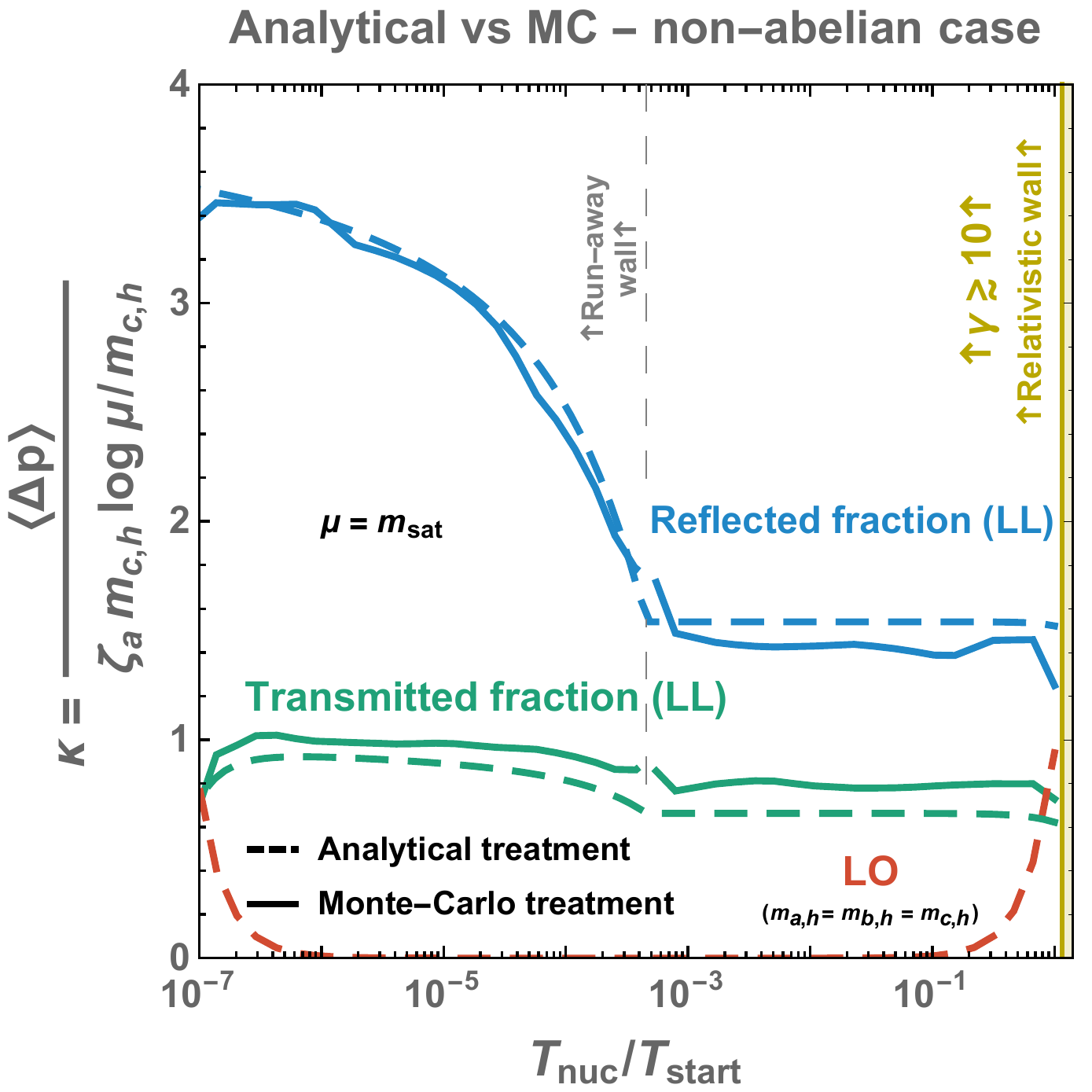}}}
\end{adjustbox}
\caption{\it \small
\label{fig:DeltaP_ana_vs_MC}
Average exchanged momentum $\left<\Delta p \right>=\left< \Delta p_\R \right> + \left< \Delta p_\T \right>$ where the resummed pressure $\mathcal{P}_\LL$ is proportional to $\gamma\, \Tnuc^3 \left<\Delta p\right>$, see Eq.~\eqref{eq:PNLO}.
Dashed lines use the \textbf{analytical estimate} in Eq.~\eqref{eq:rough_estimate}, continuous lines the \textbf{Monte-Carlo simulation} in Sec.~\ref{sec:MC}.
We can see that the MC simulation validates the analytical Sudakov resummation leading to Eq.~\eqref{eq:Deltap_resummation} at the percent level.
The blue lines shows the contributions from the emitted boson which are too soft to enter the broken phase and are reflected by the wall boundary, while the green lines show the contribution from the transmitted ones.
The red dashed line show the contribution due to particle `a' getting a mass in the broken phase, see LO pressure in Sec.~\ref{sec:LO_friction_pressure}.
\textbf{Left}: IR cut-off set to the thermal mass $\mu = \alpha^{1/2} T_{\rm nuc}$ (abelian scenario).
\textbf{Right}: IR cut-off set to $m_{\rm sat}$, see Eq.~\eqref{eq:mu_msc}, accounting for the backreaction of the emitted vector bosons on their dispersion relation (non-abelian scenario). 
}
\end{figure}

From plugging the fixed kinematic boundaries in Eq.~\eqref{eq:range_x_K_perp} inside Eq.~\eqref{eq:def_dPE} and Eq.~\eqref{eq:Deltap_resummation}, in the limit $\mu \ll m_{c,h} \ll E_{a}$ and $m_{c,s} \ll m_{c,h}$, we obtain
\begin{equation}
\label{eq:rough_estimate}
\left< \Delta p \right> = \left< \Delta p_\R \right> + \left< \Delta p_\T \right>, 
\end{equation}
where $\left< \Delta p_\R \right>$ and $\left< \Delta p_\T \right>$ are the contributions from reflected and transmitted vector bosons
\begin{align}
\label{eq:Delta_p_R}
\left< \Delta p_\R \right> &\simeq \zeta_a
\int_{0}^{E_a}\frac{dk_{\perp}^2}{k_{\perp}^2} \,
\int_{\frac{\sqrt{k_{\perp}^2+\mu^2}}{E_{a}}}^{\frac{\sqrt{k_{\perp}^2+m_{c,h}^2}}{E_{a}}}\frac{dx}{x}\,\frac{k_{\perp}^4}{(k_{\perp}^2+\mu^2)^2}\, \left(\frac{m_{c,h}^2-m_{c,s}^2}{k_{\perp}^2+m_{c,h}^2}\right)^2 \times 2xE_a  \\[0.2cm]
&\simeq 4 \,\zeta_a\,m_{c,h} \,\text{ln} \dfrac{m_{c,h}}{\mu} ,
\label{eq:Delta_p_R_2}
\end{align}
and 
\begin{align}
\label{eq:Delta_p_T}
\left< \Delta p_\T \right> &\simeq \zeta_a
\int_{0}^{E_{a}^2}\frac{dk_{\perp}^2}{k_{\perp}^2} \,
\int_{\frac{\sqrt{k_{\perp}^2+m_{c,h}^2}}{E_{a}}}^{1}\frac{dx}{x}\,\frac{k_{\perp}^4}{(k_{\perp}^2+\mu^2)^2}\, \left(\frac{m_{c,h}^2-m_{c,s}^2}{k_{\perp}^2+m_{c,h}^2}\right)^2 \times \frac{k_{\perp}^2 + m_{c,h}^2}{2\,x\,E_{a}} \\[0.2cm]
&\simeq \zeta_a\,m_{c,h} \,\text{ln} \dfrac{m_{c,h}}{\mu} ,
\label{eq:Delta_p_T_2}
\end{align}
where in Eqs.~(\ref{eq:Delta_p_R}) and~(\ref{eq:Delta_p_T}) we have used the splitting probability of Eq.~(\ref{eq:dPE}), and in Eqs.~(\ref{eq:Delta_p_R_2}) and~(\ref{eq:Delta_p_T_2}) we have expanded the results of the integrals for $\mu \ll m_{c,h} \ll E_a$.
We conclude that the contribution from soft emitted vector bosons which are reflected at the wall boundary is at least comparable to the contribution from the transmitted ones (as anticipated in the fixed-order calculation in \cite{Vanvlasselaer:2020niz}).
For parameters of physical interest, $\Delta p_\R$ never never reaches its asymptotic behaviour of Eq.~(\ref{eq:Delta_p_R_2}), but stays slightly smaller, as we will study next.
We thus write a ready-to-use approximated resummed analytical result, sum of the reflected and transmitted contributions, as
\begin{equation}
\left< \Delta p \right> = \kappa\,\zeta_a\,m_{c,h} \,\text{ln} \dfrac{m_{c,h}}{\mu}, \qquad \textrm{with}\quad \kappa \approx 4.
\label{eq:Delta_p_analytical}
\end{equation}
In Fig.~\ref{fig:DeltaP_mu}, we show the analytical estimate for $\kappa$ evaluated with the IR cut-off $\mu$, either set by the thermal mass $\alpha^{1/2} T_{\rm nuc}$, or by the screening mass $m_{\rm sat}$ of phase-space-saturated boson bath, see Eq.~\eqref{eq:mu_msc}.
In Fig.~\ref{fig:DeltaP_ana_vs_MC}, we confront the analytical treatment with the numerical one, based on a MC algorithm, which is discussed in the next section, Sec.~\ref{sec:MC}.

In App.~\ref{app:validity_approximations}, we compute different corrections to Eq.~(\ref{eq:Delta_p_analytical}) due to properly computing the mode function of particle $c$, and to not expanding the various square root functions leading to Eqs.~\eqref{eq:Delta_p_R} and \eqref{eq:Delta_p_T}.
We summarize the list of corrections in Table~\ref{tab:corrections} and Fig.~\ref{fig:Error_estimates}.
We also anticipate that Eq.~\eqref{eq:Delta_p_analytical} neglects additional corrections due to the presence of the thin layer of reflected vector bosons in front of the bubble walls, discussed in Sec.~\ref{sec:reflected_bosons}, and which we leave for further studies.

The case where the vector boson does not acquire a mass in the broken phase, $m_{c,h}=m_{c,s}$, is discussed in App.~\ref{app:massless_scenario}, where we show that NLO effects are only $\mathcal{O}(\zeta_a)$ corrections to the LO pressure in Sec.~\ref{sec:LO_friction_pressure}, and therefore they can be safely neglected. 

\subsection{Fate of reflected vector bosons}
\label{sec:reflected_bosons}

Previously we have discussed the possibility for radiated $c$ particles to be reflected by the wall boundary whenever their would-be momentum in the hidden phase $p_{c, h}$, defined by Eq.~\eqref{eq:pci_h}, is negative. 
We now discuss possible corrections to $\left< \Delta p_\R \right>$ in Eq.~\eqref{eq:rough_estimate} due to the presence of those particles right in front of the wall. 
\begin{itemize}
\item[$\diamond$]
We expect the momentum of the reflected $c$ particles to change due to scatterings with the incoming $a$ particles in the thermal bath.
At a distance of the wall set by their mean free path in the plasma frame, see App.~\ref{app:fate_reflected_particles}
\begin{equation}
l_p \sim \frac{p_{c,z}^2}{\alpha^2 \Tnuc^3},
\end{equation}
where $p_{c,z} \lesssim m_{c,h}$ is their typical momentum, the $c$ particles are expected to come back in direction of the wall.
If their new momentum $p_{c,z}$ is larger than $m_{c,h}$, they are transmitted to the symmetric phase.
Otherwise, if $p_{c,z} < m_{c,h}$ they are reflected again and so on.
Those cycles of multiple reflections ended up by transmission are expected to give corrections to $\left< \Delta p \right>$ in Eq.~\eqref{eq:Delta_p_analytical}.
\item[$\diamond$]
The presence of this population of reflected $c$ particles is expected to give corrections to the dispersion relation and to the computation of $m_{\rm sat}$ in Eq.~\eqref{eq:kperp_satur}.
\item[$\diamond$]
Due to successive scatterings with the reflected $c$ particles, the incoming $a$ particles are expected to lose a small fraction of their momentum $E_a$.
The exchanged momentum $\left< \Delta p \right>$ in Eq.~\eqref{eq:rough_estimate} is independent of $E_a$ in the limit $E_a \gg m_{c, h}$ but goes to zero in the limit $E_a \sim m_{c, h}$.
The latter only arises  in the region of extreme supercooling (e.g. $T_{\rm nuc}/T_{\rm start} \sim 10^{-7}$ if $\langle \phi \rangle = 1$~TeV), see purple vertical line in Fig.~\ref{fig:Xtrem_SC_limit}.
We expect the depletion of $E_a$ due to scatterings with reflected $c$ particles, to shift the position of this purple vertical  line to the right.
\item[$\diamond$]
The successive scatterings of the incoming $a$ particles with the reflected $c$ particles, which act as a medium by itself, is expected to induce further splitting radiations, in addition to the one induced by the bubble wall.
The presence of reflected particles then induces further splitting radiations, which in turn induce further reflected particles and so on. 
\end{itemize}
Contenting ourselves with qualitative comments, we leave the quantitative study of those new effects for future works.

\FloatBarrier

\section{Monte Carlo simulation}
\label{sec:MC}

In this section, we numerically generate the shower of emitted bosons with a Monte-Carlo (MC) simulation and compute the resulting momentum exchanged with the wall.

\subsection{The Sudakov form factor}

\paragraph{The survival probability.}

A necessary ingredient to realize a MC simulation is the probability $P_{\rm NE}(k_1,\,k_2)$ of non-emission in the interval $k_1 \leq k_\perp \leq k_2$ \cite{Ellis:1991qj,Peskin:1995ev,Schwartz:2013pla,Larkoski:2017fip}.
The latter can be derived from the probability of non-emission in the infinitesimal interval $[k_1^2,\,k_2^2+dk_{\perp}^2]$
\begin{equation}
dP_{\rm NE} = 1 - dP_E,
\end{equation}
as follows. $dP_E$ is given in Eq.~\eqref{eq:dPE}.
We consider the finite interval $[k_1^2,\, k_2^2]$ which we divide in $N$ small intervals of length $dk_{\perp}^2 = (k_1^2 - k_2^2)/N$.
The probability of non-emission in $[k_1^2,\,  k_2^2]$ reads
\begin{align}
P_{\rm NE}(k_1,\,k_2,\,E_a) & = \lim_{N\to +\infty} \prod_{n=1}^{N} \left(  1 -  \frac{\zeta_a}{2} \, \frac{dk_{\perp}^2}{k_{\perp,\, n}^2}  \,\frac{m_{c,h}^4}{(k_{\perp,\,n}^2+m_{c,h}^2)^2}\ln{\frac{E_{a}^2}{k_{\perp,\,n}^2+m_{c,s}^2}} \right) \notag\\
&= \exp \left( -  \frac{\zeta_a}{2}  \int_{k_1^2}^{ k_2^2} \frac{{dk_{\perp}^{\!2}}}{{k_{\perp}}^{\!2}}  \,\frac{m_{c,h}^4}{(k_{\perp}^{2}+m_{c,h}^2)^2}\ln{\frac{E_{a}^2}{{k_{\perp}}^{\!2}+m_{c,s}^2}}   \right) \notag \\\label{eq:Sudakov_fac}
&=\exp \left[ - P_{\rm E}(k_1,\,k_2)  \right],
\end{align}
where $P_{\rm E}(k_1,\,k_2)$, is the perturbative probability to emit a vector boson with transverse momentum $k_{\perp}$ in the interval $k_\perp \in [k_1,\,k_2]$
\begin{align}
\label{eq:P_E_1}
&P_{\rm E}(k_1,\,k_2) = F(k_2) - F(k_1),
\end{align}
with
\begin{align}
F(k)=\, &\frac{\zeta_a}{2} \left(\frac{m_{c,h}^2 -m_{c,s}^2}{m_{c,h}^2}\right)^2\Bigg[\left(\frac{m_{c,h}^2}{k^2+m_{c,h}^2}+  \ln{\left(\frac{k^2}{k^2+m_{c,h}^2}\right)}\right)\ln{\left(\frac{E_a^2}{k^2+m_{c,s}^2}\right)} \notag 
\\
&\qquad \qquad
+\ln\left(1+\frac{k^2}{m_{c,s}^2}\right)\ln\left(\frac{k^2}{m_{c,h}^2-m_{c,s}^2}\right)+ \frac{m_{c,h}^2}{m_{c,h}^2-m_{c,s}^2}\ln{\left( \frac{k^2+m_{c,s}^2}{k^2+m_{c,h}^2} \right)}\\
&\qquad\qquad\qquad\qquad\qquad
+\text{PolyLog}_2\left(-\frac{k^2}{m_{c,s}^2} \right) 
+\text{PolyLog}_2\left(- \frac{k^2+m_{c,s}^2}{m_{c,h}^2-m_{c,s}^2} \right) \Bigg],\notag
\end{align}
which, in the limit where $k_1 \ll m_{c,h} \ll k_2$ and $m_{c,s} \ll m_{c,h}$, reduces to
\begin{equation}
\label{eq:P_E_2_recall}
P_E(k_1,\,k_2) \simeq  2\zeta_a\,\textrm{ln} \frac{m_{c,h}}{k_1} \,\textrm{ln} \frac{k_2}{{ m_{c,h} }}+ \zeta_a\,\textrm{ln}^2 \frac{m_{c,h}}{k_1}.
\end{equation}

\paragraph{Resummation of leading-log real and virtual corrections.}

Another way to recover the survival probability $P_{\rm NE}$ in Eq.~\eqref{eq:Sudakov_fac} is
\begin{equation}
P_{\rm NE}(k_1,\,k_2,\, E_a) = \exp \left[ P_{\rm E}(\mu,\,k_1) + P_{\rm E}(k_2,\,E_a) \right] \, \exp \left[ - P_{\rm E}(\mu,\,E_a)\right].
\end{equation}
The first exponential factor includes real emissions outside the interval $[k_1,\,k_2]$, while the second includes virtual emissions (inside the interval $[\mu,\,E_a]$), both resummed to all leading-log orders.

\subsection{The algorithm}

\begin{figure}[t]
\centering
\raisebox{0cm}{\makebox{\includegraphics[ width=0.495\textwidth, scale=1]{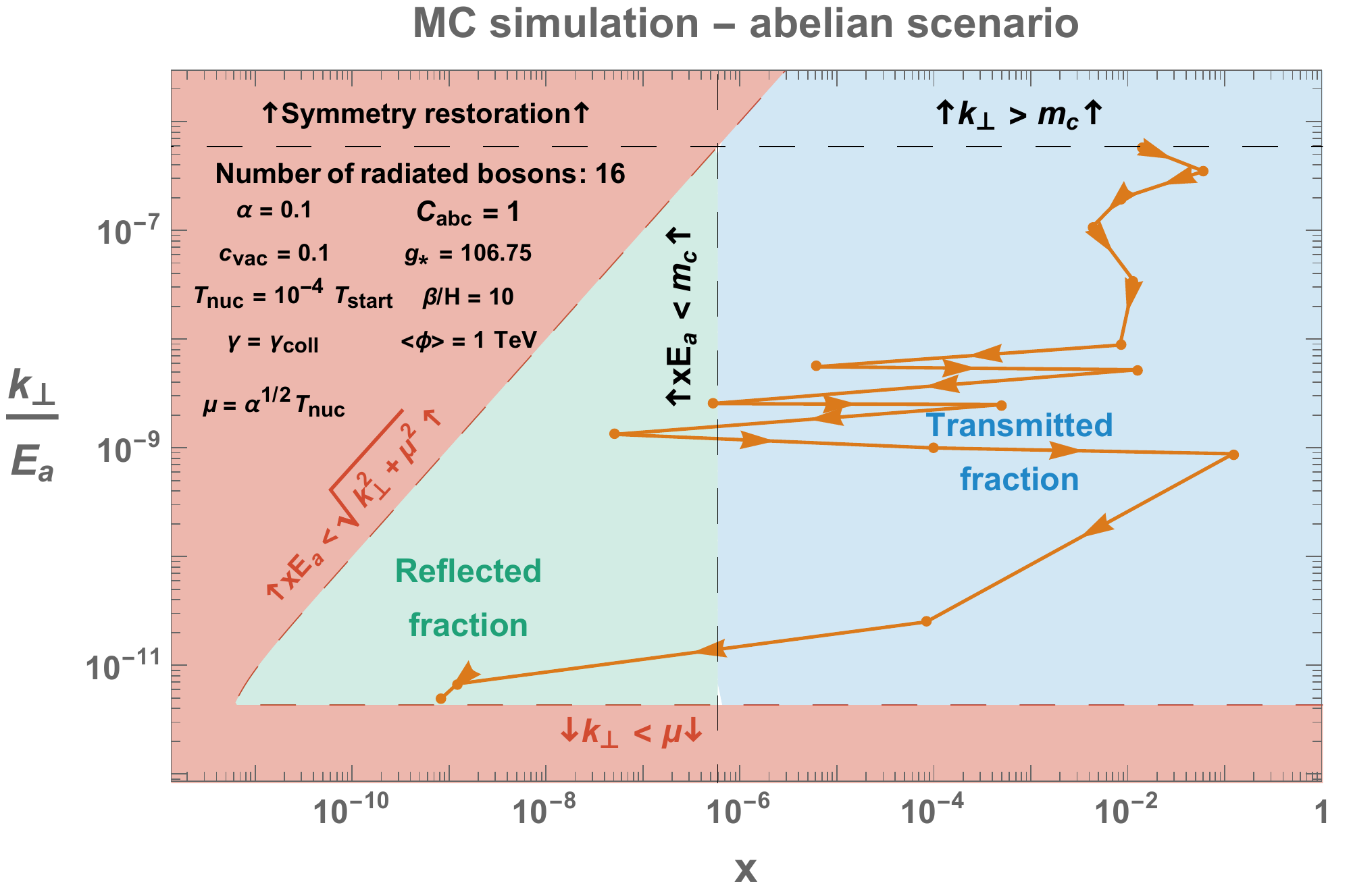}}}
\raisebox{0cm}{\makebox{\includegraphics[ width=0.495\textwidth, scale=1]{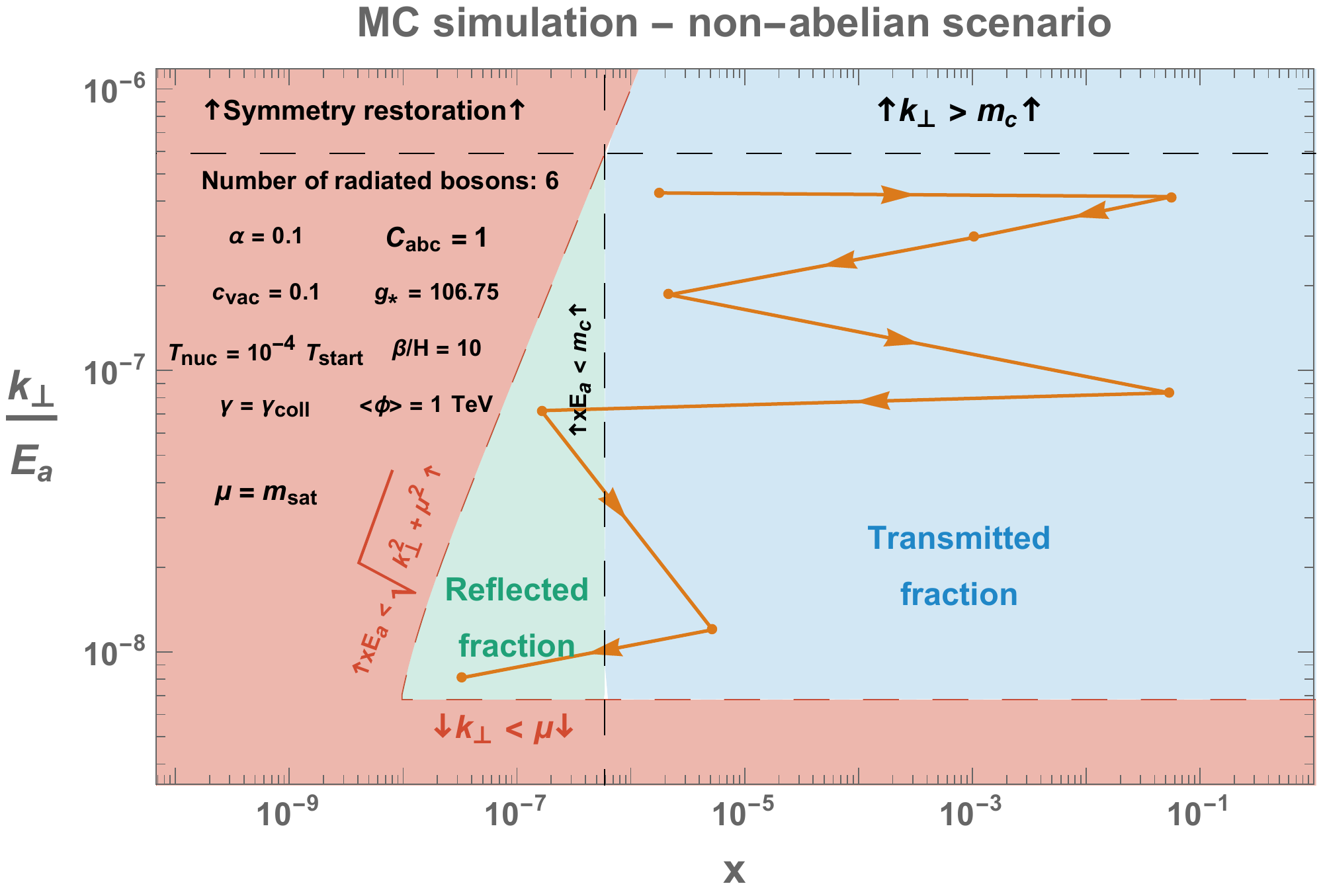}}}
\caption{\it \small  \label{fig:MC}
Trajectory in the phase space $(x,\, k_{\rm perp})$ (also called Lund plane, e.g. \cite{Larkoski:2017fip}) for two given Monte-Carlo simulations.
We set $T_{\rm nuc}/T_{\rm start}= 10^{-4}$, $\alpha \simeq 1/10$ and $\left<\phi \right>=1~\rm TeV$, which imply $m_{c,h}/E_a \simeq 6\times 10^{-7}$ and $\gamma \simeq 2 \times 10^{10}$.
\textbf{Left}: We set the IR cut-off to the thermal mass $\mu = \alpha^{1/2} T_{\rm nuc}$, cf.~Eq.~\eqref{eq:mcs_thermal}, which leads to $\mu/m_{c,h} \simeq 7 \times 10^{-6}$.
\textbf{Right}: We set the IR cut-off to the self-energy of phase-space-saturated boson bath $\mu =m_{\rm sat}$, cf.~Eq.~\eqref{eq:kperp_satur}, which leads to $\mu/m_{c,h} \simeq 0.01$.
The \textbf{orange} points linked by arrows are the kinematics variables of the successive emitted bosons generated by the MC algorithm.
The ones which pass the threshold $E_c >m_{c,h}$ will be able to enter the broken phase (\textbf{blue} region) while the others will be reflected at the wall boundary (\textbf{green} region).
In the \textbf{red} region, the emission energy is smaller than the thermal mass $\mu$, and therefore emission is kinematically forbidden.
}
\end{figure}

\paragraph{Motivation.}

In order to give support to our analytical estimate in Sec.~\ref{sec:analytical_estimate}, but also to include backreaction, see App.~\ref{app:backreaction}, we simulate the shower of emitted vector bosons with a Monte Carlo (MC) algorithm \cite{Marchesini:1987cf,Ellis:1991qj,Banfi:2004yd}, which we now describe.
\paragraph{Recipe.}
Starting from the hard scale $k_{\perp, 0} = E_{a}$, we generate the transverse momentum of the first boson $k_{\perp, 1}$ by solving the equation
\begin{equation}
\label{eq:PNE_MC}
P_{\rm NE}(k_{\perp, 1},\,k_{\perp, 0},\,E_a) = \mathcal{R},
\end{equation}
where $P_{\rm NE}$ is the Sudakov factor in Eq.~\eqref{eq:Sudakov_fac} and $\mathcal{R}$ is a random number between $0$ and $1$. 
The energy of the boson $x_1\,E_{a}$ is a pseudo-random number generated from the perturbative splitting probability in Eq.~\eqref{eq:dP_abc}, i.e. by solving
\begin{equation}
\int^1_{x_1} \frac{dx}{x} =  \mathcal{R}\, \int^1_{\sqrt{k_{\perp}^2+\mu^2}/E_{a}} \frac{dx}{x}. \label{eq:x_distrib_MC}
\end{equation}
The kinematics of the second emitted vector boson $(k_{\perp,2},\,x_2)$ are determined the same way with $k_{\perp, 0}$ replaced by $k_{\perp, 1} $, and so on.
We stop the shower whenever the transverse momentum becomes smaller than the IR cut-off, see Eq.~\eqref{eq:dP_abc}
\begin{equation}
k_{\perp, n_{\rm max}+1} < \mu.
\end{equation}
In App.~\ref{app:transverse_recoil}, we present a MC algorithm which takes into account azimuthal emission angles and transverse recoils of successive boson emissions on the momentum of the parent particle.
\paragraph{Backreaction.}
\label{par:backreaction}
The depletion of the energy-momentum of the parent particle as the emission continues, discussed in App.~\ref{app:backreaction}, is taken into account after replacing the $x$ upper boundaries of Eq.~\eqref{eq:PNE_MC} by
\begin{equation}
\label{eq:PNE_MC_backreaction}
P_{\rm NE}\big(k_{\perp, i},\,k_{\perp, i-1},\,\big(1-\sum_{j<i}x_j\big)E_a\big) = \mathcal{R},
\end{equation}
and the $x$ upper boundaries of Eq.~\eqref{eq:x_distrib_MC} by 
\begin{equation}
\int^{1-\sum_{j<i}x_j}_{x_i} \frac{dx}{x} =  \mathcal{R}\, \int^{1-\sum_{j<i}x_j}_{\sqrt{k_{\perp}^2+\mu^2}/E_{a}} \frac{dx}{x}. \label{eq:x_distrib_MC_backreaction}
\end{equation}
and by stopping the cascade whenever
\begin{align}
p_b < 0 \qquad &\implies \qquad 
\left(1-\sum_{i=1}^{n_{\rm max}+1} x_i\right)^2E_{a}^2 -\left( \sum_{i=1}^{n_{\rm max}+1} k_{\perp,i}\right)^{2} < 0 \label{eq:backreaction_stop}\\
\qquad &\implies \qquad 
\sum_{i=1}^{n_{\rm max}+1}\big(x_i+\frac{k_{\perp,i} }{E_a}\big) > 1. 
\end{align}
\paragraph{Results.}
Fig.~\ref{fig:MC} shows the phase space trajectory of one given MC simulation.
The resulting momentum exchanged with the wall is given by the master formula in Eq.~\eqref{eq:Delta_p}.
In Fig.~\ref{fig:DeltaP_ana_vs_MC}, we show that numerical computations based on MC shower and the analytical estimates of Sec.~\ref{sec:analytical_estimate} agree up to percent level.

\section{The bubble wall velocity} 
\label{sec:bubble_velocity}

\subsection{The final retarding pressure}
\paragraph{Non-confining PTs.}

The goal of this paper was to compute the retarding pressure at all leading-log orders (LL) for non-confining PTs, meaning PTs where particles simply acquire a mass in the broken phase.
We have obtained
\begin{equation}
\mathcal{P}=\mathcal{P}_{\LO} + \mathcal{P}_{\LL}, \label{eq:Ptot}
\end{equation}
where $ \mathcal{P}_{\LO}$ is given in Eq.~\eqref{eq:PLO_1}, which we rewrite here
\begin{equation}
\mathcal{P}_{\LO} =\sum_a g_a c_a \, \frac{\Delta m^2\,T_{\rm nuc}^2}{24},\qquad c_a = 1~(1/2)~ \textrm{for bosons (fermions)},\label{eq:PLO}
\end{equation}
and  $\mathcal{P}_{\LL}$ follows from Eqs.~\eqref{eq:PNLO_formula} and \eqref{eq:Delta_p_analytical}, 
\begin{align}
\mathcal{P}_\LL
&\simeq  \sum_a g_a \int\frac{\gamma \,d^3 p_a}{(2 \pi)^3} \frac{1}{e^{p_a/T_{\rm nuc}}\pm 1}~\left<\Delta p\right> \notag\\
\label{eq:PNLO_deltap}
&\simeq  \gamma \Tnuc^3 \frac{\zeta(3)}{\pi^2} \sum_a \nu_a g_a \,\left<\Delta p\right> \\
\label{eq:PNLO}
&\simeq \frac{\kappa\,\zeta(3)}{\pi^3}\left[\sum_{a,b,c} \nu_a g_a C_{abc}\right] \alpha\,\text{ln} \dfrac{m_{c,h}}{\mu}\,\gamma \,m_{c,h}\,T_{\rm nuc}^3 ,
\end{align}
In Eq.~\eqref{eq:PNLO_deltap} $g_a$ is the number of relativistic degrees of freedom of particle $a$ and $\nu_a = 1~(3/4)$ for bosons (fermions).
We also have boosted the phase space volume $d^3p_{a}$ to the wall frame by introducing the Lorentz factor of the wall in the plasma frame $\gamma$.
In Eq.~\eqref{eq:PNLO} $C_{abc}$ is the charge factor for the SM and can be found in \cite{Hoche:2020ysm}, and $\kappa \approx 4$.
In our numerical analysis we use $\sum_{a, b, c} \nu_a g_a C_{abc} = 100$ for simplicity.
The level of approximation leading to $\kappa \approx 4$ is studied in App.~\ref{app:validity_approximations}.
The value of the IR cut-off $\mu$ is discussed in Sec.~\ref{sec:IR_cut_off}.
In the case of abelian gauge theory, we set it to the thermal mass $\mu = \alpha^{1/2} T_{\rm nuc}$ and, in the case of non-abelian gauge theory, we set it to the screening mass $\mu = m_{\rm sat}$ resulting from phase space saturation, see  Eq.~\eqref{eq:mu_msc}.

The regime of validity of Eq.~\eqref{eq:PNLO} is $\mu \ll m_{c,h} \ll E_a$ and thin walls.
Instead, outside those regimes, we should use Eq.~\eqref{eq:PNLO_deltap} with our most refined formula for $\Delta p$, see Eq.~\eqref{eq:Delta_p_exact_app} of App.~\ref{app:validity_approximations}.
Finally, Eq.~\eqref{eq:PNLO} neglects additional effects due to the presence of the thin layer of reflected vector bosons in front of the bubble wall, see Sec.~\ref{sec:reflected_bosons}.
We leave their study for further works.

\begin{figure}[t]
\centering
\begin{adjustbox}{max width=1\linewidth,center}
\raisebox{0cm}{\makebox{\includegraphics[ width=0.6\textwidth, scale=1]{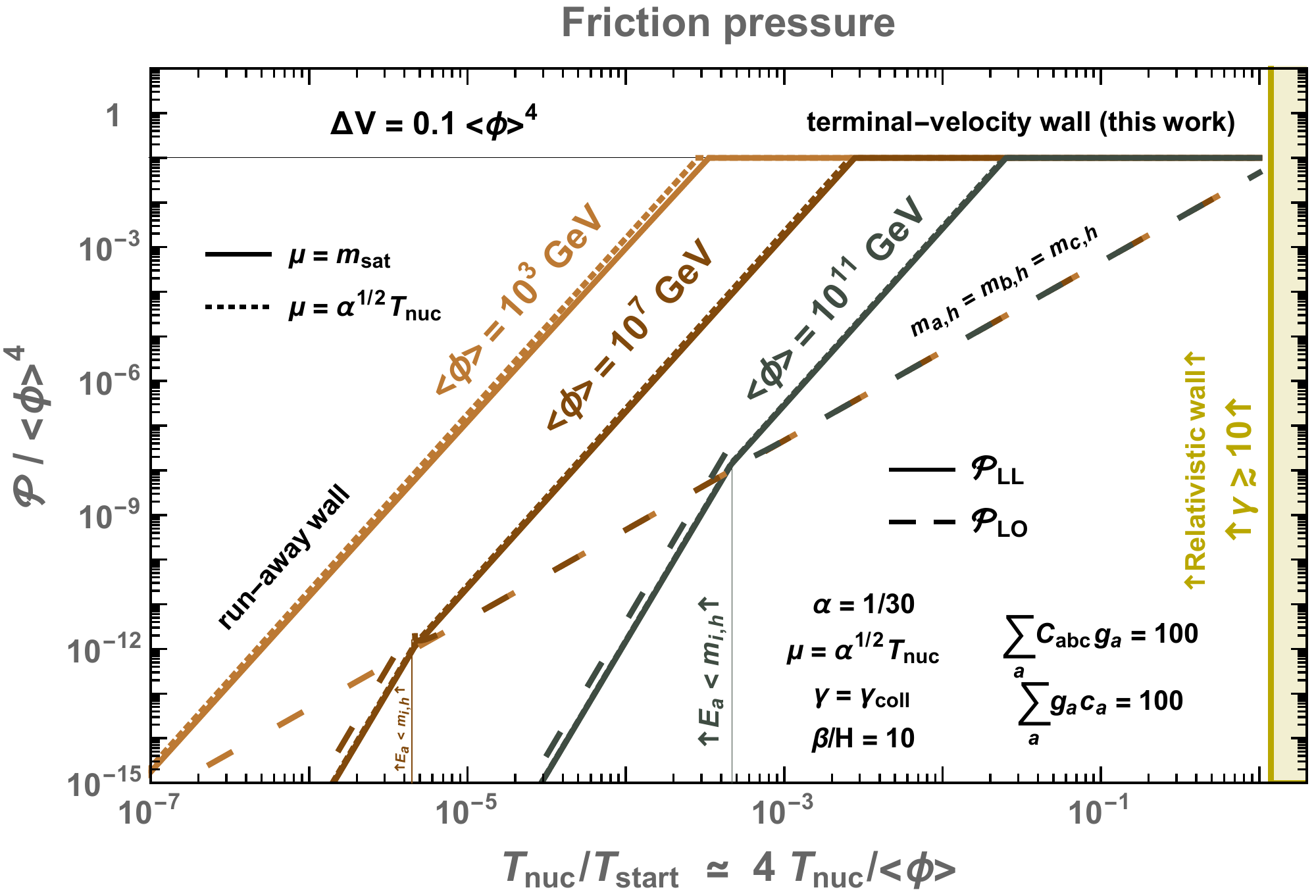}}}
\raisebox{0cm}{\makebox{\includegraphics[ width=0.6\textwidth, scale=1]{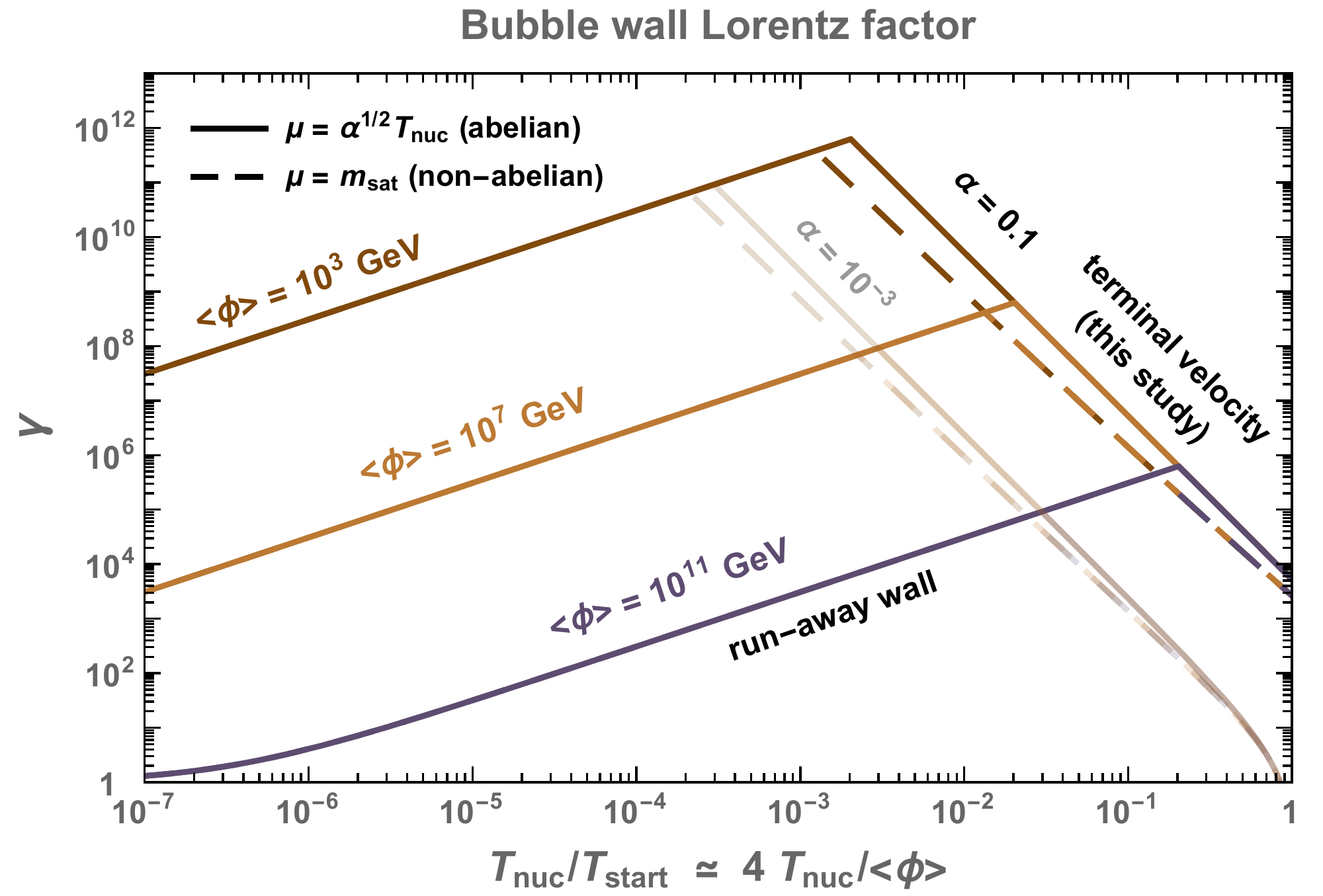}}}
\end{adjustbox}
\caption{\it \small  \label{fig:friction_pressure_bubble_wall_velocity}
\textbf{Left:} Friction pressure at bubble wall collision. The dashed line shows the leading-order contribution to the friction pressure, cf. Eq.~\eqref{eq:PLO}, while the colored solid lines show the leading-log contribution resulting from splitting radiation, cf. Eq.~\eqref{eq:PNLO}. The regime $E_a \lesssim m_{i,h}$ is treated in App.~\ref{sec:snowball}.
\textbf{Right:} Bubble wall Lorentz factor $\gamma$ computed with the results of this paper, see Eq.~\eqref{eq:gamma_final}.
On the left of the peak, bubble walls collide before they reach their terminal velocity $\gamma_{\LL}$ determined by the friction pressure.
This is the so-called run-away regime.
The associated $\gamma$ at collision is proportional to $T_{\rm nuc}$, which is the inverse of the bubble size  at nucleation time, see Eq.~\eqref{eq:gamma_run}.
On the right of the peak, bubble walls reach their terminal velocity before collision, which is where results from this study matter.
}
\end{figure}

We conclude that after having performed a leading-log Sudakov resummation, we have recovered the linear $\gamma$ increase found in \cite{Bodeker:2017cim}, in contrast to \cite{Hoche:2020ysm} in which $\mathcal{P}_\LL \propto \gamma^2$ has been found (see App.~\ref{app:comment_hoeche} for more details on this discrepancy).
In the left panel of Fig.~\ref{fig:friction_pressure_bubble_wall_velocity} we display the resulting pressure Eq.~\eqref{eq:PNLO_deltap} at the time of bubble wall collision, using the full analytical result for $\Delta p = \Delta p_\R + \Delta p_\T$ in Eqs.~\eqref{eq:Delta_p_R} and~\eqref{eq:Delta_p_T}.
In that figure, we also show the regime of extremely small $\Tnuc/T_{\rm start}$, where particles are too soft in the wall frame to enter the broken phase and so the pressure decreases faster, see App~\ref{sec:snowball} for more details.

\paragraph{Confining PTs.}
\label{par:confining_PT}

We review here the findings of Ref.~\cite{Baldes:2020kam} about the pressure in confining PTs.
For confining PTs, the friction pressure is not due to particles getting a mass but instead due to particles becoming strongly-coupled.
At small supercooling $T_{\rm nuc} \simeq T_{\rm start}$, one expects the pressure to be controlled by the formation of bound-states, and to conserve the scaling of Eq.~\eqref{eq:PLO}
\begin{equation}
\mathcal{P}_{\rm BS} \sim \sum_{i \in BS} \frac{m_{i}^2 T_{\rm nuc}^2}{24},
\end{equation}
where the sum is operated over a spectrum of bound-states of mass $m_i$ which depends on the model.
At large supercooling $T_{\rm nuc} \ll T_{\rm start}$, particles entering the bubble are separated by $d \sim 1/T_{\rm nuc}$ and therefore are far from each other compared to the confining distance $f^{-1}$ with $f \equiv \left<\phi\right>$. Since the confining force grows linearly with the distance $d$, confinement with closest neighbors through flux-tube would cost too much energy.
Instead, one expects particles entering the bubble to form flux-tube attached to the wall, where the confining scale $f$ is at its weakest value.
Those strings are expected to fragments into bound states. Besides, in order to conserve color charge, during this process, a particle must be ejected from the wall.
The conversion of the momentum $p_a$ of the incoming particles into string confining energy plus the ejection of a quark is expected to reduce the momentum of the wall by an amount
\begin{equation}
\mathcal{P}_{\rm string} \simeq \frac{E_{\rm string}^2}{2p_a} + \frac{f}{2} \simeq f,
\end{equation}
where $E_{\rm string}$ is the center-of-mass energy of the string
\begin{equation}
E_{\rm string}^2 \simeq p_a \,f.
\end{equation}
One obtains the retarding pressure
\begin{equation}
\mathcal{P}_{\rm string} \sim \frac{\zeta(3)}{\pi^2} g_{\rm TC}\, \gamma\,T_{\rm nuc}^3\,f ,
\end{equation}
where $g_{\rm TC}=g_g + \frac{3g_q}{4}$ with $g_g$ ($g_q$) the relativistic number of techni-quarks (techni-gluons).
We conclude that the retarding pressure on bubble walls of confining PTs has the same $\gamma$ scaling as non-confining PTs.

\begin{figure}[t]
\centering
\begin{adjustbox}{max width=1\linewidth,center}
\raisebox{0cm}{\makebox{\includegraphics[ width=0.7\textwidth, scale=1]{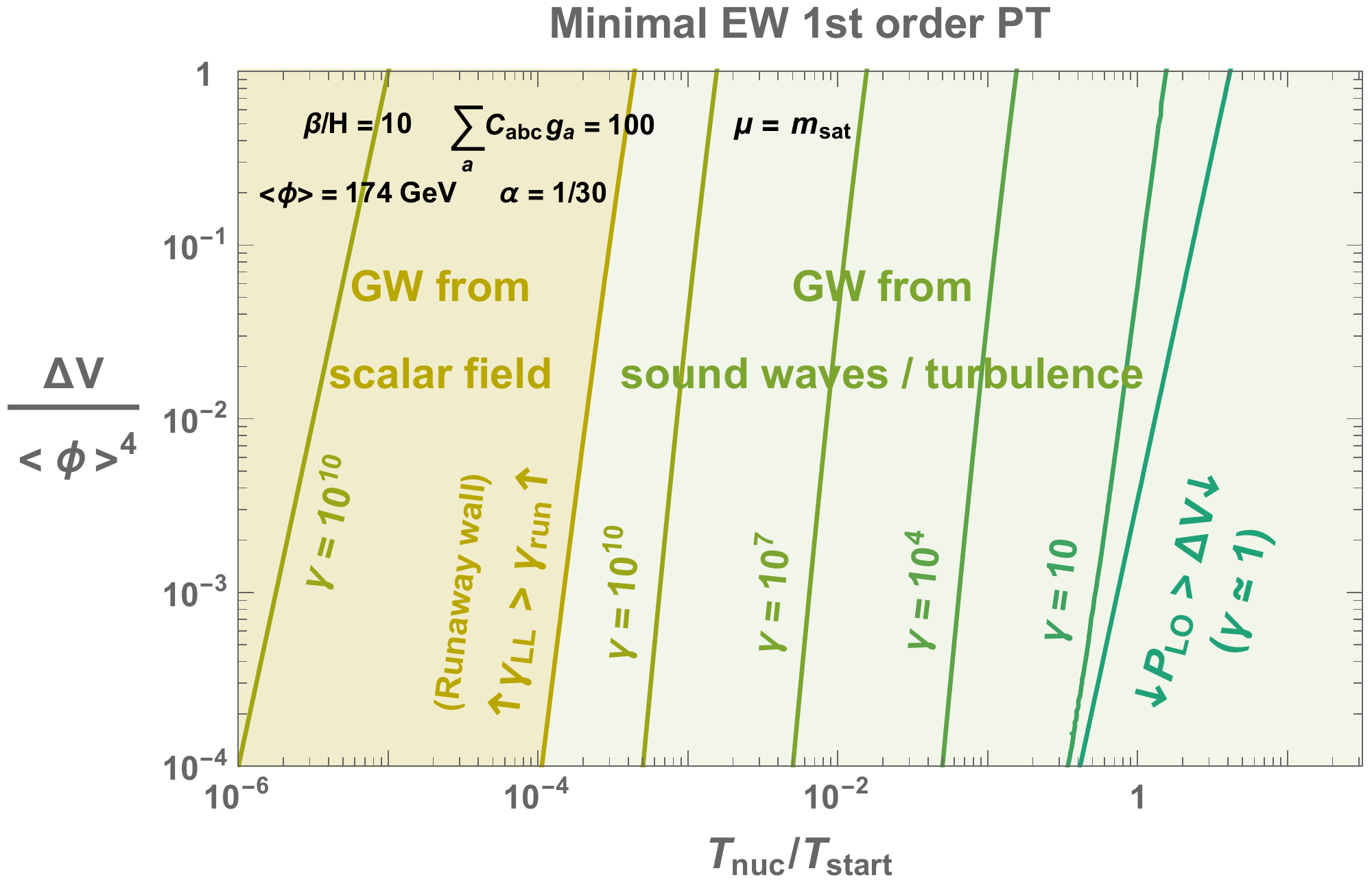}}}
\raisebox{0cm}{\makebox{\includegraphics[ width=0.7\textwidth, scale=1]{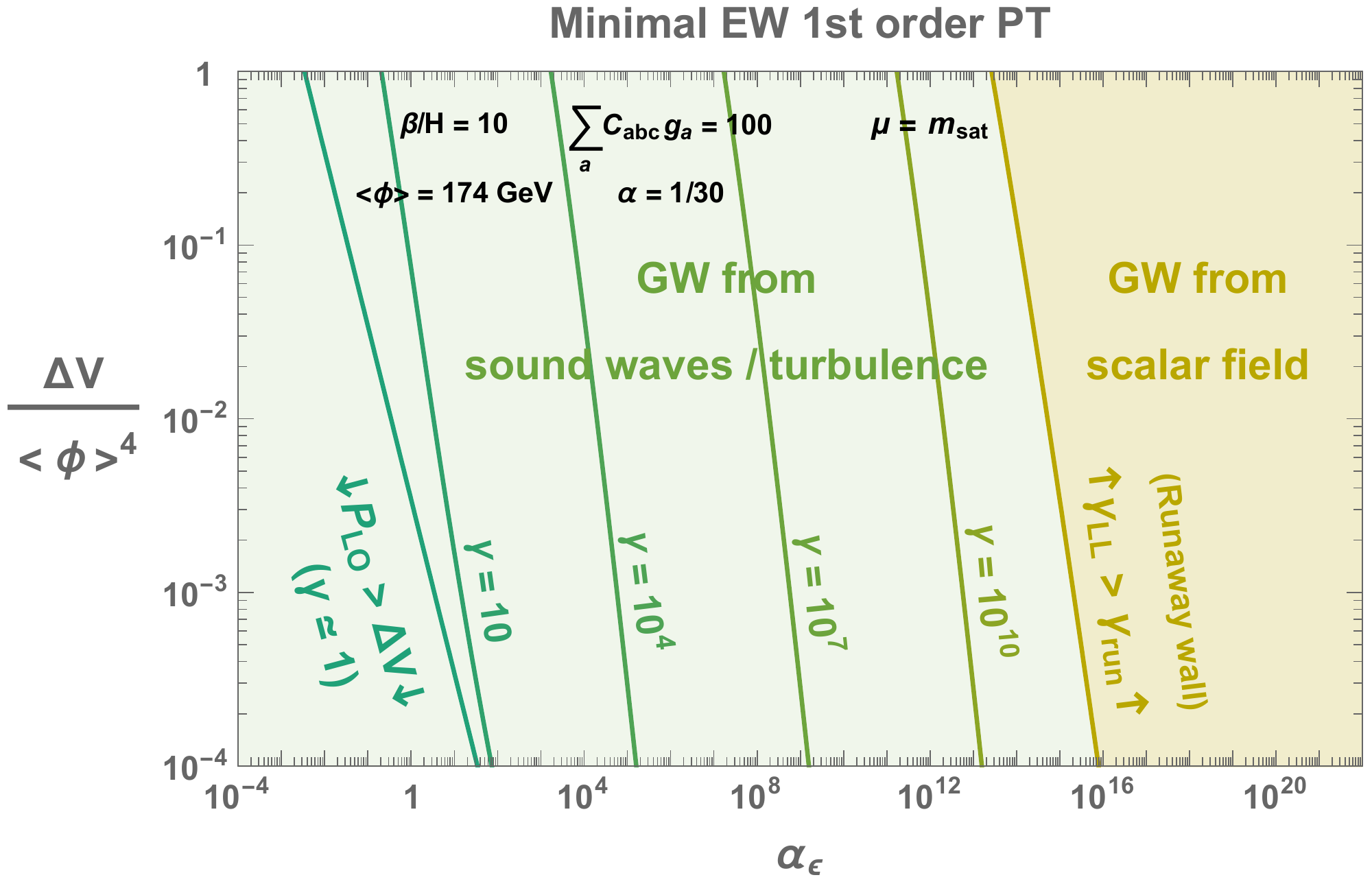}}}
\end{adjustbox}
\caption{\it \small  \label{fig:bubble_wall_velocity}
Bubble wall Lorentz factor $\gamma$ computed with the results of this paper, see Eq.~\eqref{eq:gamma_final}.
In the \textbf{yellow} region, bubble walls collide before they reach their terminal velocity $\gamma_{\LL}$.
This is the so-called run-away regime in which GW are dominantly produced by the scalar field gradient.
In contrast, in the \textbf{green} region, bubble wall reach their terminal velocity before collision and the GW signal is dominated by sound-waves and turbulence. 
In the right-hand panel, $\alpha_{\epsilon}\equiv \Delta V/\rho_{\rm rad} = \left(T_{\rm start}/T_{\rm nuc}\right)^4$ is the latent heat of the phase transition in the Bag model \cite{Espinosa:2010hh}, $\Delta V$ is the vacuum energy of the phase transition.
For $\mathcal{P}_\LO$ we use Eq.~\eqref{eq:PLO} and account for the contributions from $t$, $W^\pm$, $Z$ and $h$.
}
\end{figure}

\subsection{The terminal Lorentz factor}
\label{sec:Lorentz_factor}

Thanks to the dependence of the pressure on $\gamma$, bubble walls cannot be accelerated forever but instead they reach a terminal Lorentz factor, $\gamma_{\LL}$, when the driving pressure from the vacuum energy $\Delta V$ is compensated by the friction pressure
\begin{equation}
\Delta V = \mathcal{P}_{\LO} + \mathcal{P}_{\LL}(\gamma_{\LL}),
\end{equation}
where $ \mathcal{P}_{\LO}$ and $\mathcal{P}_{\LL}$ are given by Eq.~\eqref{eq:PLO} and Eq.~\eqref{eq:PNLO} for non-confining PTs.
Introducing $\mathcal{P}_{\LL} \equiv \gamma P_{\LL}$, we obtain 
\begin{align}
\gamma_{\LL} &= \frac{\Delta V - \mathcal{P}_{\LO}}{P_{\LL}}\label{eq:gamma_infty}\notag\\
&\simeq 3.3 \times 10^{12} \times \left(\frac{g_*}{100}\right)^{\!3/4} \left(\frac{1/30}{\alpha} \right)^{\!3/2} \left(\frac{10^{-4}\,T_{\rm start}}{T_{\rm nuc}}\right)^{\!3}\notag\\
&\qquad\qquad\quad~~ \times \left(\frac{ 4\times 10\times 100}{ \kappa  \ln (m_{c,h} / \mu) \sum_{a,b,c} \nu_a g_a C_{abc}}\right)\left( \frac{\sqrt{2\pi \alpha} \left<\phi\right>}{m_{c,h}} \right)\left(\frac{\Delta V}{0.1\left<\phi\right>^4}\right)^{\!1/4},
\end{align}
where in the second line we assumed $\Delta V \gg \mathcal{P}_{\LO}$. 
In the non-abelian scenario, the value of the IR cut-off $\mu$ depends on $\gamma$, cf. Eq.~\eqref{eq:mu_msc}, such that the computation of $\gamma_{\LL}$ in Eq.~\eqref{eq:gamma_infty} might require iterations.

Before pressures equilibrate, $\gamma$ grows linearly with the bubble radius, see App.~\ref{app:run_away_Lorentz_factor}.
It can happen that bubble wall collisions occur before $\gamma$ saturates to its terminal value $\gamma_{\LL}$.
This is the so-called run-away regime. In that case, the bubble wall Lorentz factor at the time of collision is given by (see App.~\ref{app:run_away_Lorentz_factor})
\begin{align}
\gamma_{\rm run} 
&\simeq \frac{R_{\rm coll}}{3 L_{w, {\rm tot}}}
\simeq \frac{\beta^{-1}}{c_w T_{\rm nuc}^{-1}}
\nonumber \\
&\simeq 3 \times 10^{10} \times
\left(\frac{1}{c_{\rm w}}\right) \left(\frac{100}{g_*}\right)^{\!1/4} \left(\frac{T_{\rm nuc}}{10^{-4}\,T_{\rm start}}\right)\left(\frac{10}{\beta/H_*}\right)\left(\frac{\rm TeV}{\left<\phi\right>}\right)\left(\frac{0.1\left<\phi\right>^4}{\Delta V}\right)^{\!1/4},
\label{eq:gamma_run}
\end{align}
where $L_{w, {\rm tot}} = c_{\rm w} / T_{\rm nuc}$, and in the following we use $c_{\rm w} = 1$ for simplicity.
In the general case, the Lorentz factor at collision time is given by 
\begin{equation}
\gamma_{\rm coll} \simeq \textrm{Min}\big[ \gamma_{\LL},\, \gamma_{\rm run}\big].
\label{eq:gamma_final}
\end{equation}
We show the bubble wall Lorentz factor in Fig.~\ref{fig:bubble_wall_velocity}.

\begin{figure}[t]
\centering
\begin{adjustbox}{max width=0.7\linewidth,center}
\raisebox{0cm}{\makebox{\includegraphics[ width=0.7\textwidth, scale=1]{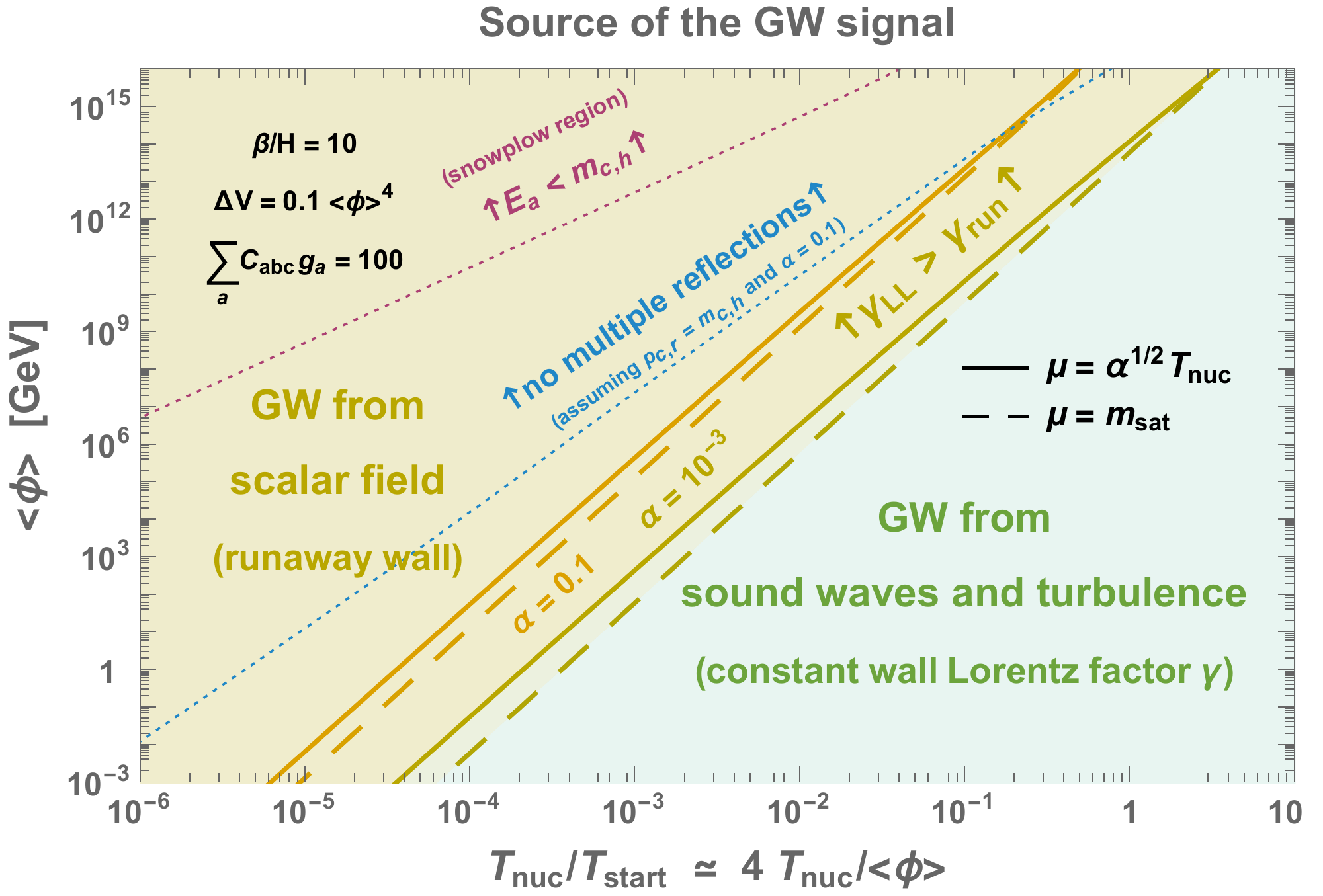}}}
\end{adjustbox}
\caption{\it \small
\label{fig:bubble_wall_velocity_phi}
In the \textbf{yellow} region, bubble walls collide before reaching the terminal Lorentz factor and GW are sourced by anisotropic stress tensor of the scalar field gradient localized at the bubble wall.
In the \textbf{green} region, bubble walls reach their terminal velocity before colliding and GW are sourced by the anisotropic stress tensor of the fluid shells dragged by bubble walls.
We vary the gauge coupling constant $\alpha$ from $10^{-3}$ to $0.1$ and the IR cut-off $\mu$ from the thermal mass $\alpha^{1/2} T_{\rm nuc}$ to the one $m_{\rm sat}$ treated in Sec.~\ref{sec:IR_cut_off}.
We assumed that the pressure $\mathcal{P}$ is given by Eq.~\eqref{eq:Ptot}.
We can see that GW are sourced by scalar field in presence of large supercooling, or when the 1stOPT occurs at very high energies $\left<\phi\right> \gtrsim 10^{10}~\rm GeV$. 
On the right of the blue dashed line drawn for $\alpha=0.1$, effects due to multiple reflections of vector bosons in front of the wall must be considered, see Sec.~\ref{sec:reflected_bosons} and App.~\ref{app:fate_reflected_particles}, but we leave them for further studies.
On the left of the purple dashed lines the pressure is smaller than in Eq.~\eqref{eq:PNLO}, see App.~\ref{sec:snowball}, so that bubble walls still run away.
}
\end{figure}

\subsection{Source of the GW signal}
\label{sec:GW_signal}

Upon comparing Eq.~\eqref{eq:gamma_infty} and Eq.~\eqref{eq:gamma_run}, we find that bubble collisions occur before the terminal velocity is reached (run-away regime) when
\begin{align}
&\gamma_{\rm run} \lesssim \gamma_{\LL} \nonumber \\[0.2cm] &\implies \quad \frac{T_{\rm nuc}}{T_{\rm start}} \lesssim 3.2 \times 10^{-4} \notag\\ 
&\qquad\qquad\qquad\quad~ \times \left( \frac{\left< \phi \right>}{\text{\rm TeV}}\frac{ \beta/H_*}{10} \frac{g_{*}}{\sum_{a,b,c} \nu_a g_a C_{abc}}\frac{4 \times 10}{\kappa \ln (m_{c,h}/\mu)}\right)^{\!1/4} \left(\frac{1/30}{\alpha} \right)^{\!3/8} \left(\frac{\Delta V}{0.1\left<\phi\right>^4}\right)^{\!1/8}.
\label{eq:run_away_wall_cond}
\end{align}
Whether the walls run away or not changes the energy budget of the expanding bubbles drastically~\cite{Espinosa:2010hh,Ellis:2019oqb}, and thus changes the dominant contribution to the GW production.
If $\gamma_{\rm run} \lesssim \gamma_{\LL} $, most of the vacuum energy is used for accelerating the bubble walls and the dominant source of GW is the anisotropic part of the stress-energy tensor of the scalar field kinetic term.
In contrast, if $\gamma_{\rm run} \gtrsim \gamma_{\LL} $, then most of the vacuum energy is converted into thermal and kinetic energy of the thermal plasma through friction, leading to a GW spectrum dominated by the contribution from sound waves and turbulence, though the fate of the highly relativistic fluid must be investigated carefully~\cite{Jinno:2019jhi,Ellis:2020awk}.\footnote{
Note that it is not yet confirmed if such relativistic and localized fluid motion successfully develops into sound waves, which we define here to be the fluid motion well approximated by a linear equation of motion $(\partial_t^2 - c_s^2 \nabla^2) \vec{v} \simeq 0$ (up to the vorticity term).
This is one of the necessary conditions for the GW enhancement from sound waves~\cite{Hindmarsh:2013xza,Hindmarsh:2015qta,Hindmarsh:2017gnf,Hindmarsh:2016lnk,Cai:2018teh,Hindmarsh:2019phv,Guo:2020grp,Wang:2021dwl}
}
We do not report the formula for the GW spectrum here but we instead refer to the reviews \cite{Caprini:2015zlo, Caprini:2019egz}.
The classification of strong 1stOPT according to their GW sources can be visualized in Fig.~\ref{fig:bubble_wall_velocity} in the case of minimal electroweak phase transition models with $\left<\phi\right> = 174~$GeV, and in Fig.~\ref{fig:bubble_wall_velocity_phi} in the case of strong 1stOPT of arbitrary scale $\left<\phi\right>$.

\FloatBarrier

\section{Summary and outlook}
\label{sec:conclusion}

Particles passing the bubble wall of a cosmological first-order phase transition undergo splitting radiation.
This is analogous to the classical radiation emitted whenever a charged particle passes from one medium into another, see e.g. Chap.~13 of~\cite{Jackson:1998nia}.
Radiation from particles in the cosmological bath exert a pressure on the bubble walls which affect their velocity, and in turn the physics of quantities that depend on it (gravity waves, dark matter, the baryon asymmetry, primordial black holes, topological defects, etc). 
In this paper we made progress in the computation of this pressure.

In Sec.~\ref{sec:NLO_pressure} and App.~\ref{app:vertex}, we reviewed the perturbative splitting probability, which contains an IR logarithmic divergence.
We improved over previous literature by discussing possible origins for the IR cutoff in Sec.~\ref{sec:IR_cut_off}, and by quantifying the effect of finite wall thickness and other approximations in App.~\ref{app:validity_approximations}.

In the regime of large gauge coupling constant and/or large supercooling, the probability can exceed unity, which calls for resummation, see Table~\ref{tab:tablePE}.
In Sec.~\ref{sec:all_order}, we performed the resummation at all leading-log orders, at both real and virtual levels, using the master formula in Eq.~\eqref{eq:mean_O}.
We found that the averaged momentum $\left< \Delta p \right>$ transferred to the wall is IR-dominated, more precisely it is dominantly due to radiated gauge bosons with energies in the ballpark of their mass in the broken phase (the order parameter), $m_{c,h}$.
We also found that the contribution from the reflected vector bosons is at least of the same order of the contribution from the transmitted ones.
We pointed out additional novel effects due to the population of reflected bosons in Sec.~\ref{sec:reflected_bosons}, and we left their detailed study to future works.
In Sec.~\ref{sec:MC}, we confirmed our analytical result using a Monte-Carlo simulation of the splitting processes, see Fig.~\ref{fig:DeltaP_ana_vs_MC}.

Based on these results, we deduced the friction pressure on the wall in Sec.~\ref{sec:bubble_velocity}.
Our final result for the pressure, with leading logs resummed, is
\beq
\mathcal{P}_\text{LL}
= \mathcal{O}(1)
\times
g^2 \,\gamma \,m_{c,h}\, \Tnuc^3 \log\left(\frac{m_{c,h}}{\mu}\right)\,,
\eeq
where $\gamma$ is the Lorentz boost of the bubble wall and $T_{\rm nuc}$ is the nucleation temperature,\footnote{
This is inconsistent with the  findings of Ref.~\cite{Hoche:2020ysm}.
We suggest that the origin of the discrepancy lies in the fact that Ref.~\cite{Hoche:2020ysm} effectively assumes energy-momentum conservation at the bubble wall, see App.~\ref{app:comment_hoeche} for more details.
}
$g$ is the gauge coupling, $m_{c,h}$ is the mass of the gauge bosons in the broken phase and $\mu$ is an IR cutoff at most of the order of a fraction of $m_{c,h}$.
This result applies as long as the energy of the incoming particles in the wall frame well exceeds the particle masses.
We provided more precise and ready-to-use expressions for $\mathcal{P}_\text{LL}$ in Eq.~\eqref{eq:PNLO} and for $\mu$ in Eq.~(\ref{eq:mu_msc}).
Our results are compatible with the friction pressure for confining phase transition, which also scales as $\mathcal{P} \propto \gamma\,f\,T_{\rm nuc}^3$ ($f$ being the confining scale), as first computed in~\cite{Baldes:2020kam} with the formalism of the gluon flux tube, see Sec.~\ref{par:confining_PT}.
In Sec.~\ref{sec:Lorentz_factor} we finally discussed implications for the terminal Lorentz factor, which we display in Fig.~\ref{fig:friction_pressure_bubble_wall_velocity} for different values of the energy scale of the PT.
In Sec.~\ref{sec:GW_signal}, we deduced
the source of the GW spectrum depending on the energy scale of the transition $\left<\phi\right>$ and on the amount of supercooling, see Fig.~\ref{fig:bubble_wall_velocity_phi}.

Our results constitute a step towards a better understanding of cosmological strong first-order phase transitions.
Future directions that may be worth a better understanding include i) effects from the fate of reflected vector bosons, like the impact on the pressure of multiple reflections, see Sec.~\ref{sec:reflected_bosons} and App.~\ref{app:fate_reflected_particles}; ii) the contribution from the longitudinal vector boson (see footnote~\ref{footnote:longi}); iii) an improved treatment of the region of very large occupation number $f_c$, for example by including the effect of Bose enhancement on the radiation (see footnote~\ref{footnote:Caputo}) and going beyond our perturbative treatment (see Sec.~\ref{sec:phase_space_sat}); iv) the effect of the multiple wall oscillations on the soft radiation, and hence on the pressure (see footnote~\ref{footnote:multiple_wall_oscillation}).

\section*{Acknowledgements}

We thank Dietrich Bodeker and Guy Moore for very helpful discussions on the effects of mutual interactions, G\'eraldine Servant for useful comments on the manuscript, and Stefan H\"oche, Andrew Long, Jessica Turner and Yikun Wang for constructive and useful correspondence about their Ref.~\cite{Hoche:2020ysm}. We thank Aleksander Azatov and Miguel Vanvlasselaer for valuable discussions on the longitudinal vector boson.
YG thanks Lorenzo Zoppi for very fruitful discussions on Sudakov resummation, Jim Talbert and Jessica Turner for correspondence, and Andrea Caputo for interesting conversations.
RJ thanks Teppei Kitahara and Yasuhiro Yamamoto for useful discussions.
FS thanks Matteo Cacciari and Diego Redigolo for useful discussions. 
YG is grateful to the Azrieli Foundation for the award of an Azrieli Fellowship.
The work of RJ is supported by Grants-in-Aid for JSPS Overseas Research Fellow (No.~201960698).
The work of RJ is supported by the Spanish Ministry for Science and Innovation under grant PID2019-110058GB-C22 and grant SEV-2016-0597 of the Severo Ochoa excellence program.
This work is supported by the Deutsche Forschungsgemeinschaft 
under Germany's Excellence Strategy -- EXC 2121 ,,Quantum Universe`` -- 390833306.
YG and FS are grateful to GGI for hospitality and partial support during the completion of this work.

\appendix

\section{Computation of the splitting radiation vertex}
\label{app:vertex}

\subsection{Splitting $ffV_T$}
We compute the transition amplitude of the splitting radiation $X(p_a) \to V(p_c)~ X(p_b)$ with $X$ being a fermion.
See \cite{Altarelli:1977zs} for the pioneering paper and \cite{Ciafaloni:2010ti,Chen:2016wkt,Chen:2019dkx,Bodeker:2017cim} for more recent derivations.
The transition amplitude reads
\begin{equation}
iV =  g \,\bar{u}(p_{a})\,\gamma^{\mu} \,u(p_{b})\, \epsilon_{\mu}(p_c).
\end{equation}
The squared amplitude averaged over fermion spins is\footnote{
The quantity which we call $|V|^2$ is actually $\frac{1}{2}\sum |V|^{2}$.
}
\begin{equation}
|V|^{2} = 2 g^2\, \epsilon(p_c)\cdot \epsilon^*(p_c)(m_am_b -p_a\cdot p_b) + 4 g^2\, (p_a \cdot \epsilon(p_c))(p_b\cdot \epsilon^*(p_c)). \label{eq:Vsqr_app}
\end{equation}
Note that since the momentum along $z$ is not conserved, the Ward identity is not satisfied, see App.~\ref{app:ward_identity}, and we cannot use the standard replacement $\sum_{\rm pol.} \epsilon_\mu \epsilon_\nu^* \to -g_{\mu\nu}+\cdots$.
Instead, we must sum over the physical polarizations $\epsilon_{+}$ and $\epsilon_{-}$ \cite{Altarelli:1977zs,Feige:2013zla}.
In the basis used for writing Eq.~\eqref{eq:pa}, \eqref{eq:pb} and \eqref{eq:pc}, the transverse polarizations of the vector boson $c$ read
\begin{align}
\label{eq:epsilon_p}
\epsilon_{+}(p_c) &= \frac{1}{\sqrt{2-2\mu_c^2}}(0,\sqrt{1-\mu_c^2-\theta^2},+i\sqrt{1-\mu_c^2},-\theta), \\
\epsilon_{-}(p_c) &= \frac{1}{\sqrt{2-2\mu_c^2}}(0,\sqrt{1-\mu_c^2-\theta^2},-i\sqrt{1-\mu_c^2},-\theta), 
\end{align}
where $\theta \equiv k_{\perp}/xE_{a}$ is the emission angle and $\mu_{c} \equiv m_c/xE_a$ is the mass fraction.
We can check that they satisfy $\epsilon(p_c)\cdot p_c = 0$. 
At leading order in $x\ll 1$, $\theta \ll 1$, we get
\begin{equation}
|V_+|^{2} \simeq 2g^2C_{abc}\frac{k_{\perp}^2}{x^2}, \qquad |V_-|^{2} \simeq 2g^2C_{abc}\frac{k_{\perp}^2}{x^2}.\label{eq:V+_V-} 
\end{equation}
where $C_{abc}$ is the charge factor of the gauge group, e.g. for $SU(N)$ we have $C_{qqg} = \frac{N^2-1}{2N}$.

\subsection{Splitting $\phi\phi V_{T}$}

When $a$ and $b$ are scalars, the transition amplitude reads 
\begin{equation}
iV =  g \,(p_a^\mu +p_b^\mu)\, \epsilon_{\mu}(p_c).
\label{eq:vertex_scalar}
\end{equation}
At leading order in $x\ll 1$, $\theta \ll 1$, we compute 
\begin{equation}
|V_+|^{2} \simeq 2g^2C_{abc}\frac{k_{\perp}^2}{x^2}, \qquad |V_-|^{2} \simeq 2g^2C_{abc}\frac{k_{\perp}^2}{x^2}. \label{eq:V+_V-_scalar} 
\end{equation}
One possible example is a dark $U(1)_D$ with $C_{\phi\phi V} = 1$.

\subsection{Splitting $V_{T} V_{T} V_{T}$}

When $a$ and $b$ are bosons, the transition amplitude reads 
\begin{equation}
iV =  -g \left( -(p_a +p_c) \cdot \epsilon^*_b (\epsilon_a \cdot \epsilon_c^*) + (p_c - p_b) \cdot \epsilon_a (\epsilon_c^* \cdot \epsilon_b^*) +(p_a +p_b) \cdot \epsilon^*_c (\epsilon_a \cdot \epsilon_b^*)  \right),
\label{eq:vertex_boson}
\end{equation}
where the parameters related to the gauge structure are implicit.
In the basis used for writing Eq.~\eqref{eq:pa}, \eqref{eq:pb} and \eqref{eq:pc}, the transverse polarizations of the vector boson $a$, $b$ and $c$ read
\begin{align}
\label{eq:epsilon_T_a}
&\epsilon_{a}^\pm = \frac{1}{\sqrt{2-2(\mu_a x)^2}}(0,\sqrt{1-(\mu_a x)^2},\pm i \sqrt{1-(\mu_a x)^2},0),  \\
\label{eq:epsilon_T_b}
&\epsilon_{b}^\pm = \frac{1}{\sqrt{2-2(\mu_b\chi)^2}}(0,\sqrt{1-(\mu_b\chi)^2-(\theta\chi)^2},\pm i \sqrt{1-(\mu_b\chi)^2},\theta\chi),  \\
\label{eq:epsilon_T_c}
&\epsilon_{c}^\pm = \frac{1}{\sqrt{2-2\mu_c^2}}(0,\sqrt{1-\mu_c^2-\theta^2},\pm i\sqrt{1-\mu_c^2},-\theta), 
\end{align}
where $\pm$ are the helicities of $a$, $b$ and $c$, $\chi \equiv x/(1-x)$, $\theta \equiv k_{\perp}/xE_{a}$ is the emission angle and $\mu_{a,b,c} \equiv m_{a,b,c}/xE_{a}$ are the mass fractions.
We compute 
\begin{align}
&|V_{+++}|^{2} \simeq 2g^2C_{abc}\frac{k_{\perp}^2}{x^2}, \qquad \qquad |V_{++-}|^{2} \simeq 2g^2C_{abc}\frac{k_{\perp}^2}{x^2},\qquad \qquad  |V_{-++}|^{2} = 0,\\
&|V_{---}|^{2} \simeq 2g^2C_{abc}\frac{k_{\perp}^2}{x^2}, \qquad\qquad  |V_{--+}|^{2} \simeq 2g^2C_{abc}\frac{k_{\perp}^2}{x^2},\qquad \qquad |V_{+--}|^{2} = 0 , \label{eq:V+_V-_boson_bef_sum} 
\end{align}
where $C_{abc}$ encapsulates the parameters related to the gauge structure, e.g. for $SU(N)$ we have $C_{ggg} = N$.
Upon averaging over initial polarizations, we obtain
\begin{equation}
|V_+|^{2} \simeq 2g^2C_{abc}\frac{k_{\perp}^2}{x^2}, \qquad |V_-|^{2} \simeq 2g^2C_{abc}\frac{k_{\perp}^2}{x^2}. \label{eq:V+_V-_boson} 
\end{equation}
where $\pm$ stands for the helicity of $c$.

\section{On the validity of the diverse approximations}
\label{app:validity_approximations}

\subsection{Mode functions}
\label{app:mode_function}

\subsubsection{Finite wall thickness: WKB method}
\label{app:finite_wall_thickness}

\paragraph{Outside the wall.}

As discussed in the next paragraph, far away from the wall, the mode functions are expected to be a superposition of plane waves, solutions of the Klein-Gordon equation $\chi^{''}(z) + (p^{z2}(z)/\hbar^2) \chi(z) = 0$.

\paragraph{Inside the wall.}
\label{sec:finite_wall_thickness}

\begin{figure}[h!]
\centering
\begin{adjustbox}{max width=1\linewidth,center}
\raisebox{0cm}{\makebox{\includegraphics[ width=0.5\textwidth, scale=1]{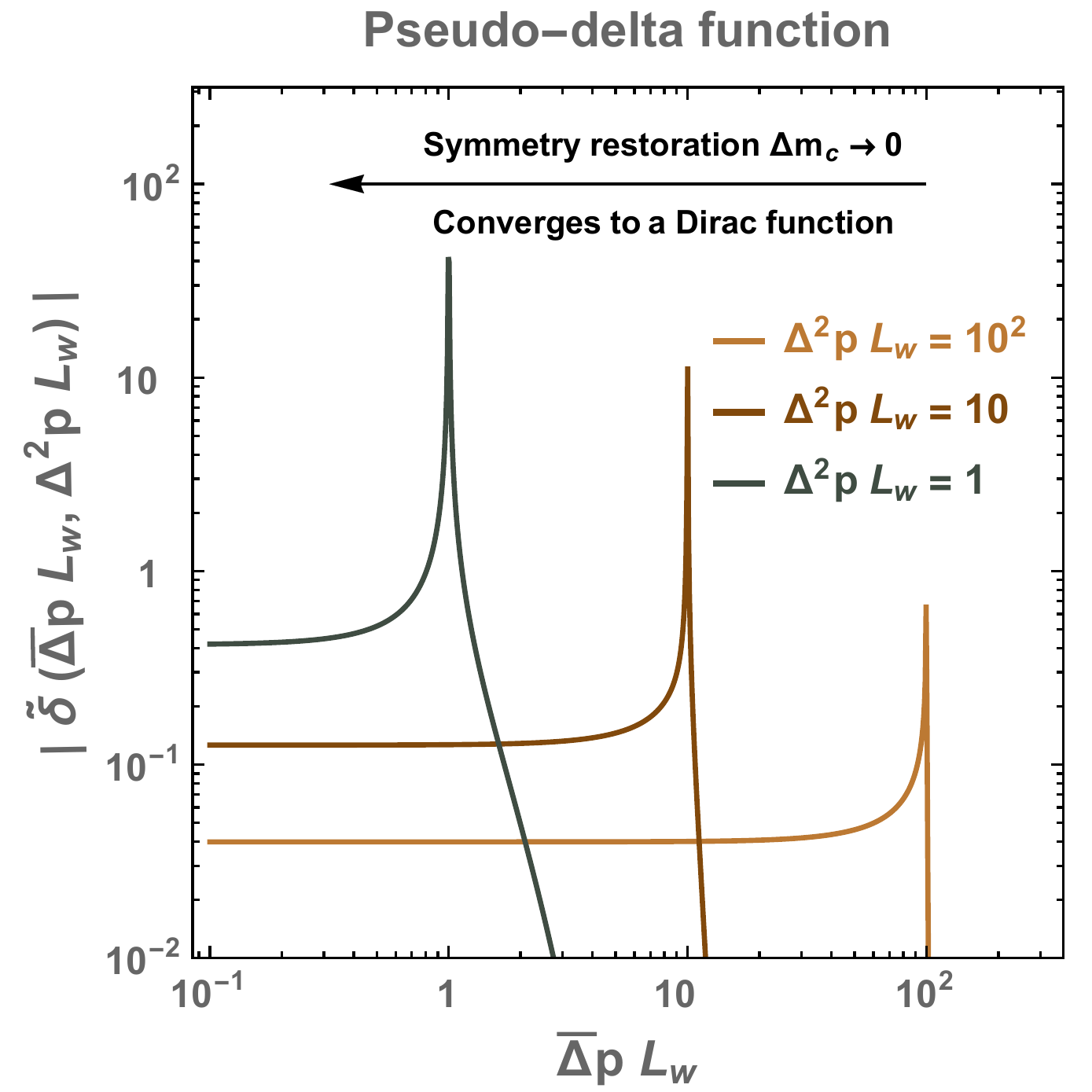}}}
\end{adjustbox}
\caption{\it \small  \label{fig:pseudo_delta_function}
The function $\tilde{\delta}(\bar{\Delta} pL_{\rm w},\,\Delta^2 p L_{\rm w}) $ tends to the Dirac $\delta$ function for $\Delta^2 p \equiv \Delta m_c^2/4xE_a \to 0$.}
\end{figure}

\begin{figure}[h!]
\centering
\begin{adjustbox}{max width=1\linewidth,center}
\raisebox{0cm}{\makebox{\includegraphics[ width=0.5\textwidth, scale=1]{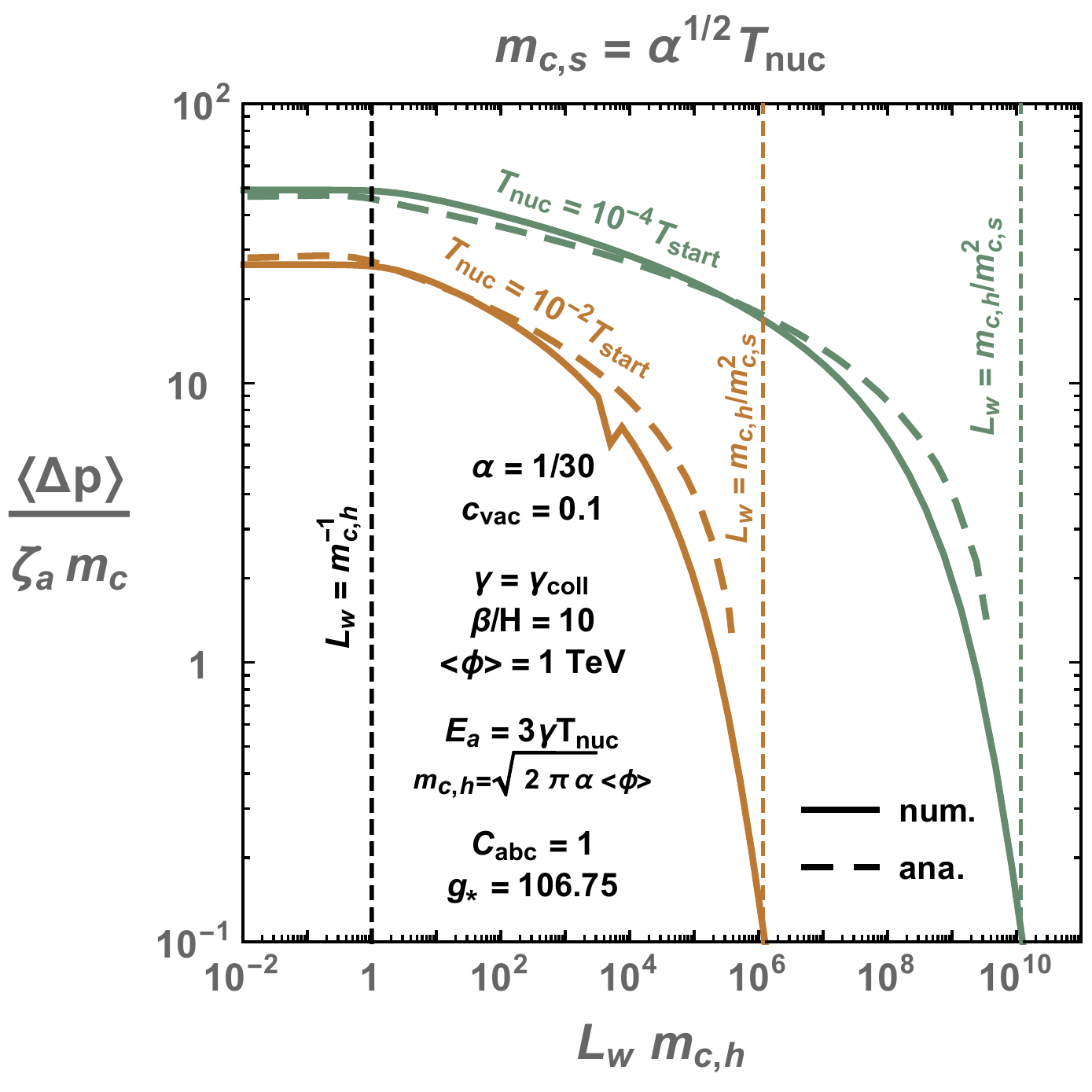}}}
\raisebox{0cm}{\makebox{\includegraphics[ width=0.5\textwidth, scale=1]{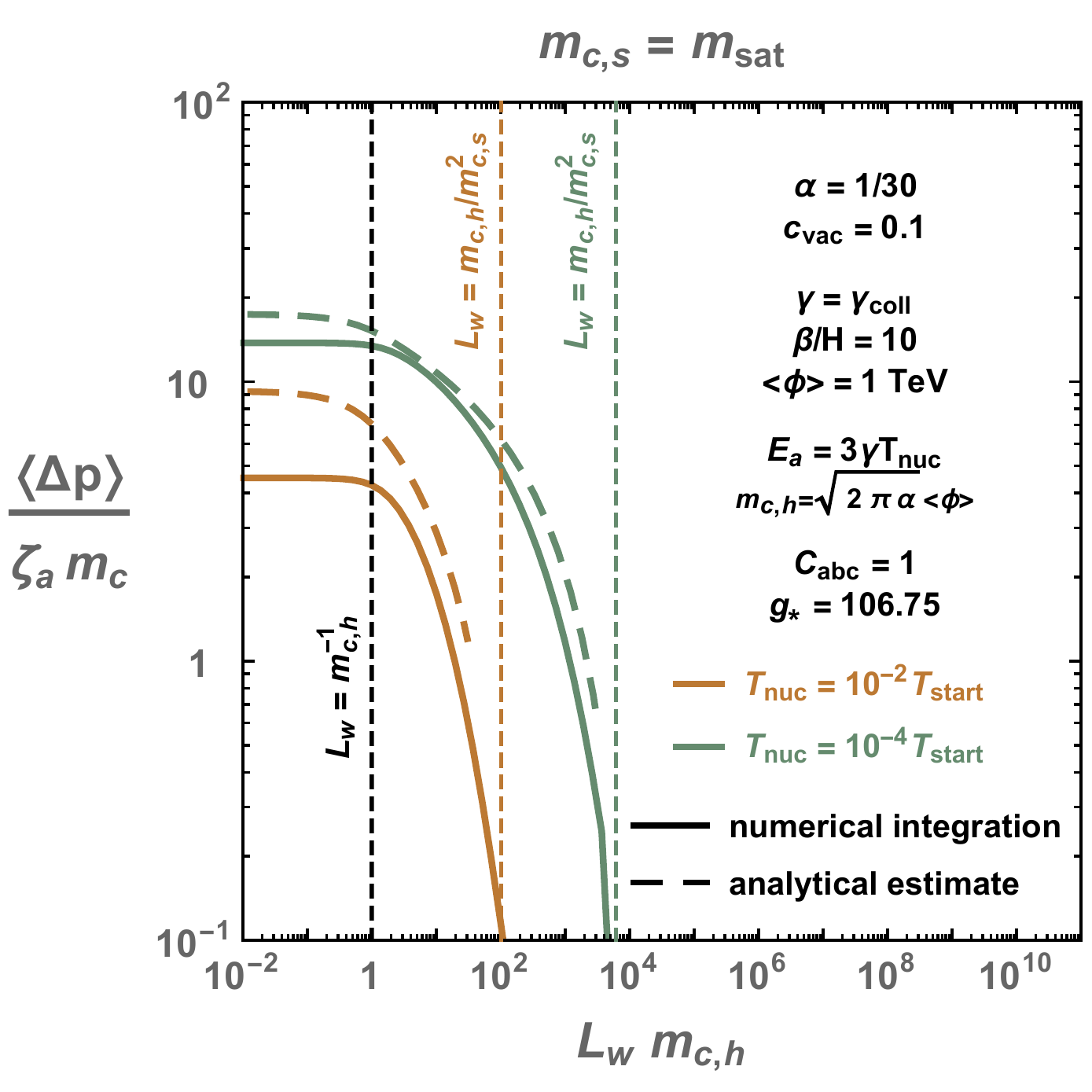}}}
\end{adjustbox}
\caption{\it \small
\label{fig:finite_wall_DeltapL}
The exchanged momentum $\left< \Delta p\right>$ becomes first logarithmically suppressed when the wall thickness satisfies $L_{\rm w} \gtrsim m_{c,h}^{-1}$ and then exponentially suppressed around $L_{\rm w} \sim m_{c,h}/m_{c,s}^{2}$.
Solid lines show the numerically integrated result in Eq.~\eqref{eq:Deltap_finite_wall_thickness} while dashed lines show the analytical estimate in Eq.~\eqref{eq:Deltap_finite_wall_thickness_approx}.
We fix the IR cut-off $\mu$ (or equivalently the mass $m_{c,s}$ in the symmetric phase) to the thermal mass $\alpha^{1/2} T_{\rm nuc}$ in the \textbf{left} panel, see Eq.~\eqref{eq:mcs_thermal}, and to the screening mass $m_{\rm sat}$ of phase-space-saturated non-abelian vector boson bath, in the \textbf{right} panel, see Eq.~\eqref{eq:kperp_satur}.
}
\end{figure}

Instead, inside the wall we first propose using the WKB approximation, which is valid in the limit $\hbar  p^{z'}/p^{z2} \ll 1$.
It consists of injecting $\chi(z)=e^{\frac{i}{\hbar}(S_0+S_1\hbar + S_2  \hbar^2 +\cdots )}$ in the differential equation and of matching the terms which are of the same order in $\hbar$.
Then we solve the infinite set of equations order by order in $\hbar$. We obtain 
\begin{equation}
\chi(z) = \sqrt{\frac{p^z_s}{p^z(z)}}\exp \left[\frac{i}{\hbar}\int dz \,p^z  + i \hbar \int dz\, \frac{1}{p^z}\left[\frac{3}{8}\left( \frac{p^{z'}}{p^z} \right)^2 - \frac{p^{z''}}{4p^z}\right]+\cdots\right].
\label{eq:chi_c_WKB}
\end{equation}
At first order in $\hbar$ and upon neglecting the prefactor, the ${\cal M}$-matrix defined in Eq.~\eqref{eq:M_matrix} becomes
\begin{equation}
\mathcal{M} \simeq V \int dz \exp \left( i \int_0^z dz' \Delta p(z')  \right),\qquad \Delta p(z) \equiv p_a^z (z) - p_b^z (z) - p_c^z (z) \simeq \frac{k_{\perp}^2+m_c^2(z)}{2xE_a} , \label{eq:M_matrix_WKB_tanh}
\end{equation}
where we have used that $V$ is $z$-independent, see Eq.~\eqref{eq:vertex_BM}.
We assume a tanh wall profile
\begin{equation}
m_c^2(z) = \frac{m_{c,h}^2+m_{c,s}^2}{2} + \frac{m_{c,h}^2-m_{c,s}^2}{2} \tanh(z/L_{\rm w}),
\end{equation}
where $L_{\rm w}$ is the wall thickness.\footnote{
\label{footnote:multiple_wall_oscillation}
The tanh wall profile which we consider here, accounts for the initial rising part of the wall, which interpolates between $m_{c,s}$ and $m_{c,h}$.
However, it neglects the subsequent multiple wall oscillations around $m_{c,h}$.
The thickness of the rising part of the wall is expected to be $L_{\rm w} \simeq c_{\rm vac}^{-1/2} \left<\phi\right>^{-1}$ which for supercooled phase transition can be much smaller than the total thickness of the wall accounting for the multiple oscillations, $L_{\rm w}^{\rm tot} \simeq \Tnuc$, cf. App.~A of \cite{Baldes:2020kam}.
We leave the study of the impact of multiple wall oscillations on particle splitting for future studies.
}
We then obtain the following WKB phase
\begin{equation}
\Delta p(z) = \bar{\Delta} p + \Delta^2 p\,\tanh(z/L_{\rm w}), \label{eq:WKBphase_tanh}
\end{equation}
with
\begin{align}
&\bar{\Delta} p \equiv \frac{\Delta p(+\infty) + \Delta p(-\infty)}{2} \simeq \frac{1}{2xE_a}\left(k_{\perp}^2 + \frac{m_{c,h}^2+m_{c,s}^2}{2}\right) ,\\
& \Delta^2 p \equiv \frac{\Delta p(+\infty) - \Delta p(-\infty)}{2} \simeq \frac{\Delta m_c^2}{4xE_a} \quad \text{with} \quad \Delta m_c^2  \equiv m_{c,h}^2-m_{c,s}^2.
\end{align}
From injecting Eq.~\eqref{eq:WKBphase_tanh} into Eq.~\eqref{eq:M_matrix_WKB_tanh}, we obtain \cite{Bodeker:2017cim}
\begin{equation}
\mathcal{M} \simeq  2\pi V \tilde{\delta}(\bar{\Delta} pL_{\rm w},\,\Delta^2 p L_{\rm w}),
\end{equation}
with
\begin{align}
\tilde{\delta}(\bar{\Delta} pL_{\rm w},\,\Delta^2 p L_{\rm w}) \equiv L_{\rm w}  e^{-i\Delta^2 p\,L_{\rm w}\ln{2} } \times \frac{\Gamma\left(\frac{i}{2}(\bar{\Delta} p -\Delta^2 p)L_{\rm w}\right)\Gamma\left(-\frac{i}{2}(\bar{\Delta} p + \Delta^2 p)L_{\rm w}\right)}{4\pi \Gamma\left(-i\Delta^2 p\,L_{\rm w}\right)}.
\end{align}
As expected, in the limit of vanishing mass difference $\Delta^2 p L_{\rm w}  \to 0$, the function $\tilde{\delta}(\bar{\Delta} pL_{\rm w},\,\Delta^2 p L_{\rm w})$ approaches the Dirac $\delta$ function, see Fig.~\ref{fig:pseudo_delta_function} 
\begin{equation}
\tilde{\delta}(\bar{\Delta} pL_{\rm w},\,\Delta^2 p L_{\rm w})  \xrightarrow[\Delta^2 p\, L_{\rm w}\to 0]{}   \delta(\bar{\Delta} p),
\end{equation}
and conservation of momentum along $z$ is restored.
Using that $|\Gamma(iy)|^2 = \pi/[y \,{\rm sh}{(\pi y)}]$ for real $y$ where ${\rm sh}{\,x}$ is the sinus hyperbolic function \cite{abramowitz1988handbook}, we obtain
\begin{equation}
|\mathcal{M}|^2 = \frac{\pi |V|^2\Delta^2 p L_{\rm w} }{\bar{\Delta} p^2 - \Delta^2 p^2} \times {\rm sh}{\left(\pi \Delta^2 p L_{\rm w}\right)} \Big/ {\rm sh}{\left(\frac{\pi}{2}(\bar{\Delta} p -\Delta^2 p)L_{\rm w}  \right)} {\rm sh}{\left(\frac{\pi}{2}(\bar{\Delta} p +\Delta^2 p)L_{\rm w}  \right)}.
\end{equation}
The splitting probability in Eq.~\eqref{eq:dP_abc_0} reduces to
\begin{equation}
dP_{a \to bc} =\zeta_a\frac{d k_{\perp}^2}{k_{\perp}^2}  \frac{d x}{x}~\Pi(k_\perp) \,\mathcal{W}_{\rm wall}(L_{\rm w},\,k_{\perp}, \,x),
\label{eq:dP_abc_mcs_finite_wall}
\end{equation}
where $\Pi(k_{\perp})$ is defined in Eq.~(\ref{eq:IR_UV_suppression_factor}) and
\begin{align}
\mathcal{W}_{\rm wall} &\equiv
\textrm{shc}\left(\pi \Delta^2 p L_{\rm w}\right) \!\Big/\; \textrm{shc}{\left(\frac{\pi}{2}(\bar{\Delta} p -\Delta^2 p)L_{\rm w}  \right)} \textrm{shc}{\left(\frac{\pi}{2}(\bar{\Delta} p +\Delta^2 p)L_{\rm w}\right)} \notag \\[0.2cm]
&= \textrm{shc}{\left(\frac{\pi L_{\rm w}}{4x E_a} \Delta m_c^2  \right)} \Bigg/\; \textrm{shc}{\left(\frac{\pi L_{\rm w}}{4x E_a} (k_\perp^2 + m_{c,s}^2)  \right)} \textrm{shc}{\left(\frac{\pi L_{\rm w}}{4x E_a} (k_\perp^2 + m_{c,h}^2)  \right)},
\label{eq:Wwall}
\end{align}
with $\textrm{shc}(x) \equiv {\rm sh}{x}/x$. We compute the average exchanged momentum, cf. Eq.~\eqref{eq:Deltap_resummation}
\begin{equation}
\left< \Delta p \right> = \int dP_{a \to bc}\, \Delta p_i ,
\label{eq:Deltap_finite_wall_thickness}
\end{equation}
with $dP_{a \to bc}$ in Eq.~\eqref{eq:dP_abc_mcs_finite_wall} and $\Delta p_i$ in Eq.~\eqref{eq_Delta_p_sum_Delta_pi}  sums over both transmitted and reflected bosons.
We did not find an analytical expression for $\left< \Delta p \right>$ in Eq.~\eqref{eq:Deltap_finite_wall_thickness} so we report its numerical value in Fig.~\ref{fig:finite_wall_DeltapL}.
We nonetheless find an analytical approximation ($E_i(x)$ is the exponential integral special function),
\begin{equation}
\left< \Delta p \right>
\simeq 2\,\zeta_a\,m_{c,h} \,\left[ E_i\left(-\frac{\pi}{2} L_{\rm w} m_{c,h} \right) - E_i\left(-\frac{\pi}{2}\frac{L_{\rm w} m_{c,s}^2}{m_{c,h}} \right)\right] ,
\label{eq:Deltap_finite_wall_thickness_approx}
\end{equation}
that is valid up to $\mathcal{O}(50\%)$. We identify three behaviors for $\left< \Delta p \right>$, 
\begin{equation}
\left< \Delta p \right>  \simeq 2\,\zeta_a\,m_{c,h} \times  \left\{
                \begin{array}{ll}
                 \displaystyle 2\,\text{ln} \left(\frac{m_{c,h}}{\mu}\right) \qquad\qquad \textrm{for}\quad  L_{\rm w} \lesssim m_{c,h}^{-1}, \\[0.5cm]
                   \displaystyle \ln \left( \frac{2m_{c,h}}{\pi L_{\rm w} m_{c,s}^2} \right) ~\qquad\;\,  \textrm{for}\quad m_{c,h}^{-1} \lesssim  L_{\rm w} \lesssim m_{c,h}/m_{c,s}^2, \\[0.6cm]
                  \displaystyle \exp{\left( -\frac{\pi L_{\rm w} m_{c,s}^2}{2m_{c,h}} \right)} \Bigg/ \frac{\pi L_{\rm w} m_{c,s}^2}{2m_{c,h}}~\, \qquad \textrm{for}\quad m_{c,h}/m_{c,s}^2  \lesssim L_{\rm w}.
                \end{array}
              \right.
\label{eq:wall_thickness_matters}
\end{equation}
We conclude that $\left< \Delta p \right>$ decreases logarithmically as soon as $L_{\rm w} \gtrsim m_{c,h}^{-1}$ and exponentially around $L_{\rm w} \sim  m_{c,h}/m_{c,s}^{2}$, see Fig.~\ref{fig:finite_wall_DeltapL}.
We remind the reader that the scaling of the pressure with the Lorentz boost $\gamma$ arises from the phase space integration and not from $\Delta p$, see e.g. Eq.~\eqref{eq:PNLO_deltap}.
We also comment that Eq.~\eqref{eq:wall_thickness_matters} includes both the reflected and the transmitted contributions, so it properly  includes the case where an emitted particle is softer than $m_{c,h}$ and is thus reflected.

In concrete scenarios we expect the wall thickness $L_{\rm w}$ to be of the order of the inverse mass $m_{\phi}^{-1}$ of the scalar field driving the phase transition, for which we estimate $m_{\phi}^2\left< \phi\right>^2\!/2 \simeq \Delta V \equiv c_{\rm vac} \left< \phi\right>^4$ where $\Delta V$ is the vacuum energy difference, which implies
\begin{equation}
L_{\rm w} \simeq c_{\rm vac}^{-1/2} \left< \phi\right>^{-1} = \frac{1.4}{m_{c,h}}\left(\frac{\alpha}{1/30}\frac{0.1}{c_{\rm vac}}\right)^{\!1/2}.
\label{eq:wall_thickness}
\end{equation}
Therefore, we expect the finite wall thickness to bring only logarithmic corrections for $ m_{c,s} \ll m_{c,h}$, see Eq.~\eqref{eq:wall_thickness_matters}.
However, in the regime  $ m_{c,s} \simeq m_{c,h}$, we expect the friction pressure to receive an exponential suppression factor, see Eq.~\eqref{eq:wall_thickness_matters}, in addition to the UV suppression factor contained in $\Pi(k_{\perp})$ in Eq.~\eqref{eq:IR_UV_suppression_factor}.
In Fig.~\ref{fig:DeltaP_mu}, with dotted lines we show the impact of the wall thickness on $\left< \Delta p\right>$, assuming that Eq.~\eqref{eq:wall_thickness} holds.

\paragraph{Thin-wall limit.}

In light of the preceding paragraph, in the regime $m_{c,s} \ll m_{c,h}$, effects coming from the mode functions inside the wall bring at most logarithmic corrections and we can safely approximate the wall profile by a Heaviside function.
This is the subject of the next section, Sec.~\ref{sec:step_potential_1D}. 

\paragraph{WKB breaks down.}

In order to further motivate the use of a Heaviside function, we would like to comment about two difficulties which we have to deal with if we rely on the WKB method, and which arise when the particle $c$ is reflected
\begin{equation}
p_{c,s} \lesssim m_c.
\end{equation}
First, the hypothesis of the WKB approximation, 
\begin{equation}
\hbar  p^{z'}/p^{z2} \lesssim 1 \quad \implies \quad p_c > L_{\rm w}^{-1} \sim m_c,
\end{equation}
breaks down at the turning point, $p_c(z) = 0$, that is present when a particle is reflected, which implies that the WKB expansion $\chi(z)=e^{\frac{i}{\hbar}(S_0+S_1\hbar + S_2  \hbar^2 +\cdots )}$ cannot be used.
Second, the prefactor $\sqrt{p_s^z/p^z(z)}$ of $\chi_c(z)$ in Eq.~\eqref{eq:chi_c_WKB} becomes infinite at the turning point.
Regularization of the WKB solution near turning points has a well-known solution based on the Airy equation \cite{hall2013quantum}.
In order to avoid those difficulties and thanks to fact the the thin wall approximation is a good one (see Eq.~\eqref{eq:wall_thickness_matters} and discussion around it), in this paper we decide to not use the WKB approximation for computing the mode functions, see next section for more details.

\begin{figure}[t]
\centering
\begin{adjustbox}{max width=1\linewidth,center}
\raisebox{0cm}{\makebox{\includegraphics[ width=0.6\textwidth, scale=1]{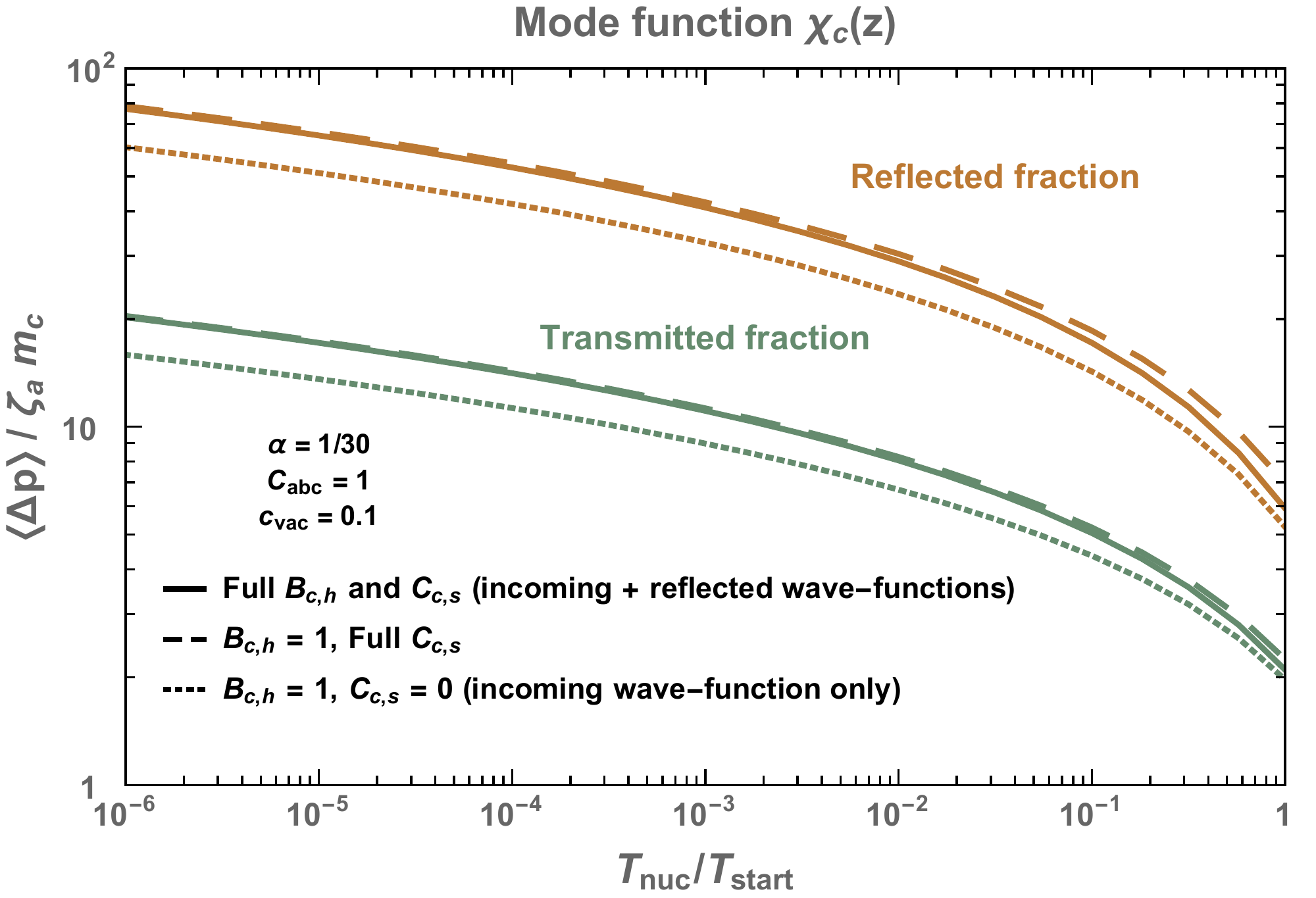}}}
\end{adjustbox}
\caption{\it \small
\label{fig:Delta_p_mode_function}
We compare the approximation $\chi_{c,s}(z) = e^{i p_{c,s} z}$ and $\chi_{c,h}(z) = e^{i p_{c,h} z}$ chosen in the main text, cf. Eq.~\eqref{eq:chi_z}, to the one including the reflected wave-function $\chi_{c,s}(z) = e^{i p_{c,s} z} + C_{c,s}e^{-i p_{c,s} z} $ and $\chi_{c,h}(z) = B_{c,h} e^{i p_{c,h} z} $, see Eq.~\eqref{eq:mode_function_app}.
}
\end{figure}
\subsubsection{Step potential in 1D}
\label{sec:step_potential_1D}

\paragraph{Klein-Gordon equation in presence of a step potential.}

As motivated by the preceding paragraph, we choose to model the wall by a Heaviside function
\begin{equation}
m_j(z) = \left\{
                \begin{array}{ll}
                 0 \qquad ~\;\, \text{when}~z<0, \\
                  m_j \qquad \text{when}~z\geq 0,
                \end{array}
              \right.
\end{equation}
with  $j = a,b,c$.
From solving the Klein-Gordon equation on each sides, we obtain
\begin{equation}
\begin{split}
\chi_{j,s}(z) &= B_{j,s} \, e^{i p_{j,s} z} + C_{j,s} \, e^{-i p_{j,s} z} \qquad \text{when}~z<0, \\
\chi_{j,h}(z) &= B_{j,h} \, e^{i p_{j,h} z} \qquad \qquad  \qquad \quad~~\; \text{when}~z\geq 0, \label{eq:mode_function_app}
\end{split}
\end{equation}
with
\begin{equation}
p_{j,h} = \sqrt{p_{j,s}^2-m_j^2}.
\end{equation}
Note that in contrast to \cite{Bodeker:2017cim,Azatov:2021ifm, Hoche:2020ysm}, we have included the plane wave $e^{-i p_{j,s} z}$ moving in the symmetric phase direction.
We normalize the incoming wave-functions to $B_{j,s} = 1$.
Then we impose continuity of the wave-functions and of their derivative in $z=0$ and we get 
\begin{equation}
C_{j,s} = \frac{p_{j,s} - \sqrt{p_{j,s}^2 - m_j^2}}{p_{j,s} + \sqrt{p_{j,s}^2 - m_i^2}}, \qquad
B_{j,h}= \frac{2p_{j,s}}{p_{j,s} + \sqrt{p_{j,s}^2 - m_j^2}}. \label{eq:Bh_Cs}
\end{equation}
We recover that in the high-energy limit $p_j \gg m_j$, the particle $j$ gets transmitted $C_{j,s} =0$ and $B_{j,h} =1$, while in the low-energy limit $p_j \ll m_j$, the particle $j$ gets reflected  $C_{j,s} = -1$ and $B_{j,h} = 0$. 

\paragraph{${\cal M}$-matrix.}

For relativistic walls, the particles $a$ and $b$ always satisfy $p_a \gg m_a$ and  $p_b \gg m_b$, such that
\begin{equation}
C_{j,s} =0, \quad B_{j,h} =1,\quad {\rm for}~j=1,2.
\end{equation}
The ${\cal M}$-matrix in Eq.~\eqref{eq:M_matrix_2} becomes
\begin{align}
\mathcal{M} &\simeq V_s \int_{-\infty}^0 dz\,e^{iz\frac{A_s}{2E_a} +\epsilon z} + V_s \,C_{c,s}\,\int_{-\infty}^0 dz\,e^{iz\frac{A_{\rm s,r}}{2E_a} +\epsilon z}  + V_h\,B_{c,h}\, \int_{0}^{\infty} dz\,e^{iz\frac{A_h}{2E_a}- \epsilon z} \notag \\[0.1cm]
&= 2iE_a \left(  \frac{V_hB_{c,h}}{A_h} -\frac{V_s C_{c,s}}{A_{s,r}} - \frac{V_s}{A_s}  \right),
\label{eq:M_matrix_mode_function}
\end{align}
with 
\begin{align}
A_h\simeq\frac{m_c^2 + k_{\perp}^2}{x},  \qquad  A_{s,r} \simeq \frac{ k_{\perp}^2}{x} +2x E_a, \qquad A_s \simeq \frac{ k_{\perp}^2}{x}.
\end{align}

\paragraph{Validity of the approximation in the main text.}

In the main text, for the sake of simplicity we neglect the reflected wave-function $C_{c,s}=0$ and we approximate $B_{c,h} = 1$ in Eq.~\eqref{eq:mode_function_app}, see $\chi_c(z)$ in Eq.~\eqref{eq:chi_z} and the corresponding $M$-matrix in Eq.~\eqref{eq:M_matrix_2}.
In Fig.~\ref{fig:Delta_p_mode_function}, we show that the latter approximation underestimates the value of $\left<\Delta p\right>$ computed from Eq.~\eqref{eq:M_matrix_mode_function}, by $\sim 20~\%$. 
The validity of the approximation ($C_{c,s}=0$, $B_{c,h} = 1$)  was expected since among the three terms of Eq.~\eqref{eq:M_matrix_mode_function}, the last one dominates over the others.
In Table~\ref{tab:corrections} of the next section, we compare the error due to the simplification of the mode function to other sources of error, discussed in App.~\ref{app:beyond-relativistic-limit} and in the main text.

\begin{table}[h]
\centering
\begin{adjustbox}{max width=1\linewidth,center}
\begin{tabular}{|c|cc|}
\hline
& & \\
\begin{tabular}[c]{@{}c@{}}
Error estimates for the different approximations\\
in the \textbf{analytical} treatment 
\end{tabular} 
& $\dfrac{\left<\Delta p\right>_R}{\zeta_a m_{c,h}\ln{m_{c,h}/\mu}}$ & $\dfrac{\left<\Delta p\right>_T}{\zeta_a m_{c,h}\ln{m_{c,h}/\mu}}$   \\ 
& & \\
\hline
\begin{tabular}[c]{@{}c@{}}Simple analytical expression in main text, cf. Eq.~\eqref{eq:Delta_p_R} and Eq.~\eqref{eq:Delta_p_T} \\ Transmission and reflection coefficient $B_h =1$, $C_s = 0$ \\ + relativistic,soft, collinear limit    \\
\textcolor{myLightBrown}{\textbf{(Simple analytical)}}   \end{tabular}         & 3.30~ (+0\%)                                          & 0.91~ (+0\%)                                        \\
& & \\
$B_h$ given by  Eq.~\eqref{eq:Bh_Cs}  and $C_s = 0$                 &    4.34~ (+27\%)             & 1.12~ (+21\%)  \\
& & \\
$B_h$ and $C_s$ given by  Eq.~\eqref{eq:Bh_Cs}                     &     4.01 ~(-11\%)                                    &  1.10 ~(-2.0\%) \\
& & \\
Full phase space factor  $\frac{1}{E_c} \to \frac{1}{p_c}$, cf. Eq.~\eqref{eq:phase_space_app}               &  4.15 ~(+3.3\%)    &2.12 ~(+63\%)\\
& & \\
Vertex function beyond the soft-collinear limit, cf. Eq.~\eqref{eq:Vsqr_app}                 &   4.06~ (-2.3\%)                   &    2.10~ (-1.1\%)                                  \\
& & \\
WKB phase $A_s$ $A_{h}$ and $A_{sr}$  beyond the relativistic-soft-collinear limit, cf. Eqs.~\eqref{eq:Ah_exact}, \eqref{eq:As_exact}, \eqref{eq:Asr_exact}       &  3.00~(-30\%)   &   1.33 ~(-45\%)   \\
& & \\
\begin{tabular}[c]{@{}c@{}}
Momentum exchange $\Delta p_z$  in Eq.~\eqref{eq_Delta_p_sum_Delta_pi} beyond the relativistic-soft-collinear limit \\
\textcolor{myBrown}{\textbf{(Refined analytical)}}      
\end{tabular}     &  2.96~ (-1.3\%)                  &  1.51~ (+12\%)   \\
& & \\
\begin{tabular}[c]{@{}c@{}} Finite wall thickness, see Eq.~\eqref{eq:Deltap_finite_wall_thickness} with $L_{\rm w}$ assumed from Eq.~\eqref{eq:wall_thickness}.  
\end{tabular}      &  2.88~ (-2.6\%)                  &  1.49~ (-1.4\%)      \\   
& & \\ 
\begin{tabular}[c]{@{}c@{}}
Total 
\end{tabular}      &  2.88~ (-13\%)                  &  1.49 ~(+48\%)  \\
& & \\
\begin{tabular}[c]{@{}c@{}} Fate of reflected vector bosons, multiple wall oscillations, \\   longitudinal component of the vector boson, \\possible Bose enhancement, beyond perturbative treatment   \\
\end{tabular}     & \qquad \qquad \qquad \qquad \qquad \qquad left for future studies           &    
\\ & & \\ 
\hline
\end{tabular}
\end{adjustbox}
\caption{\it \small
In the table above, we refine the \textcolor{myLightBrown}{\textbf{simplest}} analytical estimate, line after line from top to bottom until we obtain the \textcolor{myBrown}{\textbf{most refined}} one.
The percents $\%$ show the relative differences between two consecutive lines, except for the ones on the line entitled `total' which shows the relative difference between the \textcolor{myLightBrown}{\textbf{simplest}} and the \textcolor{myBrown}{\textbf{most refined}} estimates.
The relative difference between the analytical and numerical treatment can be appreciated in Fig.~\ref{fig:Error_estimates}.
The values of the present table were evaluated for $\mu = 10^{-4}m_{c,h}$ and $\left<\phi\right> = \rm TeV$ as well as the same values of the parameters as in Fig.~\eqref{fig:Error_estimates}.
Corrections due to the presence of reflected bosons in front of the wall, see Sec.~\ref{sec:bubble_velocity}, due to multiple wall oscillations, see footnote~\ref{footnote:multiple_wall_oscillation}, due to the longitudinal vector boson, see footnote~\ref{footnote:longi}, due to Bose enhancement, see footnote~\ref{footnote:Caputo}, and due to non-perturbative effects, see Sec.~\ref{sec:IR_cut_off}, are left for further works.
}
\label{tab:corrections}
\end{table}

\begin{figure}[h!]
\centering
\begin{adjustbox}{max width=1\linewidth,center}
\raisebox{0cm}{\makebox{\includegraphics[ width=0.5\textwidth, scale=1]{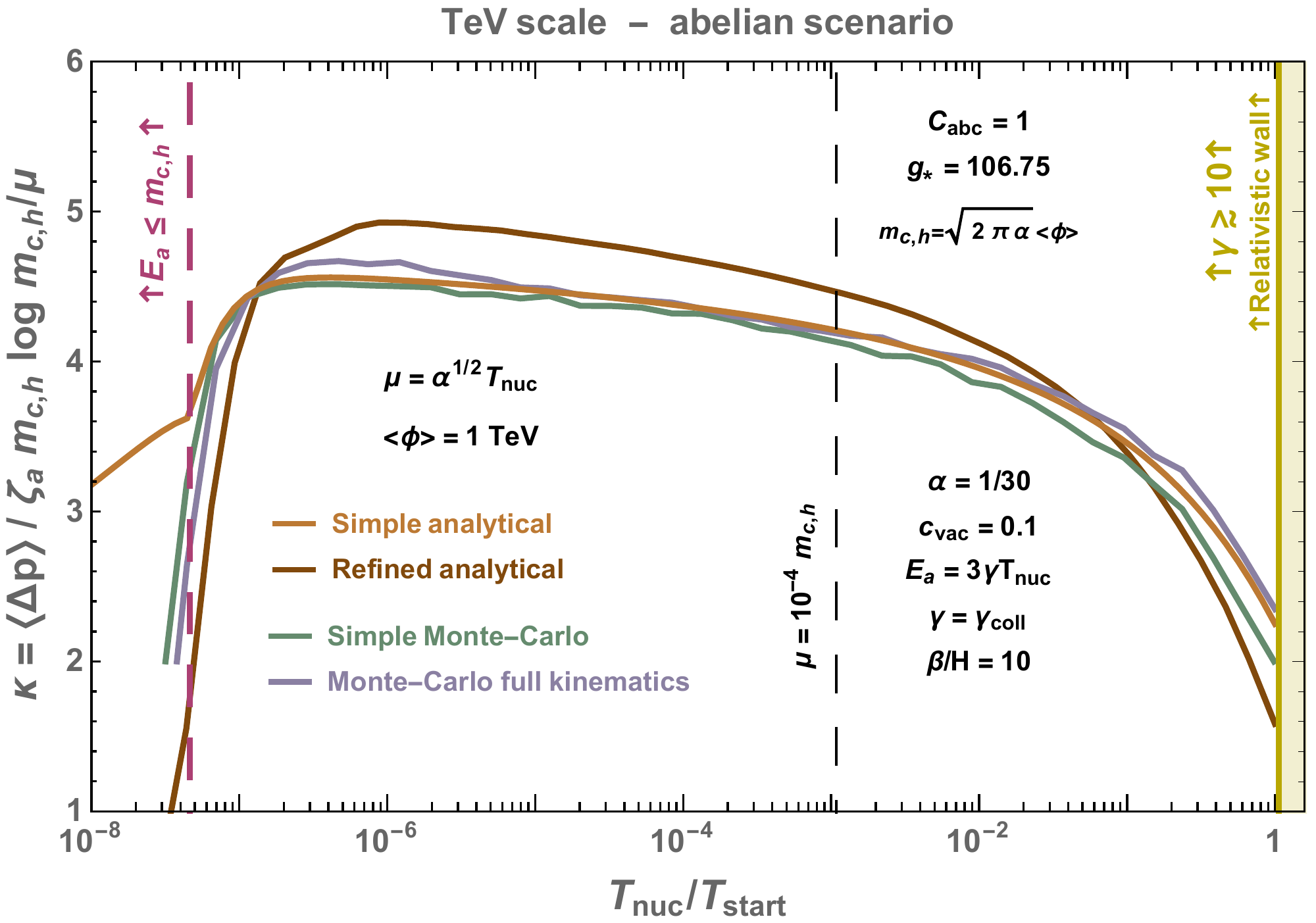}}}
\raisebox{0cm}{\makebox{\includegraphics[ width=0.5\textwidth, scale=1]{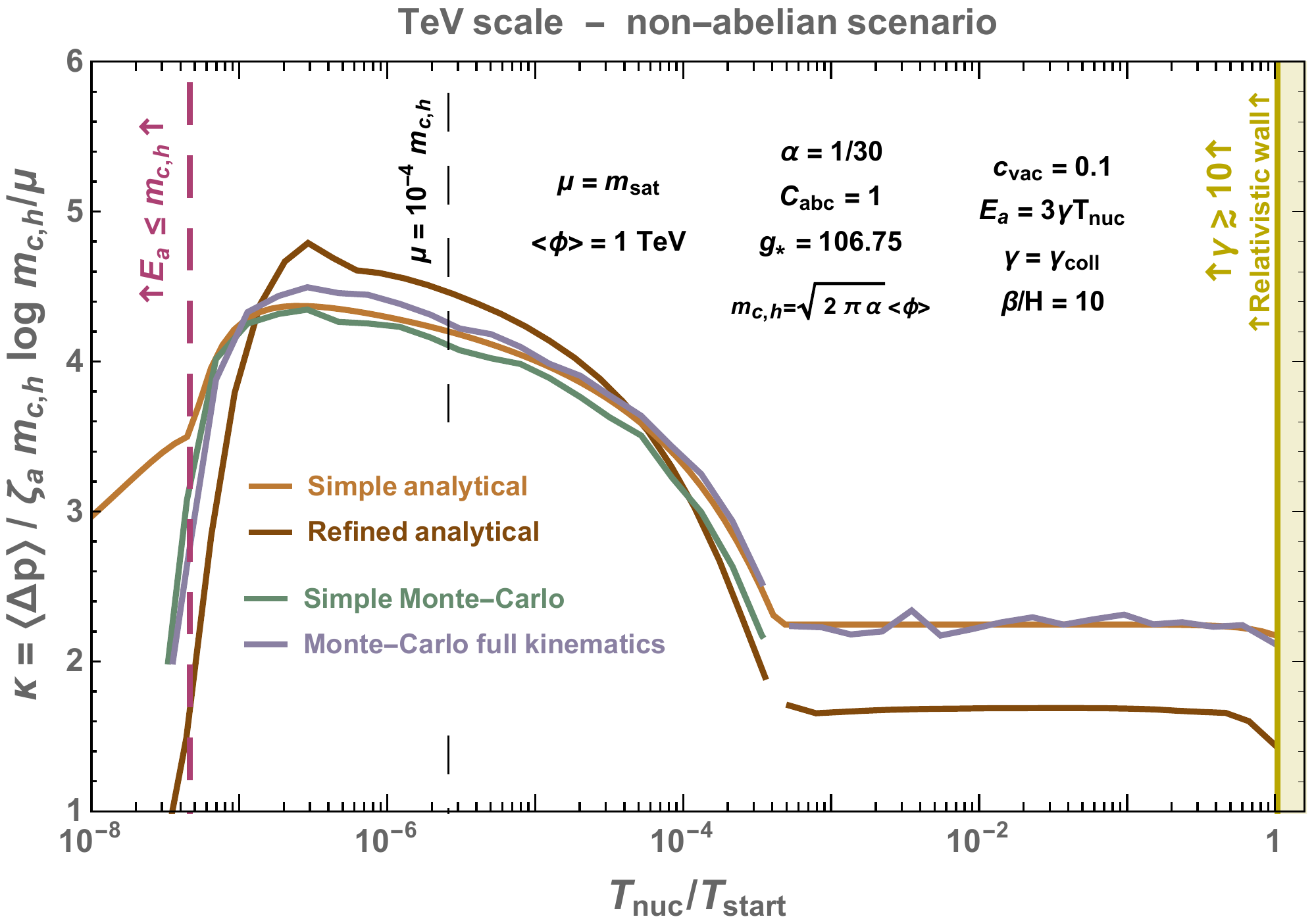}}}
\end{adjustbox}
\begin{adjustbox}{max width=1\linewidth,center}
\raisebox{0cm}{\makebox{\includegraphics[ width=0.5\textwidth, scale=1]{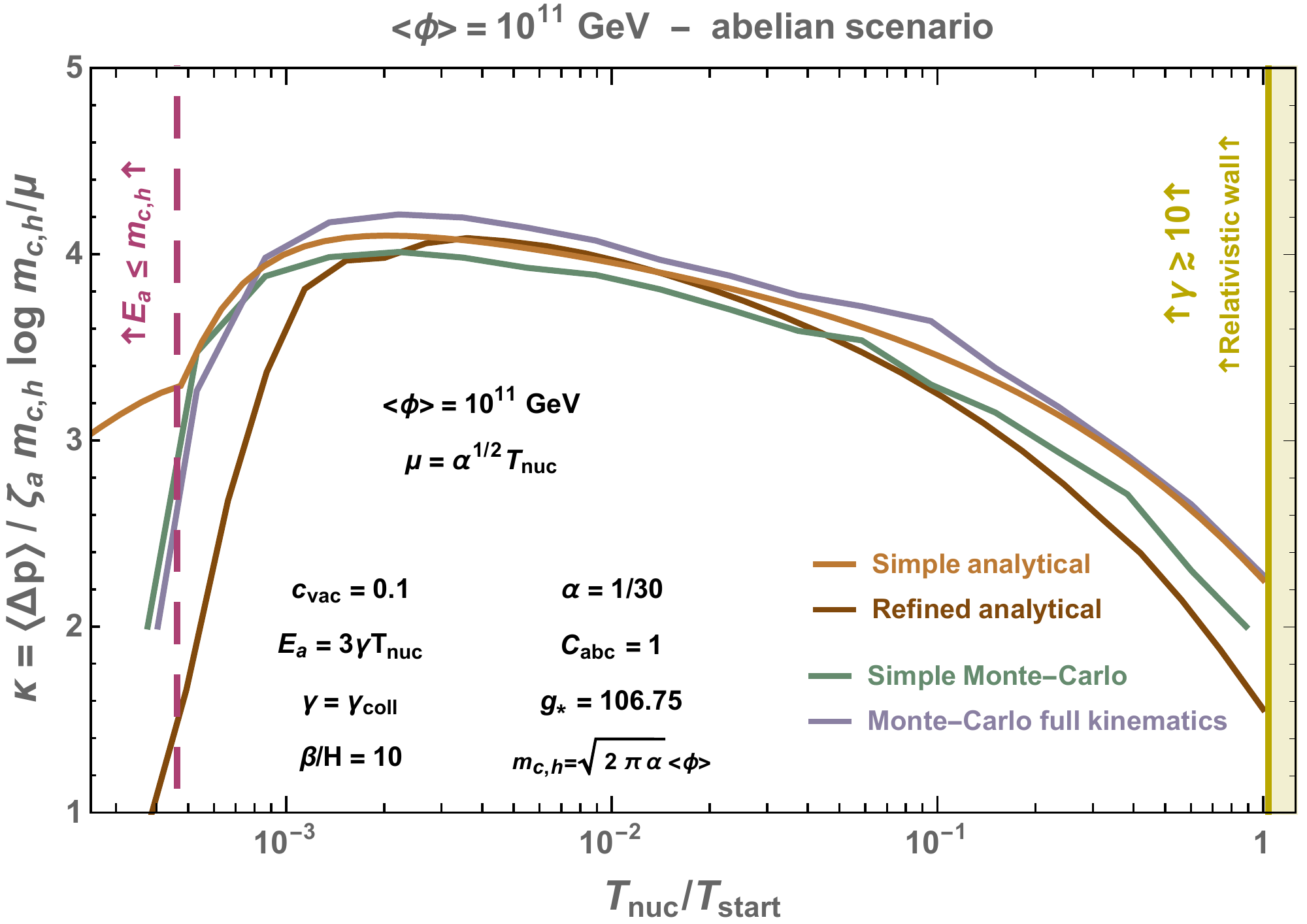}}}
\raisebox{0cm}{\makebox{\includegraphics[ width=0.5\textwidth, scale=1]{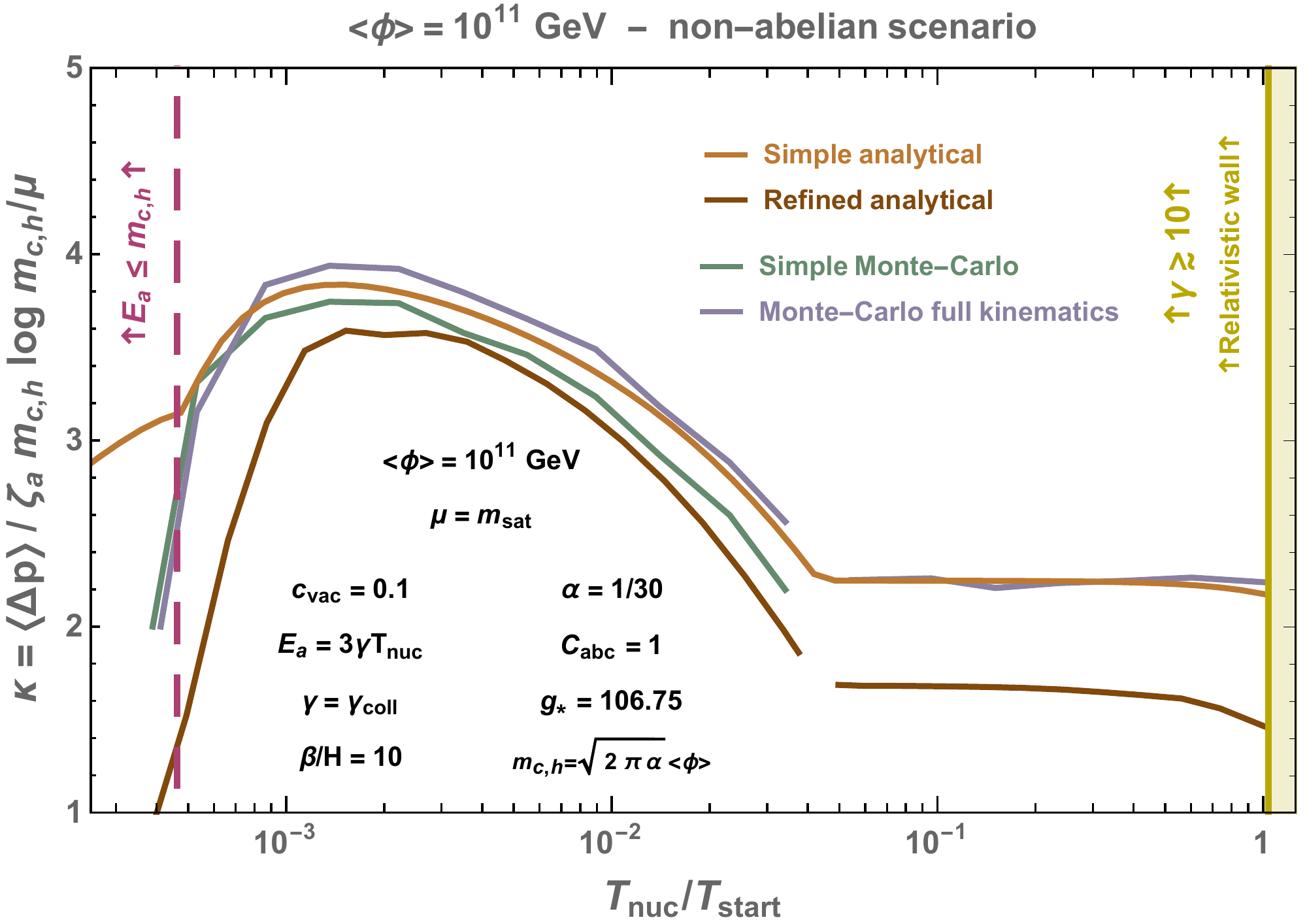}}}
\end{adjustbox}
\caption{\it \small
\label{fig:Error_estimates}
We compare different schemes for computing the exchange momentum $\Delta p$ induced by splitting radiation resummed at all leading-log orders, both at the analytical and at the numerical level.
The analytical estimates shown in \textcolor{myLightBrown}{\textbf{orange}} and  \textcolor{myBrown}{\textbf{brown}} correspond to the 1st and 7th level of correction of Table \ref{tab:corrections}.
The \textcolor{mygreen}{\textbf{green}} and \textcolor{mygray}{\textbf{gray}} lines show the results from the Monte-Carlo simulations, see Sec.~\ref{sec:MC}, using either the soft-collinear-single-plane-emission-no-transverse-recoil limit in Eq.~\eqref{eq_Delta_p_sum_Delta_pi} (green line) or the more refined prescription presented in App.~\ref{app:transverse_recoil} (gray line).
The most refined schemes of our work, shown in \textcolor{myBrown}{\textbf{brown}} and \textcolor{mygray}{\textbf{gray}} lines, do not account for the same corrections, and so they are complementary to each other.
}
\end{figure}
\label{tab:examplenf}

\subsection{Beyond the relativistic-soft-collinear limit}
\label{app:beyond-relativistic-limit}

In order to write the perturbative splitting probability in Eq.~\eqref{eq:dP_abc_mcs}, in addition to the simplified mode-function discussed in the previous section, we have assumed the relativistic limit for the phase space factor
\begin{equation}
\frac{1}{2p_a^z}
\frac{1}{2p_b^z}
\frac{1}{2p_c^z} \rightarrow \left(\frac{1}{2E_a}\right)^2
\frac{1}{2E_c}, \label{eq:phase_space_app}
\end{equation}
the soft and collinear limit $x \ll 1$, $k_\perp \ll E_a$ for the vertex function in Eq.~\eqref{eq:vertex_BM}, and the relativistic, soft and collinear limit for the phase of the mode function in Eq.~\eqref{eq:A_def} 
\begin{equation}
(1-x)E_a \gg \sqrt{m_b^2+k_\perp^2}, \quad x E_a \gg \sqrt{m_c^2+k_\perp^2}, \quad x \equiv E_{c} /E_a \ll 1.
\end{equation}
While the relativistic limit is a very good approximation for particles $a$ and $b$, it becomes incorrect for particle $c$ when $E_c$ is close to $m_{c,h}$ or $m_{c,s}$ depending on whether $c$ is transmitted or reflected. 
In this appendix, we compute the error caused by all these approximations.
Our most precise analytical calculation of the exchanged momentum follows from, cf. Eq.~\eqref{eq:Deltap_resummation}
\begin{equation}
\left<\Delta p_{\rm R,T} \right> = \int dP_{a \to bc} \,\Delta p\,\, \Theta(\pm p_{c,h}^2), \label{eq:Delta_p_exact_app}
\end{equation}
with the perturbative splitting probability given by
\begin{align}
\int dP_{a \to bc}
&=
\int \frac{d^2 k_{\perp}}{(2 \pi)^2} \int \frac{dE_c}{2 \pi}
~
\frac{1}{2p_a^z}
\frac{1}{2p_b^z}
\frac{1}{2p_c^z}
~
|{\cal M}|^2\,\mathcal{W}_{\rm wall}(L_{\rm w},\,k_{\perp}, \,x).
\label{eq:dP_abc_app}
\end{align}
where
\begin{equation}
\mathcal{M} =  2iE_a \left( \frac{V_hB_{c,h}}{A_h} -\frac{V_s C_{c,s}}{A_{s,r}} - \frac{V_s}{A_s}  \right).
\label{eq:M_matrix_app}
\end{equation}
with $B_h$, $C_s$ given by Eq.~\eqref{eq:Bh_Cs} and
\begin{align}
&A_h = 2iE_a\left(E_a - \sqrt{(1-x)^2E_a^2 - m_{b,h}^2 - k_\perp^2}- \sqrt{x^2E_a^2 - m_{c,h}^2 - k_\perp^2}\right), \label{eq:Ah_exact}\\
&A_s = 2iE_a\left(E_a - \sqrt{(1-x)^2E_a^2 - m_{b,s}^2 - k_\perp^2}- \sqrt{x^2E_a^2 - m_{c,s}^2 - k_\perp^2} \right),\label{eq:As_exact} \\
&A_{s,r} = 2iE_a\left(E_a - \sqrt{(1-x)^2E_a^2 - m_{b,s}^2 - k_\perp^2} +\sqrt{x^2E_a^2 - m_{c,s}^2 - k_\perp^2} \right).\label{eq:Asr_exact}
\end{align}
The factor $\mathcal{W}_{\rm wall}$, which is defined in Eq.~\eqref{eq:Wwall}, accounts for the finite wall thickness $L_{\rm w}$.
Since its impact is already studied in the Sec.~\ref{app:finite_wall_thickness}, in this subsection we set it to $\mathcal{W}_{\rm wall}=1$. 
The differences between the full analytical result in Eq.~\eqref{eq:Delta_p_exact_app} and the simplified one in Eq.~\eqref{eq:Delta_p_R}, \eqref{eq:Delta_p_T} are listed in Table~\ref{tab:corrections} along with the respective errors.
For $\mu = 10^{-4} m_{c,h}$ and $\left<\phi\right> = \rm TeV$, the simplified formula in Eqs~\eqref{eq:Delta_p_R} and~\eqref{eq:Delta_p_T} only underestimates the total $\left< \Delta p \right> = \left< \Delta p_R \right> + \left< \Delta p_T \right>$ by O(5\%).
The corrections for other values of  $\mu/m_{c,h}$ (or $T_{\rm nuc}/T_{\rm start}$) are shown in Fig.~\ref{fig:Error_estimates}.

\subsection{Azimuthal angle and transverse recoil}
\label{app:transverse_recoil}

\paragraph{Azimuthal angle.}

The formula for $\Delta p$ given in Eq.~\eqref{eq:Delta_p}, which we rewrite here
\begin{equation}
\Delta p  = E_{a} - \sqrt{(1-X)^2E_{a}^2 - m_{b,h}^2 -K_{\perp}^{2}} - \sum_{i=1}^n p_{c_i}^z, \label{eq:Delta_p_app}
\end{equation}
with
\begin{equation}
X =  \sum_{i=1}^n x_i, \qquad \text{and} \qquad K_{\perp} = \sum_{i=1}^n k_{\perp,i}, \label{eq:def_X_Kperp_app}
\end{equation}
assumes that the successive emissions occur in the same $(xz)$ plane. Instead, Eq.~\eqref{eq:def_X_Kperp_app} should be replaced by
\begin{equation}
K_{\perp} = \sum_{i=1}^n k_{\perp,i} \qquad \rightarrow \qquad \vec{K}_{\perp} = \sum_{i=1}^n \vec{k}_{\perp,i}, \quad \text{with} \quad  \vec{k}_{\perp,i} = \vec{k}_{\perp,i} (\cos{\phi_i},\,\sin{\phi}_i,\,0) .\label{eq:def_X_Kperp_app_2}
\end{equation}
with the azimuthal angle $\phi_i$ generated randomly at each emission
\begin{equation}
\phi_i = \mathcal{R}_{2\pi},
\end{equation}
where $\mathcal{R}_{2\pi}$ is a random number between $0$ and $2\pi$.

\paragraph{Transverse recoil.}

Also the formula for $p_{c_i}^z$ in Eq.~\eqref{eq:pci_exact} does not account for the recoils of the successive emission on the transverse momentum of the parent particle. Indeed, $k_{\perp,i}$ is the transverse momentum relative to the actual emitter and because of the successive recoils, it must differ from the absolute transverse momentum $\tilde{k}_{\perp,i}$ relative to the initial incoming momentum $p_a$ in Eq.~\eqref{eq:pa}.
Instead, upon taking into account successive transverse recoils, Eq.~\eqref{eq:pci_exact} becomes
\begin{equation}
p_{c_i}^z= \sqrt{x_i^2 E_{a}^2 - m_{c,h}^2 - \tilde{k}_{\perp, \, i}^{2}}\,\Theta(p_{c_i,h}^2)  - \sqrt{x_i^2 E_{a}^2 - m_{c,s}^2 - \tilde{k}_{\perp, \, i}^{2}} \,\Theta(-p_{c_i,h}^2). \label{eq:pci_exact_app}
\end{equation}
with 
\begin{equation}
p_{c_i,h}^2 =x_i^2 E_{a}^2 - m_{c,h}^2-  \tilde{k}_{\perp, \, i}^{2},\qquad \tilde{\vec{k}}_{\perp, \, i} =  \vec{k}_{\perp, \, i} - \sum_{j=1}^{i-1} \vec{k}_{\perp,j}.
\end{equation}
The recoils on the energy of the parent particle, $x_i  \rightarrow x_i\left(1 - \sum_{j=1}^{i-1} x_j \right)$, were already discussed in Sec.~\ref{par:backreaction} and App.~\ref{app:backreaction}.
In Fig.~\ref{fig:Error_estimates}, we compare the exchanged momentum $\left< \Delta p \right>$ calculated with Monte-Carlo simulations, using Eq.~\eqref{eq_Delta_p_sum_Delta_pi} (green line) which assumes the soft-collinear limit, single plane emission and which neglects transverse recoil, with the MC using Eq.~\eqref{eq:Delta_p_app}, \eqref{eq:def_X_Kperp_app_2} and \eqref{eq:pci_exact_app} (gray line).
We conclude that transverse recoil and azimuthal angle can be safely neglected.

\section{$3\to 2$ vector boson scattering.}
\label{sec:3to2_scattering}

\begin{figure}[t]
\centering
\begin{adjustbox}{max width=1\linewidth,scale=1,center}
\begin{tikzpicture}

\begin{feynman}
\vertex (a) {\(a\)};
\vertex [right=2. cm of a] (b);
\vertex [right=2.cm of b] (ab);
\vertex [above=0.01cm of ab] (c) {\(b\)};
\vertex [below right=1.8cm of b] (d);

\vertex [below=1. cm of a] (e){\(a\)};
\vertex [right=2. cm of e] (f);
\vertex [right=2. cm of f] (fg);
\vertex [above=0.1 cm of fg] (g){\(b\)};

\vertex [below=1.25 cm of e] (h){\(a\)};
\vertex [right=2. cm of h] (i);
\vertex [right=2. cm of i] (ij);
\vertex [below=0.1 cm of ij] (j){\(b\)};

\vertex [below=0.001 cm of d] (d1);
\vertex [below=0.001 cm of d] (d2);
\vertex [below=0.001 cm of d] (d4);
\vertex [below=0.8 cm of d] (n);
\vertex [below=0.001 cm of n] (d3);
\vertex [below=0.001 cm of n] (d5);

\vertex [right=1 cm of d] (l){\(c\)};
\vertex [right=1 cm of n] (m){\(c\)};

\diagram*{
 (a) -- [fermion] (b),
 (b) -- [fermion] (c),
 (b) -- [gluon, edge label=\(c\)] (d1),
 (e) -- [fermion] (f),
 (f) -- [fermion] (g),
 (f) -- [gluon, edge label'=\(c\)] (d2),
 (h) -- [fermion] (i),
 (i) -- [fermion] (j),
 (i) -- [gluon, edge label=\(c\)] (d3),
 (d4) -- [gluon] (l),
 (d5) -- [gluon] (m),
 (d4) -- [gluon] (d5),
};
\end{feynman}

\end{tikzpicture}
\end{adjustbox}
\caption{\it \small \label{fig:3to2}
3-to-2 scattering between emitted vector bosons which is expected to deplete the boson abundance and to possibly provide an IR cut-off $\mu$ for the emission.
}
\end{figure}
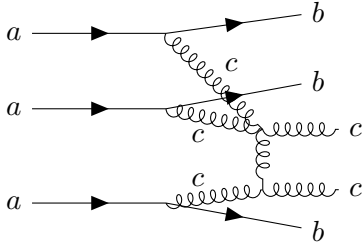

\subsection{Estimation of the scattering rate}

In the case of a non-abelian gauge theory, when the density of the emitted IR vector bosons becomes large, one must account for the possibility of scatterings that decrease the number of initial particles, because they can deplete the number of vector bosons and in turn lower the pressure on the wall.
As an example of such scatterings, here we consider $3\to 2$ processes, see Fig.~\ref{fig:3to2}. They become important if the rate $\Gamma_{3 \to 2}$ is larger than the time it takes for these particles to cross the wall $L_{\rm w}^{-1}$
\begin{equation}
\Gamma_{3\to 2} L_{\rm w} \gtrsim 1.
\end{equation}
The rate reads 
\begin{equation}
\Gamma_{3\to 2} = \int \frac{d^3 \vec{k}_1}{(2\pi)^32k_1^0}\frac{d^3 \vec{k}_2}{(2\pi)^32k_2^0}\frac{d^3 \vec{k}_3}{(2\pi)^32k_3^0} \; f_g(k_1) f_g(k_2)f_g(k_3) |\mathcal{M}|^2 \Bigg/ \int \frac{d^3 \vec{k}}{(2\pi)^3} \; f_g(k), \label{eq:gamma_32_def}
\end{equation}
where $\mathcal{M}$ is the 5-vector-boson scattering amplitude \cite{Berends:1981rb}.
From plugging Eq.~\eqref{eq:occupation_number} into Eq.~\eqref{eq:gamma_32_def}, we obtain
\begin{equation}
\Gamma_{3\to 2} = \left(\sum_a g_a\zeta_a \frac{\zeta(3)}{\pi^2} \gamma T_{\rm nuc}^3\right)^2 \frac{1}{(2\pi)^6(2E_a)^3}\frac{\int \prod_{i=1}^3 \frac{d k_{\perp,i}^2}{ k_{\perp,i}^2} \frac{d x_i}{x_i^2}\,\Pi(k_{\perp, i})|\mathcal{M}|^2}{\int \frac{d k_{\perp}^2}{ k_{\perp}^2} \frac{d x}{x}\,\Pi(k_{\perp})}. \label{eq:gamma_32_intermediate}
\end{equation}
Assuming that the five gluons are soft and collinear with
\begin{equation}
p_i \cdot p_j = k_{\perp,1}^2 + \mu^2, \qquad \forall i\neq j \in [1\cdots 5],
\end{equation}
at tree-level order \cite{Berends:1981rb}, we obtain 
\begin{equation}
|\mathcal{M}|^2 \simeq  \frac{g^6 N^2}{N^2 -1} \frac{60}{k_{\perp,1}^2 +\mu^2}. \label{eq:3to2xs}
\end{equation} 
Upon integrating Eq.~\eqref{eq:gamma_32_intermediate} over the range in Eq.~\eqref{eq:range_x_K_perp} with $m_c(z)=\mu$, we obtain the scattering rate
\begin{equation}
\Gamma_{3\to 2} \simeq \left(\sum g_a C_{abc}\right)^2 \gamma^2 T_{\rm nuc}^6 \times\frac{16}{9\pi^9 l} \frac{\alpha^5}{\mu^5},
\end{equation}
where
\begin{equation}
l\equiv \ln{\frac{E_a}{m_{c,h}}}\ln{\frac{m_{c,h}}{\mu}}.
\end{equation}

\subsection{Impact on the IR cut-off}

As argued above, the value of $\mu=\mu_{3\to2}$ for which the population of vector boson is depleted is found after requiring that $3\to 2$ processes happen within a wall length $L_{\rm w} \simeq c_{\rm vac}^{-1/2} \langle \phi \rangle^{-1}$.
So we determine $\mu_{3\to2}$ from
\begin{equation}
\Gamma_{3\to 2} L_{\rm w} = 1.
\end{equation}
We obtain
\begin{equation}
\fontsize{10pt}{0pt}
\mu_{3\to2} \simeq 
\left\{
                \begin{array}{ll}
                  \displaystyle 0.001\,m_{c,h}\left( \frac{\gamma}{\gamma_{\rm run}}\frac{10}{\beta/H_*} \frac{\sum_{a,b,c} g_a \,C_{abc} }{2 g_{*}} \frac{\rm TeV}{\left<\phi\right>} \frac{5}{\sqrt{l}} \right)^{\!2/5}\hspace{-0.15cm}\left(\frac{\alpha}{1/30}\right)^{\!1/2}\hspace{-0.1cm}\left(\frac{\Delta V}{0.1\left< \phi\right>^4} \right)^{\!3/10}\hspace{-0.1cm}\left(\dfrac{T_{\rm nuc}}{10^{-4}T_{\rm start}}  \right)^{\!8/5}\hspace{-0.5cm}\\[0.5cm] \quad \text{(run-away)}, \\[0.2cm]
                 \displaystyle 0.01 \,m_{c,h}\left(\frac{\gamma}{\gamma_{\LL}} \frac{\sum_{a,b,c} g_a \,C_{abc} \ln{100} }{2 g_{*}\ln{m_{c,h}/\mu_{3\to2}}}\frac{10}{\sqrt{l}}\right)^{\!2/5} \left( \frac{1/30}{\alpha}\right)^{\!1/10} \left(\frac{\Delta V}{0.1\left< \phi\right>^4}\right)^{\!1/2} \\[0.5cm] \quad \text{(terminal-velocity~walls)}.
                \end{array}
              \right.
\label{eq:mu_3to2}
\end{equation}
We find that $\mu_{3\to2}$ is always smaller than the scale $m_{\rm sat}$ in Eq.~\eqref{eq:kperp_satur} below which phase space is saturated and perturbation theory breaks down.
Therefore, the IR cut-off in Eq.~\eqref{eq:mu_3to2}, which relies on perturbation theory and where additionally in Eq.~\eqref{eq:3to2xs} only tree-level has been included, is not trustable and in this paper for non-abelian gauge theories we instead rely on the more conservative IR cut-off $m_{\rm sat}$ in Eq.~\eqref{eq:kperp_satur}.

\section{Fate of the reflected $c$ particles}
\label{app:fate_reflected_particles}

In this appendix, we compute the typical distance $l$ from the wall beyond which the reflected $c$ particles, cf. Sec.~\ref{sec:reflected_bosons}, have exchanged enough momentum with the incoming $a$ particles in order to come back in the direction of the wall.

\subsection{Mean free path}

\paragraph{Elastic cross-section.}

\begin{figure}[h]
\begin{center}
\includegraphics[width=0.4\columnwidth]{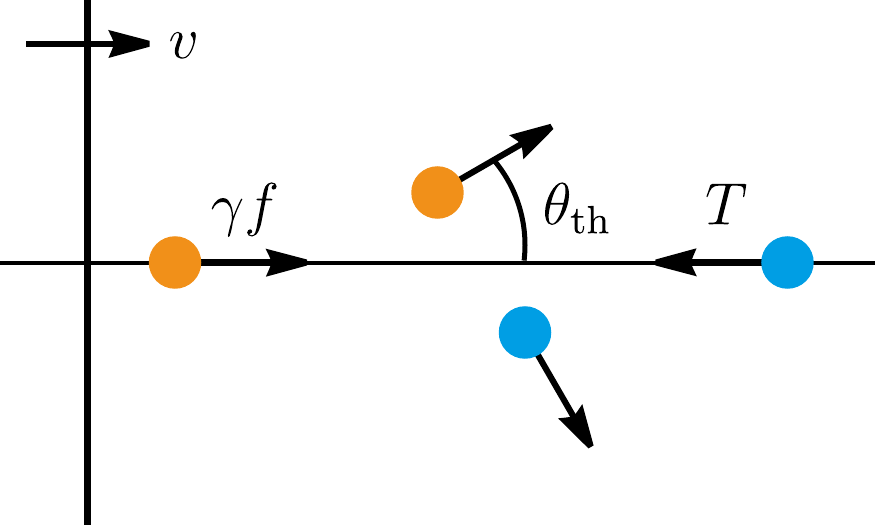}
\hskip 0.5cm
\includegraphics[width=0.4\columnwidth]{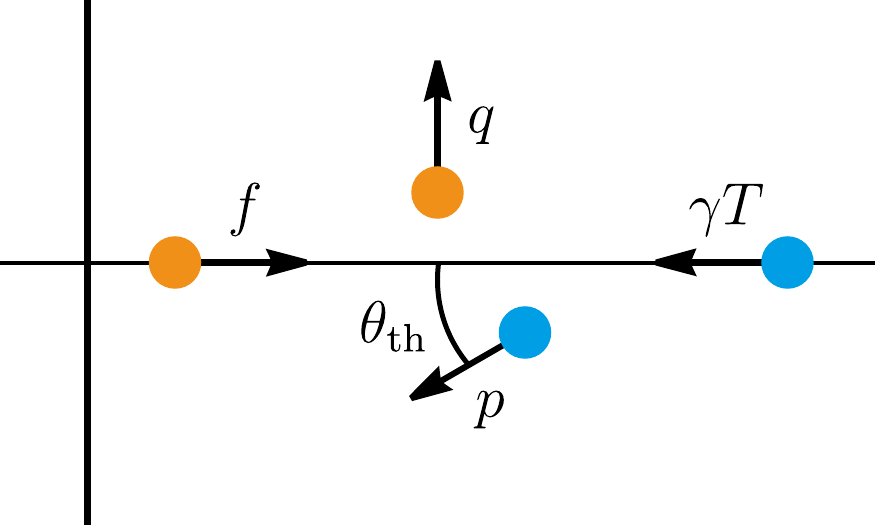}
\caption{\small
(Left) Plasma frame paramaterization.
(Right) Wall frame parameterization.
}\label{fig:reflected_particles_sketch}
\end{center}
\end{figure}

We assume that the reflected $c$ particles scatter elastically with the $a$ particle with the differential cross section
\begin{align}
\frac{d \sigma}{d (-t)}
&\sim
\frac{\alpha^2}{(- t)^2},
\end{align}
where $t$ is the usual Mandelstram variable, with $(- t) > 0$.
We approximate the $c$ particles to be massless and denote by $f$ their momentum in the wall frame, see Fig.~\ref{fig:reflected_particles_sketch}-right.

\paragraph{Plasma frame.}

The threshold value for the deflection angle $\theta$ is given by the condition that the velocity of the scattered $c$ particle in the $z$ direction becomes equal to the wall velocity,
\begin{align}
\cos \theta_{\rm th}
&=
v,
\end{align}
see Fig.~\ref{fig:reflected_particles_sketch}-left.
The Mandelstam variable $t$ corresponding to this deflection angle reads
\begin{align}
(- t_{\rm th})
&\sim
(\gamma f)^2 (1 - \cos \theta_{\rm th})
\sim
\gamma^2 (1 - v) f^2
\sim 
f^2.
\end{align}
The relevant cross section is given by integrating $(- t)$ above this threshold value.
Since the cross section is dominated by small $(- t)$ values, it becomes
\begin{align}
\sigma
&\sim
\int_{(- t_{\rm th})} d (- t)
\frac{d \sigma}{d (- t)}
\sim
\frac{\alpha^2}{(- t_{\rm th})}
\sim
\frac{\alpha^2}{f^2}.
\end{align}
Multiplying by the number density of the incoming particles $n\sim (\zeta(3) / \pi^2) g_* T^3$, we obtain the typical length scale in the plasma frame, after which the $c$ particles come back in the direction of the wall
\begin{align}
l_p
&\sim
\frac{1}{n \sigma}
\sim \frac{\pi^2}{g_*}
\frac{f^2}{\alpha^2 T^3}.
\end{align}

\paragraph{Wall frame.}

In the wall frame, the threshold value of the deflection angle $\theta$ beyond which a given scattered $c$ particle comes back in the direction of the wall, is given by the condition that $c$ moves perpendicular to the wall, see Fig.~\ref{fig:reflected_particles_sketch}-right.
Energy and momentum conservation for this case becomes
\begin{align}
\gamma T + f = p + q
~~&:~~
{\rm energy},
\\
\gamma T - f = p \cos \theta_{\rm th}
~~&:~~
z {\rm ~momentum},
\\
p \sin \theta_{\rm th} = q
~~&:~~
x, y {\rm ~momentum}.
\end{align}
The solution for $\gamma T \gg f$ is 
\begin{align}
p
&\sim \gamma T - f,
~~~~~~
q \sim f,
~~~~~~
\theta_{\rm th}
\sim \frac{f}{\gamma T}.
\end{align}
The Mandelstam $t$ for this deflection angle becomes
\begin{align}
(- t_{\rm th})
&\sim (\gamma T)^2 (1 - \cos \theta_{\rm th})
\sim f^2,
\end{align}
which is consistent with the plasma frame calculation.
The relevant cross section reads
\begin{align}
\sigma
&\sim \frac{\alpha^2}{(- t_{\rm th})}
\sim \frac{\alpha^2}{f^2}.
\end{align}
Multiplying by the number density of the incoming particles $n \sim \frac{\zeta(3)}{\pi^2} g_* \gamma T^3$, we obtain the mean free path in the wall frame 
\begin{align}
l_w
&\sim \frac{1}{n \sigma}
\sim \frac{1}{\gamma}  \frac{\pi^2}{g_*} \frac{f^2}{\alpha^2T^3}.
\end{align}
The extra $1 / \gamma$ is because the thermal bath is Lorentz contracted.

\paragraph{Account for multiple scatterings.}

We pursue the discussion in the plasma frame.
The average deflection angle squared $\braket{\theta^2}$ after $N$ scatterings $\theta_1, \cdots, \theta_N$, can be determined with a random walk approach.
Assuming that scatterings are independent from each other, we can write  $\braket{\theta^2} = \braket{(\theta_1 + \cdots + \theta_N)^2} = \braket{\theta_1^2} + \cdots + \braket{\theta_N^2}$.
Thus, we can simply calculate the average deflection angle squared for each, and sum up.
Noting that $(- t) \sim \gamma^2 f^2 \theta^2$ and $d \sigma / d (- t) \sim \alpha^2 / (- t)^2$, the average deflection squared per unit distance in the $z$ direction reads
\begin{align}
\frac{d\braket{\theta^2}}{dz}
&\sim n \int_{\theta_{\rm cut}^2} d(\theta^2)~\theta^2 \frac{d \sigma}{d \theta^2} \sim \frac{g_*}{\pi^2}\frac{\alpha^2T^3}{\gamma^2 f^2} \ln \frac{1}{\theta_{\rm cut}^2}.
\end{align}
The cutoff angle $\theta_{\rm cut}$ is related to the IR cut-off $\mu$ discussed in Sec.~\ref{sec:IR_cut_off} by $\theta_{\rm cut} = \mu/E_a$.
Since the particle gets caught by the wall once the deflection accumulates to $\theta_{\rm th} \sim \sqrt{1 - v} \sim 1 / \gamma$, the mean free path becomes
\begin{align}
l_p
&\sim \theta_{\rm th}^2 \Big/ \frac{d\braket{\theta^2}}{dz}
\sim \frac{\pi^2}{g_*}\frac{1}{\ln \frac{1}{\theta_{\rm cut}^2}} \frac{f^2}{\alpha^2T^3}.
\end{align}
We conclude that upon taking multiple scatterings into account, the mean free path $l$ only receives a logarithmic correction such that the big picture remains unchanged.

\subsection{Motivation for further studies}

A given reflected $c$ particle gets scattered by incoming $a$ particles before traveling the mean bubble separation when
\begin{equation}
l_p ~ \lesssim ~ \beta^{-1},
\end{equation}
which implies 
\begin{equation}
\frac{T_{\rm nuc}}{T_{\rm start}} ~\gtrsim ~ 7 \times 10^{-6} ~\alpha^{-1/3}  \left( \frac{f}{m_{c,h}} \right)^{2/3}\left( \frac{\left<\phi\right>}{\rm TeV} \right)^{1/3} \left( \frac{\beta/H}{10} \right)^{1/3}\left( \frac{100}{g_*} \right)^{1/3} \left( \frac{10}{\ln \frac{1}{\theta_{\rm cut}^2}} \right)^{1/3} \left( \frac{0.1}{\Delta V/\left< \phi\right>^4} \right)^{1/12} .\label{eq:TnucOTstart_mean_free_path_bubble_sep}
\end{equation}
We show this condition with a dashed blue line in Fig.~\ref{fig:bubble_wall_velocity_phi}.
We conclude that as soon as $T_{\rm nuc}/T_{\rm start}$ is larger than what is indicated in Eq.~\eqref{eq:TnucOTstart_mean_free_path_bubble_sep} (basically most of the cosmological first-order phase transitions considered in the literature), the interactions of reflected $c$ particles with $a$ particles, and the possible associated corrections to the friction pressure, discussed in Sec.~\ref{sec:reflected_bosons}, should be considered. We leave the quantitative study of such effects for further works.

\section{Massless vector boson scenario}
\label{app:massless_scenario}

In this section, we consider the emitted vector boson to remain massless in the broken phase, up to expected thermal effects. We find that the contributions to the friction pressure are equal to the LO one, up to $\mathcal{O}(\zeta_a)$. 

\paragraph{Perturbative splitting probability.}

We suppose that the vector boson mass is phase-independent and given by the thermal contribution $m_{c,h} = m_{c,s} =\mu$ while $m_{a,h} \neq m_{a,s} =0$ and $m_{b,h} \neq m_{b,s}=0$.
The WKB suppression factor introduced along Eq.~\eqref{eq:A_def}, in the soft-collinear limit, reads
\begin{align}
\left(\frac{1}{A_h} - \frac{1}{A_s}\right)^2 &=     \left(\frac{1}{-m_{a,h}^2+\frac{k_{\perp}^2+m_{b,h}^2}{1-x}+\frac{k_{\perp}^2+\mu^2}{x}} - \frac{1}{\frac{k_{\perp}^2}{1-x}+\frac{k_{\perp}^2+\mu^2}{x}}\right)^{\!2} \\
&= \frac{x^4\big((m_{b,h}^2-m_{a,h}^2)+m_{a,h}^2 x\big)^2}{(k_{\perp}^2+\mu^2)^2(k_{\perp}^2 + \mu^2+x (m_{b,h}^2-m_{a,h}^2)+x^2 m_{a,h}^2)^2},
\end{align}
so that the splitting probability in Eq.~\eqref{eq:dP_abc} becomes
\begin{align}
dP_{a \to bc} =\zeta_a\frac{d k_{\perp}^2}{k_{\perp}^2}  d x\,x\,\frac{k_{\perp}^4}{(k_{\perp}^2+\mu^2)^2}\, \frac{\big((m_{b,h}^2-m_{a,h}^2)+m_{a,h}^2 x\big)^2}{(k_{\perp}^2+ \mu^2+ x (m_{b,h}^2-m_{a,h}^2)+x^2 m_{a,h}^2)^2}.
\label{eq:dP_abc_massless}
\end{align}
\subsection{Case $m_{b}=m_{a}$}
We consider the case where $m_{b,h}=m_{a,h}$ in Eq.~\eqref{eq:dP_abc_massless}.

\paragraph{Exchanged momentum averaged over resummed distribution.}

Since emitted vector bosons do not acquire a mass in the broken phase, they are never reflected against the wall boundary.
The exchanged momentum in the soft $X,\,x_i \ll 1$ and collinear $K_{\perp},\,k_{\perp,i}\ll 1$ limit, reads
\begin{align}
\Delta p  &\simeq E_{a}- \sqrt{(1-X)^2E_{a}^2 - m_{b,h}^2 -K_{\perp}^{2}} - \sum_{i=1}^n \sqrt{x_i^2 E_{a}^2  - k_{\perp, \, i}^{2}} \label{eq:Delta_p_massless_app}\\
&\simeq \Delta p_1 + \Delta p_2 + \Delta p _3, \label{eq:Delta_p1_2_3}
\end{align}
with 
\begin{equation}
\Delta p_1 = \frac{m_{b,h}^2}{2E_a},\qquad \Delta p_2 = \frac{ \big(\sum_{i=1}^n k_{\perp,\,i}\big)^2}{2E_a}, \qquad \Delta p_3 = \sum_{i=1}^n \frac{k_{\perp,i}^{2}}{2x_iE_a}.
\end{equation}
where we recall that $K_{\perp} \equiv \sum_{i=1}^n k_{\perp,i}$.
We average over the resummed splitting distribution in Eq.~\eqref{eq:mean_O}.
At first, since $\Delta p_1$ is independent of the splitting kinematics, we have
\begin{equation}
\left< \Delta p_1  \right> = \Delta p_1=\frac{m_{b,h}^2}{2E_a}.
\end{equation}
This coincides with the LO order piece in Eq.~\eqref{eq:pLO_momentum}.
Next we compute
\begin{align}
\left< K_{\perp,\,i}K_{\perp,\,j} \right> &=  \sum_{n=0}^{\infty}\frac{1}{n!}\left[ \prod_{j=1}^{n} \int dP_{E,\,j}\right] K_{\perp,\,i} K_{\perp,\,j} \exp\left[-\int dP_{E} \right] \notag\\
&=\left[\int dP_{E}\, K_{\perp}\right]^2  \sum_{n=0}^{\infty}\frac{1}{n!}\left[\int dP_{E}\right]^{n-2} \exp\left[- \int dP_{E} \right] \notag \\
&= \left[\frac{\int dP_{E}\, K_{\perp} }{\int dP_{E}}\right]^2,
\end{align}
which implies
\begin{equation}
\left< \Delta p_2 \right> = \frac{1}{2E_a}\left[
\int dP_{E}\, k_{\perp}^2 + \frac{n(n-1)}{n^2}  \left(\int dP_{E}\, k_{\perp} \right)^2\right], \label{eq:delta_p_2_proof_app}
\end{equation}
where $n =\int dP_{E}$ is the mean number of emitted bosons.
We compute 
\begin{align}
\int dP_{E}\, k_{\perp}^2 \simeq \zeta_a \frac{m_{b,h}^2}{2}, \qquad \left(\int dP_{E}\, k_{\perp}\right)^2 \simeq 2\zeta_a m_{b,h}^2, \qquad n \simeq \zeta_a \ln^2 \frac{m_{b,h}}{\mu},
\end{align}
which in the limit $n \gg 1$, implying
\begin{equation}
\left< \Delta p_2 \right> \simeq \zeta_a \frac{5m_{b,h}^2}{4E_a}.
\end{equation}
Finally, we use Eq.~\eqref{eq:Deltap_resummation} to compute 
\begin{align}
\left<\Delta p_3\right>  \simeq \zeta_a
\int_{\mu^2}^{E_{a}^2}\frac{dk_{\perp}^2}{k_{\perp}^2} \,
\int_{\frac{\sqrt{k_{\perp}^2+\mu^2}}{E_{a}}}^{1}dx\,x^3\frac{m_{b,h}^4}{(k_{\perp}^2+x^2m_{b,h}^2)^2}\, \frac{k_{\perp}}{2x E_a} \simeq \zeta_a \frac{m_{b,h}^2}{2E_a}.
\end{align}

\paragraph{Final results.}

Hence, we conclude that in the limit where the vector boson is massless and $m_a=m_b$, the exchanged momentum in Eq.~\eqref{eq:Delta_p1_2_3} is equal, up to $O(\zeta_a)$, to the LO contribution
\begin{equation}
\left< \Delta p \right>\simeq \Delta p_{\LO} \left(1 + 3 \zeta_a  \right), \qquad \text{with} \quad \Delta p_{\LO} \simeq \frac{m_{b,h}^2}{2E_a}.
\end{equation}
This is in contrast to \cite{Hoche:2020ysm}, which, in the massless vector boson limit, have found $\left< \Delta p \right> \simeq \zeta_a E_a$, see App.~\ref{app:comment_hoeche}.

\subsection{Case $m_{b,h} > m_{a,h}$}

We suppose $m_{b,h} > m_{a,h}$ in Eq.~\eqref{eq:dP_abc_massless}.

\paragraph{Computations.}

As in Eq.~\eqref{eq:Delta_p1_2_3}, the exchange momentum can be written $\Delta p=\Delta p_1+\Delta p_2+\Delta p_3$ with 
\begin{equation}
\Delta p_1 = \left\{
                \begin{array}{ll}
                \displaystyle \frac{m_{b,h}^2}{2E_a} \quad \text{if~}n\text{~odd,} \\[0.25cm]
                \displaystyle \frac{m_{a,h}^2}{2E_a} \quad \text{if~}n\text{~even,}
                \end{array}
              \right.
\qquad \Delta p_2 = \frac{ \big(\sum_{i=1}^n k_{\perp,\,i}\big)^2}{2E_a}, \qquad \Delta p_3 = \sum_{i=1}^n \frac{k_{\perp,i}^{2}}{2x_iE_a}.
\end{equation}
The asymptotic state particles are $a \to bc$ if the number of splitting $n$ is odd while they are $a\to ac$ if $n$ is even.
Therefore, Eq.~\eqref{eq:mean_O} becomes 
\begin{equation}
\left< \Delta p_1 \right>=\left[\frac{m_{a,h}^2}{2E_a}\cosh{n} + \frac{m_{b,h}^2}{2E_a}\sinh{n} \right]e^{-n} = \frac{m_{a,h}^2+m_{b,h}^2}{4E_a} + \frac{m_{a,h}^2-m_{b,h}^2}{4E_a}e^{-2n}. \label{eq:Delta_p1_mb}
\end{equation}
where $n$ is the mean number of emitted bosons
\begin{equation}
n=\int dP_{E} \simeq \zeta_a \ln^2 \frac{\sqrt{m_{b,h}^2-m_{a,h}^2}}{\mu}, \label{eq:n_def_mB_app}
\end{equation}
with $\mu$ given by the vector boson thermal mass $\mu = m_{c,s}$ or larger.
Next, we use Eq.~\eqref{eq:delta_p_2_proof_app} and Eq.~\eqref{eq:Deltap_resummation} to compute
\begin{align}
\int dP_{E}\, k_{\perp}^2 \simeq \zeta_a (m_{b,h}^2-m_{a,h}^2),\qquad\quad \left(\int dP_{E}\, k_{\perp}\right)^2 \simeq 10\zeta_a (m_{b,h}^2-m_{a,h}^2),
\end{align}
which in the limit $n \gg 1$, implies
\begin{equation}
\left< \Delta p_2 \right> \simeq \zeta_a \frac{11(m_{b,h}^2-m_{a,h}^2)}{2E_a}.
\label{eq:Delta_p2_mb}
\end{equation}
Finally, we use Eq.~\eqref{eq:Deltap_resummation} to compute
\begin{equation}
\left<\Delta p_3\right>  \simeq \zeta_a \frac{m_{b,h}^2-m_{a,h}^2}{E_a}\ln{\frac{\sqrt{m_{b,h}^2-m_{a,h}^2}}{\mu}},
\label{eq:Delta_p3_mb}
\end{equation}
with $\mu$ given by the vector boson thermal mass $\mu = m_{c,s}$ or larger.
\paragraph{Final result.}
Summing up Eq.~\eqref{eq:Delta_p1_mb}, \eqref{eq:Delta_p2_mb} and \eqref{eq:Delta_p3_mb}, we obtain
\begin{equation}
\left< \Delta p \right>\simeq \Delta p_{\LO} + \frac{m_{b,h}^2 - m_{a,h}^2}{4E_a} \left(1-e^{-2n} + 22 \zeta_a  + 4\sqrt{n\zeta_a}  \right) \qquad \text{with} \quad \Delta p_{\LO} \simeq \frac{m_{b,h}^2}{2E_a}, \label{eq:Delta_p_mb_larger_ma}
\end{equation}
and where $n$ is given by Eq.~\eqref{eq:n_def_mB_app}.
Again, we conclude that the contributions to the exchanged momentum coming from the emission of a massless vector boson, in the case $m_{b,h} \geq m_{a,h}$, are only $O(\zeta_a)$ corrections to the LO result.

In the case where $m_{a,h} > m_{b,h}$, the expression in Eq.~\eqref{eq:dP_abc_massless} has a pole when $\Delta p_{z,h}=A_h/2E_a=0$, corresponding to the possibility in the broken phase for $a$ to decay to $bc$, on-shell and without the need of the presence of any wall.
After subtracting that pole, we expect the NLO correction to the exchange momentum $\left< \Delta p \right>$ to be of the same order as Eq.~\eqref{eq:Delta_p_mb_larger_ma}. 

\section{Comment on \cite{Hoche:2020ysm}}
\label{app:comment_hoeche}

\subsection{Violation of the Ward identity}
\label{app:ward_identity}

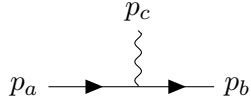
\begin{figure}[t]
\centering
{\begin{tikzpicture}
\begin{feynman}
    \vertex (a1) {\(\)};
    \vertex[right=1.5cm of a1] (a2){\(p_c\)};
    \vertex[right=1cm of a2] (a3){\(\)};

    \vertex[below=1cm of a1] (b1) {\(p_a\)};
    \vertex[right=1.5cm of b1] (b2);
    \vertex[right=1cm of b2] (b3){\(p_b\)};

    \diagram*{
       {
      },
      (b2) -- [boson] (a2),
      (b1) -- [fermion] (b2),
      (b2) -- [fermion] (b3),
    };

\end{feynman}
\end{tikzpicture}
}
\caption{\it \small \label{fig:vertex}  Splitting radiation vertex}
\end{figure}

Let us consider the splitting radiation vertex in Fig.~\eqref{fig:vertex}
\begin{equation}
{\cal M} = \bar{u}(p_b) \slashed{\epsilon}(p_c) u(p_a) .
\end{equation}
The Ward identity can be stated as
\begin{equation}
\mathcal{M} ~ \text{is invariant under}~\epsilon_{\lambda}^\mu(p_c,
\,n) \to \epsilon_{\lambda}^\mu(p_c,
\,n) + p_c^\mu.
\end{equation}
We make clear the conditions for the Ward identity to hold.
Consider a new transition matrix ${\cal M}'$ in which the polarization vector has been shifted by $\epsilon^\mu(p_c) \rightarrow \epsilon^\mu(p_c) + p_c^\mu$.
The difference from the original transition matrix is given by
\begin{equation}
{\cal M}' -  {\cal M} = \bar{u}(p_b) \slashed{p}_c u(p_a) .
\end{equation}
From eliminating $\slashed{p}_c$ with the momentum conservation condition
\begin{equation}
p_a = p_b+p_c, \label{eq:momentum_conservation}
\end{equation}
and from using
\begin{equation}
(\slashed{p}_a-m)u(p_a)=0,\qquad \bar{u}(p_b)(\slashed{p}_b-m)=0, \label{eq:on_shell_a_b}
\end{equation}
we get
\begin{equation}
{\cal M}' -  {\cal M} = \bar{u}(p_b) (\slashed{p}_a-\slashed{p}_b) u(p_a) = \bar{u}(p_b) (m-m) u(p_a) = 0.
\end{equation}

We conclude that there are two types of scenarios in which the Ward identity cannot be satisfied.
The first one is when the particles $a$ and/or $b$ are off-shell, in which case $(\slashed{p}_a-m)u(p_a)\neq 0$ and/or $\bar{u}(p_b)(\slashed{p}_b-m)\neq0$, in Eq.~\eqref{eq:on_shell_a_b}.
This can be the case for QCD splitting functions \`{a} la Altarelli and Parisi \cite{Altarelli:1977zs}, where one leg must be attached to a larger Feynman diagram and the momentum of the corresponding internal line is off-shell.
In this case, the Ward identity is only satisfied at the level of the larger Feynman diagram.
The second scenario  is when 4-momentum is not conserved $p_a \neq p_b + p_c$, in Eq.~\eqref{eq:momentum_conservation}.
This is the scenario of the calculation presented in this paper, where the existence of the wall boundary separating the two phases breaks spontaneously translation invariance and prevent momentum conservation in the $z$ direction.

As we explain in the next section, we point out that the computation in~\cite{Hoche:2020ysm} satisfies the Ward identity out of its regime of validity, thus possibly suggesting an avenue to explore to understand why the splitting probability in~\cite{Hoche:2020ysm} does not vanish in the limit where the symmetry is restored, as it instead should.

\subsection{Splitting at first order}
\label{sec:splitting_firstorder_Hoeche}

In App.~B of v3 of \cite{Hoche:2020ysm}, the matrix element squared $|{\cal M}|^2$ for splitting radiation across the wall is computed in the case of scalar QED. 
The matrix element at leading order in presence of the wall reads
\begin{equation}
\mathcal{M}_{a \to bc}^{(0)} = 2iE_a\left(\frac{V_h}{A_h} - \frac{V_s}{A_s}\right),
\label{eq:Matobc0}
\end{equation}
with the vertex functions in the symmetric and broken phase being
\begin{equation}
V_s = ig(p_{a,s} + p_{b,s})_\mu \epsilon_{\lambda}^\mu, \qquad V_h = ig(p_{a,h}+p_{b,h})_\mu \epsilon_{\lambda}^\mu.
\label{eq:vertices}
\end{equation}
Until now, there is no difference with us, see Eq.~\eqref{eq:M_matrix_2} and Eq.~\eqref{eq:vertex_scalar}.
The authors of~\cite{Hoche:2020ysm} then define $p_{b,s}$ and $p_{a,h}$ via
\begin{equation}
\label{eq:US_momenta}
p_{a,s} = p_{b,s} + p_{c,s},\quad \text{and} \quad p_{a,h} = p_{b,h} + p_{c,h}, 
\end{equation}
i.e. by assuming energy-momentum conservation (so in particular momentum conservation along $z$) in the two phases separately.
Note that this implies that at least one among the three momenta, in each of the $s$ and $h$ phases, is off-shell.
Then, using a classical current formalism, they obtain the following WKB phases
\begin{equation}
\label{eq:US_phases}
A_s = -2p_{a,s}p_{c,s}, \qquad \text{and} \qquad A_h = -2p_{b,h}p_{c,h}.
\end{equation}
By plugging Eq.~\eqref{eq:US_momenta} into the vertices Eq.~\eqref{eq:vertices}, and then the vertices and the phases~\eqref{eq:US_phases} into the matrix element Eq.~\eqref{eq:Matobc0}, one finds that the matrix elements does respect the Ward identity, if $p_{c,s} = p_{c,h} = p_c$ with $p_c^2 = 0$ as assumed in~\cite{Hoche:2020ysm}
\begin{align}
\mathcal{M}_{a \to bc}^{(0)}|_{\epsilon_\lambda^\mu \to p_c^\mu}
&=
gE_a\left(\frac{(p_{a,h} p_c + p_{b,h} p_c)}{p_{b,h} p_c} - \frac{(p_{a,s} p_c + p_{b,s} p_c)}{p_{a,s} p_c}\right)
\nonumber \\
&=
gE_a\left(\frac{p_{a,h} p_c}{p_{b,h} p_c} - \frac{p_{b,s} p_c}{p_{a,s} p_c}\right)
\nonumber \\
&=
gE_a\left(\frac{(p_{b,h} + p_c) p_c}{p_{b,h} p_c} - \frac{(p_{a,s} - p_c) p_c}{p_{a,s} p_c}\right)
\nonumber \\
&=
0.
\end{align}
This signals a possible issue with the result of~\cite{Hoche:2020ysm}, because Eq.~\eqref{eq:US_momenta} implies that at least one of the momenta in each equation should be off-shell, and so the Ward identity should not have been satisfied.

The authors of~\cite{Hoche:2020ysm} then use the polarization sum rules in axial gauge with arbitrary auxiliary vector $n^\mu$,
\begin{equation}
\label{eq:pol_sum}
\sum_{\lambda} \epsilon_{\lambda}^\mu(p_c,\,n)\epsilon_\lambda^{\nu*}(p_c,\,n) = -g^{\mu\nu} + \frac{p_c^\mu n^\nu + p_c^\nu n^\mu}{p_c n}-n^2 \frac{p_c^\mu p_c^\nu}{(p_c n)^2}\,,
\end{equation}
and obtain the gauge-invariant result\footnote{
While it is nice that this result is manifestly gauge-invariant, without Ward identities the polarization sum rules of Eq.~\eqref{eq:pol_sum} would lead to a final $\left|\mathcal{M}\right|^2 $ that depends on the arbitrary gauge vector $n^\mu$.
Instead, when computing the matrix element, only the physical polarizations (corresponding to the gauge choice $n^\mu = (1,0,0,-1)$) should be used, as prescribed long ago by Altarelli and Parisi \cite{Altarelli:1977zs}, and as we do in App.~\ref{app:vertex}.
}
\begin{equation}
\label{eq:Msqr_Hoeche}
\left|\mathcal{M}_{a \to bc}^{(0)}\right|^2 = 4 E_a^2|g|^2 \left( \frac{2p_{a,s}p_{b,h}}{p_{a,s}p_c~p_{b,h}p_c} - \frac{p_{a,s}^2}{(p_{a,s}p_c)^2} - \frac{p_{b,h}^2}{(p_{b,h}p_c)^2}  \right).
\end{equation}
The perturbative splitting probability associated to Eq.~\eqref{eq:Msqr_Hoeche} then reads
\begin{equation}
\label{eq:splitting_function}
dP^{'}_{E}  =\frac{d^3p_c}{(2\pi)^32 E_c} \frac{\left|\mathcal{M}_{a \to bc}^{(0)}\right|^2}{4E_a} \simeq  \zeta_a \frac{dk_{\perp}^2}{k_{\perp}^2} dx\,x\,\left( \frac{k_\perp^2 + x m_{b,h}^2}{k_{\perp}^2 + x^2m_{b,h}^2}   \right)^2.
\end{equation}

The splitting probability in Eq.~\eqref{eq:splitting_function} does not vanish in the limit where the order parameter of the phase transition goes to zero, either $m_{b,h} \ll k_{\perp}$ or $m_{b,h} \to m_{b,s}$.
Therefore the friction pressure is left intact in the limit in which the wall disappears $\left<\phi\right> \to 0$ \cite{Vanvlasselaer:2020niz}, thus signaling a possible problem with Eq.~\eqref{eq:splitting_function}.
We suggest that the root of the problem may lie in the effective assumptions discussed earlier, that lead to unexpectedly satisfying the Ward identities.

In contrast, our splitting probability does vanish when $m_{c,h} \ll k_{\perp}$ or $m_{b,h} \ll k_{\perp}$ (but also when $m_{i,h} \to m_{i,s}$), for vector bosons $c$ which are respectively either massive, see Eq.~\eqref{eq:dP_abc}, or massless, see Eq.~\eqref{eq:dP_abc_massless}. 

\subsection{Splitting at all orders}

\paragraph{The average exchange momentum.}

In \cite{Hoche:2020ysm}, the average exchanged momentum is computed according to
\begin{equation}
\label{eq:exchange_momentum}
\left< \Delta p \right> = \int_{k_{\perp} >\mu} dP^{'}_{E}\, \frac{k_{\perp}^2}{2xE_a}\,  \exp\left[- \int_{\tilde{k}_{\perp} > k_{\perp} } dP^{'}_{E}(\tilde{k}_\perp) \right],
\end{equation}
where the perturbative splitting probability $dP^{'}_{E}$ is given by Eq.~\eqref{eq:splitting_function}.
The integrand in Eq.~\eqref{eq:exchange_momentum} is dominated by the region where $x \sim 1$ and $k_{\perp} \sim E_a$ which leads the authors of \cite{Hoche:2020ysm} to conclude that
\begin{equation}
\left< \Delta p \right> \sim \zeta_a \,E_a.
\end{equation}
\paragraph{Our comments.}
The Sudakov resummation operated in Eq.~\eqref{eq:exchange_momentum} can be obtained from the master formula in Eq.~\eqref{eq:mean_O} if we assume that the exchanged momentum in Eq.~\eqref{eq_Delta_p_sum_Delta_pi} is dominated by the largest $k_{\perp}$
\begin{equation}
\label{eq:Delta_p_Hoeche_app}
\Delta p = \underset{k_{\perp}}{\textrm{Max}}\left[ \frac{k_{\perp,1}^2}{2x_1E_a},\,\frac{k_{\perp,2}^2}{2x_2E_a},\cdots,\frac{k_{\perp,n}^2}{2x_nE_a} \right].
\end{equation}
In that case, Eq.~\eqref{eq:mean_O} becomes
\begin{align}
\left< \Delta p \right> &=  \sum_{n=0}^{\infty}\frac{1}{n!}\left[ \prod_{j=1}^{n} \int dP^{'}_{E,\,j}\right] \Delta p ~\exp\left[-\int dP^{'}_{E} \right] \notag\\
&=\left[\int dP^{'}_{E}(k_{\perp})\, \frac{k_{\perp}^2}{2x E_a} \right]  \sum_{n=1}^{\infty}\frac{1}{(n-1)!}\left[\int dP^{'}_{E}(\tilde{k}_\perp)\,  \theta\big(k_{\perp} - \tilde{k}_{\perp} \big)\right]^{n-1} \, \exp\left[- \int dP^{'}_{E} \right] \notag\\
&= \int dP^{'}_{E}(k_{\perp})\, \frac{k_{\perp}^2}{2x E_a} \, \exp\left[ \int_{\tilde{k}_{\perp}<k_{\perp}} dP^{'}_{E}(\tilde{k}_{\perp}) - \int_{\tilde{k}_{\perp}>\mu} dP^{'}_{E}(\tilde{k}_{\perp}) \right] \notag\\
&= \int_{k_{\perp}>\mu} dP^{'}_{E}(k_{\perp})\, \frac{k_{\perp}^2}{2x E_a} \, \exp\left[ -\int_{\tilde{k}_{\perp}>k_{\perp}} dP^{'}_{E}(\tilde{k}_{\perp})\right], \label{eq:Deltap_resummation_Hoeche_app}
\end{align}
which coincides with Eq.~\eqref{eq:exchange_momentum}.
We have checked that to take the max in Eq.~\eqref{eq:Delta_p_Hoeche_app} instead of the sum of the $\Delta p_i$ in Eq.~\eqref{eq_Delta_p_sum_Delta_pi} can lead to the underestimation of the final $\left< \Delta p \right>$ by a factor $\sim \mathcal{O}(30)$ in the limit of large supercooling $T_{\rm nuc}/T_{\rm start} \sim 10^{-4}$ and large $\zeta_a \sim 0.1$.

\paragraph{Conclusion.}

We conclude that the assumption $p_a = p_b+p_c$ at the level of the vertex leads the authors of \cite{Hoche:2020ysm} to Eq.~\eqref{eq:splitting_function} instead of Eq.~\eqref{eq:dP_abc} for massive vector boson, or instead of Eq.~\eqref{eq:dP_abc_massless} for massless vector boson,\footnote{The authors of \cite{Hoche:2020ysm} consider the case of a massless vector boson $m_c=0$.} which overestimates the final $\left< \Delta p \right>$ by an amount $E_a/m_c$ or $(E_a/m_b)^2$ respectively, and does not vanish in the limit $\left<\phi \right> \to 0$.
Additionally, the approximation in Eq.~\eqref{eq:Delta_p_Hoeche_app} instead of Eq.~\eqref{eq_Delta_p_sum_Delta_pi} leads the authors of \cite{Hoche:2020ysm} to underestimate $\left< \Delta p \right>$ by an amount $\sim \mathcal{O}(30)$.

\subsection{Finite wall thickness}

Finally, let us add that if the result $\left< \Delta p \right> \sim \left< k_{\perp}^2/2xE_a \right>\sim \zeta_a \,E_a$ claimed by \cite{Hoche:2020ysm} was right, the infinitely-thin-wall assumption $L_{\rm w} \Delta p \lesssim 1$ would violently break down. 
Indeed, the perturbative splitting probability used by  \cite{Hoche:2020ysm}  and given in Eq.~\eqref{eq:splitting_function} should receive the additional factor $W_{\rm wall}(L_{\rm w},\,k_{\perp},\, x)$ defined in Eq.~\eqref{eq:Wwall}, which evaluated at $k_{\perp}^2/xE_a \sim \zeta_a \,E_a$, is exponentially suppressed
\begin{equation}
W_{\rm wall} \propto e^{-\pi L_{\rm w} E_a} \sim e^{-\pi \frac{E_a}{m_{c,h}}} \ll 1.
\end{equation}

\begin{figure}[t]
\centering
\begin{adjustbox}{max width=1\linewidth,center}
\raisebox{0cm}{\makebox{\includegraphics[ width=0.7\textwidth, scale=1]{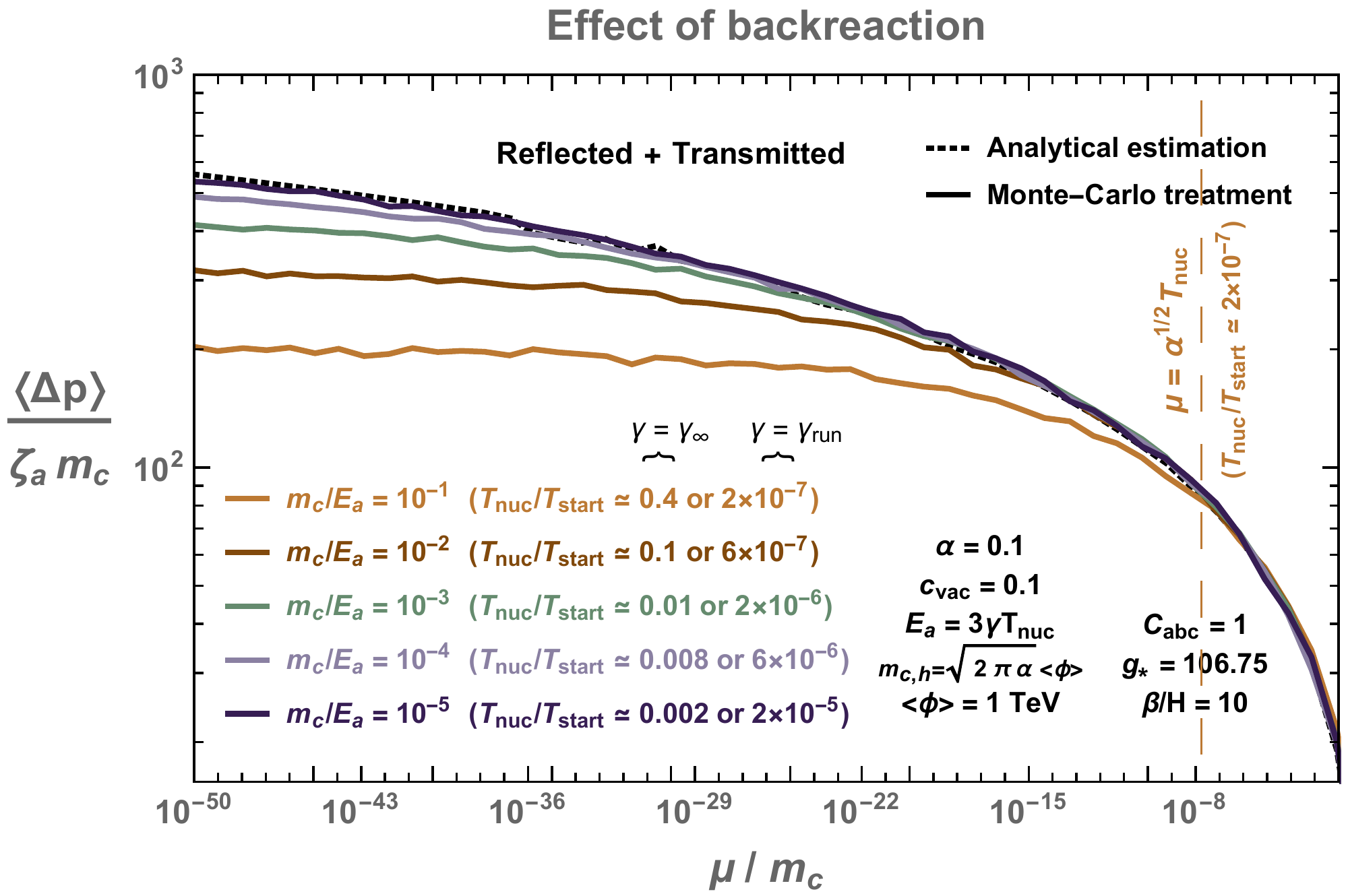}}}
\end{adjustbox}
\caption{\it \small  \label{fig:MC_simu_backreaction}
We display here the average momentum transferred to the wall $\Delta p$ as a function of the IR cut-off $\mu$, both in some units of $m_c$.
In contrast to the analytical result in \textbf{dotted},  cf. Eq~\eqref{eq:Delta_p_analytical}, which overestimates the exchanged momentum in the limit $\mu \to 0$, the MC simulation in \textbf{solid} accounts for the depletion of the incoming energy-momentum due to the successive boson emissions, see Sec.~\ref{par:backreaction}.
This implies a maximal number of boson emission and a maximal value for the exchanged momentum compatible with energy-momentum conservation.
The impact of energy depletion is higher and higher as we decrease the incoming energy $E_a/m_c$. For simplicity, $m_c = m_{c,h}$ in this plot.}
\end{figure}

\begin{figure}[t]
\centering
\begin{adjustbox}{max width=1\linewidth,center}
\raisebox{0cm}{\makebox{\includegraphics[ width=0.7\textwidth, scale=1]{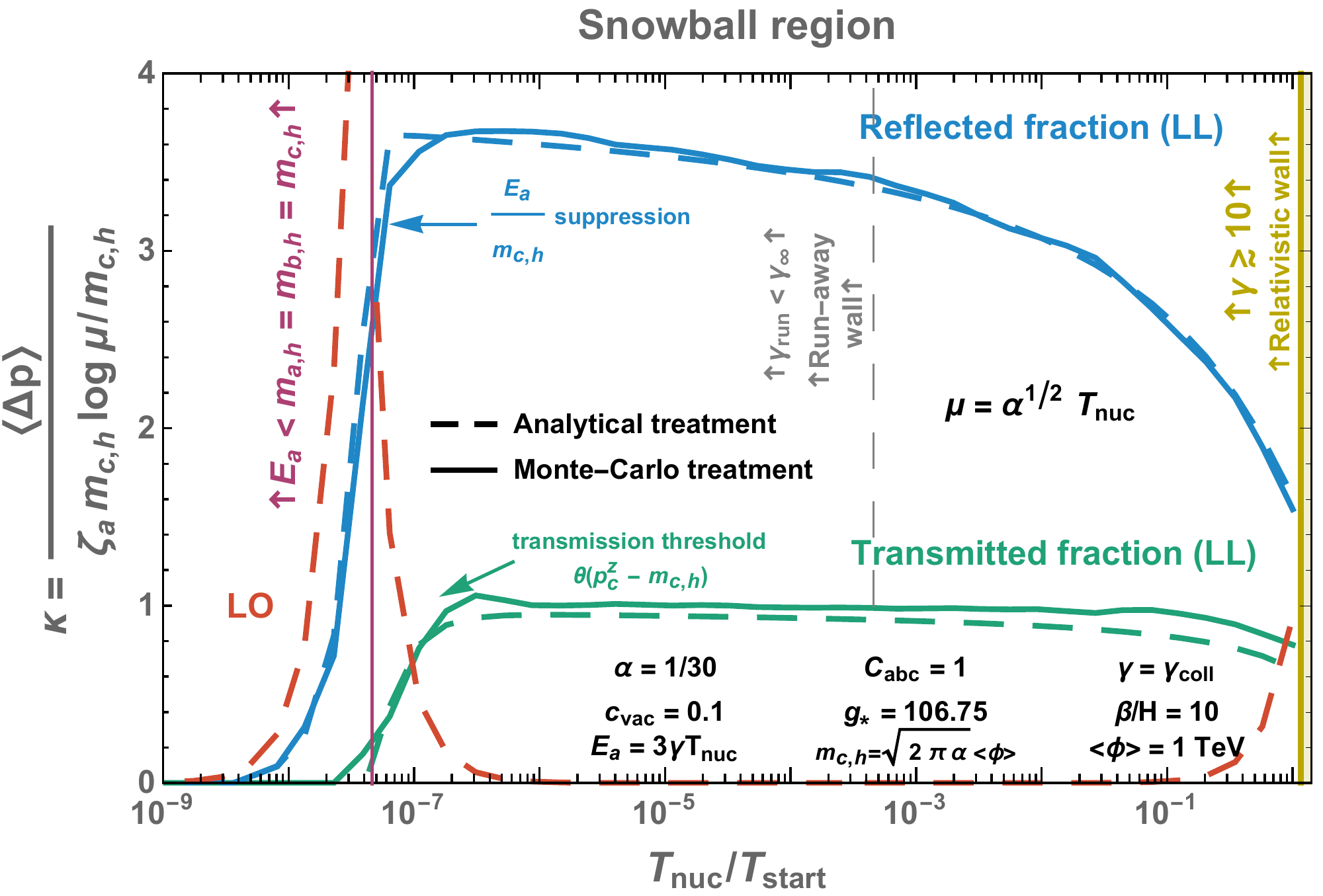}}}
\end{adjustbox}
\caption{\it \small  \label{fig:Xtrem_SC_limit}
Comparison between analytical treatment in \textbf{dashed} lines and MC treatment in \textbf{solid} lines.
The blue and green lines show the exchanged momentum induced by radiated bosons resummed at all leading-log orders, either transmitted (green) or reflected (blue) by the wall boundary.
The red dashed line shows the LO contribution assuming no splitting radiation.
In the run-away regime, we have $E_a/m_{c,h} \sim (M_{\rm pl} / \left<\phi\right>) (T_{\rm nuc}/T_{\rm start})^2$, see Eq.~\eqref{eq:thermal_initial_energy} and Eq.~\eqref{eq:gamma_run}.
Therefore, at very strong supercooling  the initial particle energy $E_a$ in the wall frame becomes smaller than the particle masses $m_{a,h}$, $m_{b,h}$ and/or $m_{c,h}$ in the broken phase, such that $a$, $b$ and/or $c$ are reflected.
}
\end{figure}

\section{Energy-momentum conservation}
\label{app:backreaction}

\subsection{Backreaction}
\label{sec:backreaction_num}

In the Sudakov resummation in Sec.~\ref{eq:sudakov_resummation}, we have neglected the depletion of the initial energy and momentum $(E_b,\,p_b)$ due to the multiple boson emissions.
When considering those effects, which we call `backreaction', Eq.~\eqref{eq:Deltap_resummation} should instead become
\begin{align}
\left< \Delta p \right> =  \sum_{n=0}^{\infty}\frac{1}{n!}\left[ \prod_{j=1}^{n} \int dP_{E,\,j} \Theta\big((1-X)^2E_a^2-K_{\perp}^2 \big) \Theta\big(1-X\big) \right] \sum_{i=1}^n \Delta p_i ~\exp\left[-\int dP_{E} \right] . \label{eq:Deltap_resummation_backreaction}
\end{align}
where $X$ and $K_{\perp}$ are defined in Eq.~\eqref{eq:def_X_Kperp}.
The $\Theta$ functions prevent the emission of more momentum and energy than what is available, i.e the momentum $p_b^z$ and energy $E_b$ in Eq.~\eqref{eq:pb} must remain positive.
We account for the effect of backreaction with our Monte-Carlo simulation, cf. Sec.~\ref{sec:MC}, in which the Theta functions in Eq.~\eqref{eq:Deltap_resummation_backreaction} enter through the boundaries of the integral in Eq.~\eqref{eq:x_distrib_MC_backreaction} and through the condition in Eq.~\eqref{eq:backreaction_stop}.
 
In Fig.~\ref{fig:MC_simu_backreaction}, we can see that in contrast to the analytical result (dotted line), in the numerical study in which the effect of energy depletion due to successive emissions is included (solid lines), the number of emissions and the resulting $\left< \Delta p\right>$ saturates to a plateau.
This could by itself regulate the logarithmic divergence if the thermal mass as well as other possible IR cut-offs were zero, see Sec.~\ref{sec:IR_cut_off}.
However, in presence of the thermally induced cut-off $\mu = \alpha^{1/2} T_{\rm nuc}$, we find that backreaction can be safely neglected.

\subsection{Snowplow region: $E_a \lesssim m_{i,h}$}
\label{sec:snowball}

The expressions for the friction pressure, $\mathcal{P}_{\LO}$ and $\mathcal{P}_{\LL}$, induced by $1\to 1$  and $1\to n$ processes, cf. Eq.~\eqref{eq:PLO} and Eq.~\eqref{eq:PNLO}, are valid in the regime $m_{a,h},\,m_{b,h} \ll E_a$ and $\mu \ll m_{c,h} \ll E_a$, respectively.
We now derive their expression in the regime where $E_a \lesssim m_{a,h},\,m_{b,h} $ and $E_a \lesssim m_{c,h}$, in which the particles $a$, $b$ and $c$ are reflected by the wall boundary.
Since the particles then accumulate in front of the wall, we call this the `snowplow region'.\footnote{
We thank Gilad Perez for suggesting the name `bulldozzer', we end up choosing `snowplow' due to proximity with the Christmas season at the time of arXiv first submission.
}
Assuming for the sake of simplicity that $a$ and $b$ are identical, the exchanged momentum between $a$ and the wall is given by $\left<\Delta p \right> = \left<2 p_a \right>$, and the friction pressure at LO in the relativistic limit is, cf. Eq.~\eqref{eq:PLO_def}
\begin{equation}
\label{eq:PLO_snowball}
E_a \lesssim m_{a,h} \qquad \implies \qquad \mathcal{P}_{\LO} \simeq \frac{\pi^2}{15} \sum_a g_a e_a~ \gamma^2 \,\Tnuc^4, \qquad e_a = 1~(7/8)~ \textrm{for bosons (fermions)}
\end{equation}
The exchanged momentum due to splitting radiation resummed at all leading-log orders is, cf. Eq.~\eqref{eq:range_x_K_perp}, Eq.~\eqref{eq:def_dPE} and Eq.~\eqref{eq:Deltap_resummation}
\begin{align}
\label{eq:rough_estimate_reflected}
\left< \Delta p \right> \simeq \int_{0}^{\frac{E_a^2}{4}}\frac{dk_{\perp}^2}{k_{\perp}^2} \,
\int_{1}^{\frac{\sqrt{k_{\perp}^2+m_{c,h}^2}}{E_{a}}}\frac{dx}{x}\,\frac{k_{\perp}^4}{(k_{\perp}^2+\mu^2)^2}\,  2xE_a  \simeq  4E_a \log \left( \frac{E_a}{\mu}\right),
\end{align}
which after injection into Eq.~\eqref{eq:PNLO_formula} with $E_a = \left< p_a\right> \simeq 2.7 \,\gamma\, \Tnuc$, leads to the friction pressure\footnote{Note that we recover the scaling $\mathcal{P}_\LL \propto \alpha \gamma^2 \,T_{\rm nuc}^4$ found in \cite{Hoche:2020ysm}, but only in the limit $E_a \lesssim m_{\rm c,h}$.}
\begin{align}
\label{eq:PLL_snowball}
E_a \lesssim m_{c,h} \qquad \implies \qquad\mathcal{P}_\LL \simeq \frac{11\,\zeta(3)}{\pi^3}\left[\sum_{a,b,c} \nu_a g_a C_{abc}\right] \alpha\,\text{ln} \left(\dfrac{E_a}{\mu}\right)\,\gamma^2 \,T_{\rm nuc}^4.
\end{align}
The exchanged momentum due to the  Eq.~\eqref{eq:rough_estimate_reflected} is shown by the dashed blue line in Fig.~\ref{fig:Xtrem_SC_limit}.
Note that the region $ E_a \lesssim m_{c,h} \sim m_{\rm a,h}$ falls in the region where interactions between reflected particles and newly incoming particles can be neglected, cf. dotted purple and blue lines in Fig.~\ref{fig:bubble_wall_velocity_phi}, such that the possible corrections to the friction pressure discussed in Sec.~\ref{sec:reflected_bosons} are not expected. 
It is interesting that a region $E_a \lesssim m_{i,h}$, where particles are too soft to penetrate inside the bubble wall, exists in the large supercooling limit (as already pointed out in \cite{Baldes:2020kam} for confining PTs).
We leave its study for further works.

\FloatBarrier

\section{Lorentz factor in the run-away regime}
\label{app:run_away_Lorentz_factor}

\subsection{Generic potential}

\paragraph{Energy conservation.}

The vacuum energy gained upon formation of a bubble of radius $R$ is
\beq
\label{eq:Ebubble}
E_\text{bubble} \simeq \frac{4}{3}\, \pi \,R^3 ~\mathcal{P},
\eeq
where $\mathcal{P}$ is the expanding pressure.
As the bubble grows, the energy stored in the bubble wall is given by
\beq
\label{eq:Ewall}
E_\text{wall} \simeq 4 \, \pi\, R^2 \,\gamma~ \sigma,
\eeq
where $\sigma$ is the surface energy of the wall in the wall frame
\begin{equation}
\sigma \equiv \int_0^\infty dr \, \left[ \frac{1}{2}(\phi^{'}(r))^2 + V(\phi(r)) - V(\phi(0)) \right].
\end{equation}
Insuring energy conservation by equating Eq.~\eqref{eq:Ebubble} and Eq.~\eqref{eq:Ewall}, we obtain that in the run-away regime, the wall Lorentz factor grows linearly with the bubble radius, e.g. \cite{Ellis:2019oqb,Cai:2020djd}
\begin{equation}
\label{eq:gamma_run_away_000}
\gamma = \frac{ \mathcal{P} }{3\,\sigma}R .
\end{equation}
In the supercooled and run-away regime, the expanding pressure is given by the vacuum energy difference $\mathcal{P} = c_{\rm vac} \left<\phi\right>^4$ while the surface tension can be estimated as
\begin{equation}
\sigma \approx L_{\rm w}^{\rm tot.} c_{\rm vac} \left<\phi\right>^4,
\end{equation}
where $L_{\rm w}^{\rm tot.}$ is the total thickness of the wall, accounting for the rising part plus the multiple oscillations around the true vacuum value, until the oscillations become negligible.
This should not be mistaken with $L_{\rm w}$ introduced earlier, which is the thickness of the rising part of the wall only, see footnote~\ref{footnote:multiple_wall_oscillation}.
The Lorentz factor of the running-away bubble wall in Eq.~\eqref{eq:gamma_run_away_000} becomes 
\begin{equation}
\label{eq:gamma_run_away_00}
\gamma = \frac{ R }{3\,L_{\rm w}^{\rm tot.}}.
\end{equation}

\paragraph{Total wall thickness.}

The wall profile can be found by sticking the space-like profile with the time-like profile, see e.g. \cite{Jinno:2019bxw}.
The former is solution of the euclidean equation of motion
\begin{equation}
\phi_{\rm E}''(s_{\rm E}) + \frac{d-1}{s_{\rm E}}\phi_{\rm E}'(s_{\rm E}) = \frac{dV}{d\phi_{\rm E}},\qquad \textrm{with} \quad\phi_{\rm E}^{'}(0)=0, \quad \textrm{and} \quad  \lim_{r \to \infty} \phi_{\rm E}(r) = 0.
\label{eq:bounce_eom}
\end{equation}
where $s_{\rm E} = \sqrt{\vec{r}^2+t_E^2} = \sqrt{\vec{r}^2-t^2}$ is the space-like light-cone coordinate, $t_E = i\,t$ is the Euclidean time and $d = 3$ or $4$.
The latter is solution of the real-time equation of motion
\begin{equation}
\label{eq:scalar-time-like}
\frac{\partial^2 \phi}{\partial s^2} +\frac{3}{s}\frac{\partial \phi}{\partial s} +\frac{\partial V}{\partial \phi} = 0,\qquad \textrm{with} \quad\phi(s)=\phi_{\rm E}(0), \quad \textrm{and} \quad \phi^{'}(0) = \phi_{\rm E}^{'}(0).
\end{equation}
where $s = \sqrt{t^2-\vec{r}^2}$ is now the time-like light-cone coordinate.
The total wall thickness $L_{\rm w}^{\rm tot.}$ is set by the sum of the characteristic length scales of the two profiles
\begin{equation}
 L_{\rm w}^{\rm tot.} = L_{\rm space-like} + L_{\rm time-like},
\end{equation}
which are themselves set by the damping terms of the two respective Eqs.~\eqref{eq:bounce_eom} and \eqref{eq:scalar-time-like}.
Due to the matching condition  $\phi^{'}(0) = \phi_{\rm E}^{'}(0)$, the two damping terms are of the same order and we conclude that
\begin{equation}
L_{\rm time-like} \simeq L_{\rm space-like} \equiv R_{\rm nuc}, 
\end{equation}
where the last equality defines the bubble radius at nucleation.
The Lorentz factor of the running-away bubble wall in Eq.~\eqref{eq:gamma_run_away_00} becomes\footnote{
In \cite{Ellis:2019oqb}, the wall tension is expressed in term of the bubble radius at nucleation $R_{\rm nuc}$, obtained from minimizing the total energy $E_\text{bubble}+E_\text{wall}$
\begin{equation}
\sigma  = R_{\rm nuc} \mathcal{P}/2, \label{eq:surface_tension_nuc}
\end{equation}
such that the Lorentz factor of the running-away bubble wall in Eq.~\eqref{eq:gamma_run_away_000} becomes 
\begin{equation}
\label{eq:gamma_run_away_1}
\gamma = \frac{ 2R }{3 R_{\rm nuc}}.
\end{equation}
However, as the bubble expands, the scalar field undergoes damped oscillations toward the true vacuum, such that Eq.~\eqref{eq:surface_tension_nuc} under-estimates the surface tension $\sigma$.
With Eq.~\eqref{eq:gamma_run_away_0}, we claim that Eq.~\eqref{eq:surface_tension_nuc} only under-estimates $\sigma$ by an $\mathcal{O}(1)$ factor.
}
\begin{equation}
\label{eq:gamma_run_away_0}
\gamma = \frac{ R }{3R_{\rm nuc}}.
\end{equation}

\subsection{Shallow potential}

The friction pressure in Eq.~\eqref{eq:Ptot} is suppressed by powers of $T_{\rm nuc}/f \ll 1$, where $f\equiv \left<\phi\right>$.
Therefore, a run-away regime at bubble collision time requires supercooled phase transitions. 
Typically, supercooled phase transitions are generated by shallow zero-temperature potentials, namely potentials where the curvature close to the false vacuum
\begin{equation}
\sqrt{V''(\phi)}\Big|_{\phi  \ll f} \simeq f\,\exp(-c/\epsilon),\qquad \epsilon \ll 1, \quad c=O(1),
\end{equation}
is much smaller than the curvature $f$ close to the true vacuum.
In that case, the tunneling exit point is very close to the false minimum and the bounce action is only sensitive to the scale $f\,\exp(-c/\epsilon)$. We expect both temperature and bubble radius at nucleation to be related to that scale, and therefore\footnote{
Note that the authors of \cite{Caprini:2019egz} have set $R_{\rm nuc} \sim f^{-1}$ instead of $R_{\rm nuc} \sim T_{\rm nuc}^{-1}$. For typical potentials leading to large supercooling, e.g. Coleman-Weinberg \cite{Baldes:2018emh} or light-dilation \cite{Bruggisser:2018mrt}, the latter choice is the correct one, see App.~A of \cite{Baldes:2020kam}.
}
\begin{equation}
R_{\rm nuc} \simeq c_{\rm w}\,T_{\rm nuc}^{-1},\label{eq:Rnuc}
\end{equation} 
where $c_{\rm w}$ is a model-dependent numerical factor.
At the time of collision, the bubble size is given by\footnote{
The factor $8 \pi$ comes from the expected number of bubbles nucleating in a volume $V$.
To derive this, one has to take into account the fact that the nucleating bubble must be in the false vacuum.
Taking the wall velocity to be unity for simplicity, and taking the nucleation rate per unit time and volume to be $\Gamma (t) = \Gamma_* e^{\beta t}$, we sum up the differential nucleation probability for time interval $[t_n, t_n + dt_n]$ to get
(expected $\#$ of bubbles) $= V \times \sum$ (prob. for the nucleation point to be in the false vacuum) $\times$ (prob. for nucleation between $[t_n, t_n + d t_n]$) $= V \times \int_{- \infty}^\infty \, e^{- \frac{4 \pi}{3} \int_{- \infty}^{t_n} dt \, (t_n - t)^3 \Gamma_* e^{\beta t}} \times \Gamma_* e^{\beta t_n} dt_n = V \times \int_{- \infty}^\infty \, e^{- 8 \pi \Gamma_* e^{\beta t_n} / \beta^4} \times \Gamma_* e^{\beta t_n} dt_n = \beta^3 V / 8 \pi$.
}
\begin{equation}
R_{\rm coll} \simeq (8\pi)^{1/3}\,v_w \beta^{-1},
\end{equation} 
where $v_w$ is the bubble wall velocity, $\beta$ is the expansion coefficient of the exponent of the nucleation rate around the typical transition time $t_*$
\begin{align}
\Gamma
&\propto
e^{\beta (t - t_*) + \cdots},
\label{eq:beta}
\end{align}
which gives the inverse of the bubble propagation time.
Therefore, in the run-away regime ($v_w =1$ and $\mathcal{P}>0$), the Lorentz factor at the time of collision in Eq.~\eqref{eq:gamma_run_away_0} becomes
\begin{equation}
\label{eq:gamma_run_away_app}
\gamma_{\rm run}  \simeq ~ \frac{T_{\rm nuc}}{c_{\rm w} \beta},
\end{equation}
which we have use in the main text, cf. Eq.~\eqref{eq:gamma_run}.

\bibliographystyle{JHEP}
\bibliography{NLO_pressure}

\end{document}